\documentclass[aps,prx,amsmath,floats,floatfix,twocolumn,
  superscriptaddress,nofootinbib,showpacs]{revtex4-1}

\pdfoutput=1 % Specifiy PDFLaTeX as per arXiv request. This has to be in the first 5 lines!
\usepackage{diagbox}
\usepackage{multirow}
\usepackage{braket}
\usepackage{amsfonts}
\usepackage{amsmath}
\usepackage{amssymb}
\usepackage{amsthm}
\usepackage{bm}
\usepackage{dcolumn}
\usepackage{epsfig}
\usepackage{graphicx}
\usepackage{graphics}
\usepackage[latin1]{inputenc}
\usepackage{latexsym}
\usepackage{rotating}
\usepackage[dvipsnames]{xcolor}
\usepackage{mathrsfs}
\usepackage{microtype}
\usepackage{verbatim}
\usepackage{url}

\usepackage[caption=false]{subfig}
\usepackage[normalem]{ulem}
\usepackage{tikz}

\usepackage[breaklinks=true]{hyperref}
\hypersetup{
    colorlinks=true,
    linkcolor=NavyBlue,
    filecolor=Magenta,
    urlcolor=NavyBlue,
    citecolor=NavyBlue
}

\usepackage{yfonts}
\usepackage{float}
\usepackage{xspace} % Sensible space treatment at end of simple macros
\usepackage{mathrsfs}
\usepackage[toc,page]{appendix}
\usepackage{siunitx}
\usepackage{array}
\usepackage[normalem]{ulem}

\newcommand{\CCK}{{\mbox{\tiny CCK}}}
\newcommand{\Hertz}{{\mbox{\tiny H}}}
\newcommand{\be}{\begin{equation}}
\newcommand{\ee}{\end{equation}}

\newcommand{\geo}{{\mbox{\tiny geo}}}
\newcommand{\typeD}{{\mbox{\tiny D}}}
\newcommand{\nonD}{{\mbox{\tiny non-D}}}

\newcommand{\matter}{{\mbox{\tiny matter}}}

\newcommand{\ba}{\begin{align}}
\newcommand{\ea}{\end{align}}

\makeatletter
\newcommand*{\rom}[1]{\expandafter\@slowromancap\romannumeral #1@}
\makeatother
\allowdisplaybreaks

\AtBeginDocument{%
    \newwrite\bibnotes
    \def\bibnotesext{Notes.bib}
    \immediate\openout\bibnotes=\jobname\bibnotesext
    \immediate\write\bibnotes{@CONTROL{REVTEX41Control}}
    \immediate\write\bibnotes{@CONTROL{%
    apsrev41Control,author="08",editor="1",pages="1",title="0",year="1"}}
    \if@filesw
    \immediate\write\@auxout{\string\citation{apsrev41Control}}%
    \fi
}
\newcommand{\UIUC}{Illinois  Center  for  Advanced  Studies  of  the  Universe \&
Department of Physics, University of Illinois at Urbana-Champaign, Urbana, Illinois 61801, USA}
\newcommand{\CAL}{Theoretical Astrophysics 350-17, California Institute of Technology, Pasadena, CA 91125, USA}
\newcommand{\AEI}{Max Planck Institute for Gravitational Physics (Albert Einstein Institute), D-14476 Potsdam, Germany}

\begin{document}

    \title{Perturbations of spinning black holes in dynamical Chern-Simons gravity \\ I. Slow rotation equations}
	
    \author{Pratik Wagle}
    \email{pratik.wagle@aei.mpg.de}
    \affiliation{\UIUC}
    \affiliation{\AEI}

    \author{Dongjun Li}
    \email{dlli@caltech.edu}
    \affiliation{\CAL}

    \author{Yanbei Chen}
    \affiliation{\CAL}
    
    \author{Nicol\'as Yunes}
    \affiliation{\UIUC}

    \date{\today}

    \begin{abstract}
    The detection of gravitational waves resulting from the coalescence of binary black holes by the LIGO-Virgo-Kagra collaboration has inaugurated a new era in gravitational physics. These gravitational waves provide a unique opportunity to test Einstein's general relativity and its modifications in the regime of extreme gravity. A significant aspect of such tests involves the study of the ringdown phase of gravitational waves from binary black-hole coalescence, which can be decomposed into a superposition of various quasinormal modes. In general relativity, the spectra of quasinormal modes depend on the mass, spin, and charge of the final black hole, but they can also be influenced by additional properties of the black hole spacetime, as well as corrections to the general theory of relativity. In this work, we focus on a specific modified theory known as dynamical Chern-Simons gravity, and we employ the modified Teukolsky formalism developed in a previous study to investigate perturbations of slowly rotating black holes admitted by the theory. Specifically, we derive the master equations for the $\Psi_0$ and $\Psi_4$ Weyl scalar perturbations that characterize the radiative part of gravitational perturbations, as well as the master equation for the scalar field perturbations. We employ metric reconstruction techniques to obtain explicit expressions for all relevant quantities. Finally, by leveraging the properties of spin-weighted spheroidal harmonics to eliminate the angular dependence from the evolution equations, we derive two, radial, second-order, ordinary differential equations for $\Psi_0$ and $\Psi_4$, respectively. These two equations are coupled to another radial, second-order, ordinary differential equation for the scalar field perturbations. 
    This work is the first attempt to derive a master equation for black holes in dynamical Chern-Simons gravity using curvature perturbations. The master equations we obtain can now be numerically integrated to obtain the quasinormal mode spectrum of slowly rotating black holes in this theory, making progress in the study of ringdown in dynamical Chern-Simons gravity.
    \end{abstract}

    \maketitle

%%%%%%%%%%%%%%%%%%%%%%%%%%%%%%%%%%%%%%%%%%%%%%%%%%%%%%%%%%%%
%%%%%%%%%%%%%%%%%%%%%%%%%%%%%%%%%%%%%%%%%%%%%%%%%%%%%%%%%%%%
\section{Introduction} 
\label{sec:introduction}

The discovery of gravitational waves (GWs) has provided a new avenue for scrutinizing the predictions and phenomenology of Einstein's theory of general relativity (GR) in regimes characterized by non-linear and dynamic gravitational effects~\cite{Abbott:2016blz, GBM:2017lvd}. GWs offer the opportunity to investigate the properties of astrophysical objects where gravity is notably intense, such as black holes (BHs) and neutron stars (NSs). In particular, GWs are often generated by the coalescence of binary BH systems, wherein two BHs orbit each other, gradually inspiraling due to the emission of GWs, and ultimately merging to produce a final BH that emits GW radiation as it settles down. GWs during this part of the coalescence, known as the ringdown phase, comprise a superposition of numerous quasinormal modes (QNMs), each with a complex eigenfrequency. By analyzing GW observations, it may thus become possible to efficiently explore the distinctive spectrum of QNMs exhibited by BHs as they ring down~\cite{Berti:2015itd, Berti:2009kk}.

QNMs are the characteristic vibrational modes of BHs that are excited when the BH is perturbed. The study of QNMs can provide important information about the fundamental properties of BHs and their surrounding spacetime. In particular, the QNM spectrum of astrophysical (i.e.,~uncharged) BHs in GR is fully determined by just two parameters: the mass and spin of the remnant BH. One promising application of QNMs is to test modified gravity theories and alternative models of compact objects~\cite{Yunes:2013dva}. The predictions of modified gravity theories may deviate from GR's and can manifest as modifications of the QNM spectrum of BHs~\cite{Cardoso:2009pk, Molina:2010fb, Blazquez-Salcedo:2016enn, Wagle:2021tam, Pierini:2021jxd, Pierini:2022eim, Cano:2020cao}. By observing the QNM spectrum of merging BHs with GW detectors, it may be possible to place constraints on these modifications to GR and its models given two or more detections of QNM frequencies with a strong signal-to-noise ratio. Recent GW detections of binary BH mergers have provided some of the most precise tests of GR in the strong-field regime~\cite{TheLIGOScientific:2016pea}. With improvements in detector technology and advancements in computational techniques, using ringdown to test modified gravity theories seems possible in the near future.

To study QNMs in GR, there are two well-established approaches within BH perturbation theory. One approach, proposed by Regge and Wheeler~\cite{Regge:PhysRev.108.1063}, perturbs the metric directly and it has been successfully applied to non-rotating and slowly rotating BHs in GR~\cite{Regge:PhysRev.108.1063, Zerilli:1971wd, Moncrief:1974am, Pani:2013pma}, as well as in modified theories of gravity~\cite{Cardoso:2009pk, Molina:2010fb, Blazquez-Salcedo:2016enn, Wagle:2021tam, Pierini:2021jxd, Pierini:2022eim, Cano:2020cao}. However, this approach has not been successfully applied to BHs with a general spin. An alternative approach uses curvature perturbations, and it was presented by Teukolsky~\cite{Teukolsky:1973ha} to study rotating Kerr BHs in GR, including their QNM spectrum and dynamical stability \cite{Teukolsky:1974yv}. This framework uses the Newman-Penrose (NP) formalism~\cite{Newman:1961qr} and considers curvature perturbations characterized by quantities known as Weyl scalars. The success of the Teukolsky formalism lies in its ability to provide a decoupled evolution equation for each of the $\Psi_0$ and $\Psi_4$ Weyl scalars, which describe transverse gravitational perturbations, and physically represent ingoing and outgoing GWs, respectively. Not only are these quantities decoupled from other gravitational degrees of freedom, their evolution equations are also separable into a radial and an angular equation~\cite{Teukolsky:1974yv, Leaver:1985ax, Berti:2009kk}.

The Teukolsky formalism in GR requires the background geometry to be algebraically of type D under the Petrov classification \cite{Petrov:2000bs}, as is the case for Schwarzschild and Kerr spacetimes in GR. This Petrov type D property implies that four of the five Weyl scalars vanish on the background. However, when considering beyond-GR (bGR) theories, the deviations introduced may lead to BHs described by non-Petrov-type-D spacetimes. For instance, rotating BHs in dynamical Chern-Simons (dCS) gravity and Einstein-dilaton Gauss-Bonnet gravity are algebraically general and classified as Petrov type I. As a consequence, the Teukolsky formalism cannot be directly applied to bGR BH spacetimes. Therefore, calculating the QNM spectra of spinning BHs in bGR theories has been an open problem for a long time and warrants new approaches. One potential resolution to this problem became available recently with the development of the modified Teukolsky formalism~\cite{Li:2022pcy, Hussain:2022ins, Cano:2023tmv, Cano:2023jbk}. This formalism, in theory, enables studying perturbations of spinning BHs in bGR theories and calculating the QNM spectra of such non-Ricci-flat, matter vacuum Petrov type I BH spacetimes. Yet, for tests with GW data, a key theoretical challenge remains to calculate the QNM spectra of BHs in modified theories, and then compare them with observations. This work aims at computing the QNM spectrum of one type of BH spacetimes. 

In this work, we restrict our attention to modifications to GR where a scalar field is non-minimally coupled to topological invariant quadratic terms in curvature. A subset of this class of theories, known as dCS gravity~\cite{Jackiw:2003pm, Alexander:2009tp}, was proposed to explain the matter-antimatter asymmetry of the universe. This is achieved by the introduction of additional parity-violating gravitational interactions, challenging a fundamental pillar of GR. Due to the quadratic nature of this theory, it is not strongly constrained using weak field tests and evades binary pulsar tests~\cite{Wagle:2018tyk} and GW polarization tests~\cite{Wagle:2019mdq}. However, early work using the metric perturbation approach shows promise that ringdown tests can be useful in constraining this modified theory of gravity \cite{Wagle:2021tam}.

We focus on slowly rotating BHs in dCS gravity to leading order in spin in this work. This calculation serves as a validation of the newly developed formalism~\cite{Li:2022pcy}, as the results herein can then be confirmed with the results obtained using the metric perturbation approach~\cite{Wagle:2021tam, Srivastava:2021imr}. We first use the formalism prescribed in~\cite{Li:2022pcy} to obtain the modified Teukolsky equation for a slowly rotating BH in dCS gravity. To leading order in spin, these BH backgrounds are described by a non-Ricci-flat, matter vacuum Petrov type D spacetime~\cite{Owen:2021eez, Li:2022pcy}. Solving the Bianchi identities, we obtain the modified Teukolsky equation first in the null NP basis. We then rewrite this equation in the coordinate basis by defining a tetrad (similar to the Kinnersly tetrad in GR). Finally, we make use of the properties of spin-weighted spheroidal harmonics to eliminate the angular dependence and obtain a radial second-order differential equation. Due to the non-minimal coupling between the scalar field and the curvature, the perturbed master equations of the scalar field and the Weyl scalars ($\Psi_0$ or $\Psi_4$) are coupled in dCS gravity. Moreover, some of these quantities also require metric reconstruction within GR \cite{Chrzanowski:1975wv, Cohen_Kegeles_1975, Keidl_Friedman_Wiseman_2007, Keidl_Shah_Friedman_Kim_Price_2010, Loutrel_Ripley_Giorgi_Pretorius_2020}, making the problem more challenging, yet still solvable, as we demonstrate here. Having obtained the master equations in this work, we will use the eigenvalue perturbation method \cite{Zimmerman:2014aha, Mark_Yang_Zimmerman_Chen_2015, Hussain:2022ins} in future work to calculate the QNM spectrum and compare it with the results using metric perturbations in dCS gravity~\cite{Wagle:2021tam, Srivastava:2021imr}. 

The remainder of this paper is organized as follows. We first present in brief the action and the field equations of dCS gravity in Sec.~\ref{sec:dCS_intro} along with the slowly rotating BH solutions in this theory. In Sec.~\ref{sec:BH_perturbation}, we present an overview of the modified Teukolsky formalism in~\cite{Li:2022pcy}, define a three-parameter expansion under the slow-rotation approximation, and calculate the NP quantities on the dCS BH background. In Sec.~\ref{sec:metric_reconstruction}, we provide a concise review of the metric reconstruction procedures in GR. In Secs.~\ref{sec:source_scalar}, \ref{sec:source_Teuk}, and \ref{sec:source_Teuk_ORG}, we calculate the source terms of the scalar field equation and the modified Teukolsky equation of $\Psi_0$ and $\Psi_4$ in the null NP basis, the results of which are summarized in Sec.~\ref{sec:executive_summary_Teuk}. In Sec.~\ref{sec:separation_variable}, we project the equations into the coordinate basis and extract their radial parts using the properties of spin-weighted spheroidal harmonics. Finally, in Sec.~\ref{sec:discuss}, we summarize our work and discuss some future avenues. Henceforth, we adopt the following conventions unless stated otherwise: we work in 4-dimensions with metric signature $(-,+,+,+)$ as in~\cite{Misner:1974qy}. For all NP quantities except the metric signature, we use the notation adapted by Chandrasekhar in~\cite{Chandrasekhar_1983}.

%%%%%%%%%%%%%%%%%%%%%%%%%%%%%%%%%%%%%%%%%%%%%%%%%%%%%%%%%%%%
%%%%%%%%%%%%%%%%%%%%%%%%%%%%%%%%%%%%%%%%%%%%%%%%%%%%%%%%%%%%
\section{BHs in dCS Gravity} 
\label{sec:dCS_intro}
	
In this section, we will present the details of the theory and the background BH spacetime used in this work.

%%%%%%%%%%%%%%%%%%%%%%%%%%%%%%%%%%%%%%%%%%%%%%%%%%%%%%%%%%%%
\subsection{dCS gravity}
In this subsection, we briefly review the dCS gravity following the discussion and the convention in~\cite{Yagi:2012ya}. A more detailed review of dCS gravity can be found in \cite{Jackiw:2003pm, Alexander:2009tp}. The action of dCS gravity is
\begin{equation} \label{eq:action}
    \begin{aligned}
    S=\;\int d^{4}x\sqrt{-g}&\Big\{\kappa_{g}R
    +\frac{\alpha}{4}\vartheta R_{\nu\mu\rho\sigma}{}^*
    R^{\mu\nu\rho\sigma} \\ 
    \;&-\frac{1}{2}\left[\nabla_{\mu}\vartheta\nabla^{\mu}\vartheta
    +2 V(\vartheta)\right]+\mathcal{L}_{\matter}\Big\}\,,
    \end{aligned}
\end{equation}
where $\kappa_{g}=\frac{1}{16\pi G}$, $^*\!R^{\mu\nu\rho\sigma}$ is the dual of the Riemann tensor,
\begin{equation} \label{eq:Riemann_dual}
    ^*\!R^{\mu\nu\rho\sigma}
    =\frac{1}{2}\epsilon^{\rho\sigma\alpha\beta}
    R^{\mu\nu}{}_{\alpha\beta}\,,
\end{equation}
and $\vartheta$ is the pseudoscalar field coupled to the Pontryagin density $P := R_{\nu\mu\rho\sigma}{}^*\!R^{\mu\nu\rho\sigma}$ via the dCS coupling constant $\alpha$. The quantity $V(\vartheta)$ is a potential for $\vartheta$, which we set to zero for the reasons explained in~\cite{Yagi:2015oca,Yagi:2013mbt,Yagi:2011xp}, along with any matter contribution $\mathcal{L}_{\matter}$ (since we will work with matter vacuum BH spacetimes) for the remainder of this work. From Eq.~\eqref{eq:action}, we find that $[\vartheta]=1$ and $[\alpha]=L^2$ in geometric units. Using these coupling constants, we can define a dimensionless parameter $\zeta$ characterizing the strength of the dCS correction to GR, where $\zeta$ is defined in~\cite{Yagi:2012ya} to be
\begin{equation} \label{eq:zeta}
    \zeta\equiv\frac{\alpha^2}{\kappa_{g}M^4}\,,
\end{equation}
with $M$ the typical mass of the system. When the system under consideration contains a single black hole, then $M$ is its mass.
When considering a binary system, then different corrections to the solutions of the field equations will scale with different (dimension-4) combinations of the two masses.
	 
Varying the action in Eq.~\eqref{eq:action} with respect to the metric and the scalar fields, respectively, we obtain,
 \begin{align}
    R_{\mu\nu}=
    & \;-\frac{\alpha}{\kappa_{g}}C_{\mu\nu}
    +\frac{1}{2\kappa_{g}} \bar{T}_{\mu \nu}^{\vartheta}\,, 
    \label{eq:EOM_R} \\
    \square\vartheta=
    & \;-\frac{\alpha}{4}R_{\nu\mu\rho\sigma}{ }^{*} 
    R^{\mu\nu\rho\sigma}\,,
    \label{eq:EOM_theta}
\end{align}
where $\square=\nabla_{\mu}\nabla^{\mu}$ is the D'Alembertian operator, and
\begin{align} 
    C^{\mu\nu}\equiv
    & \;\left(\nabla_{\sigma}\vartheta\right)
    \epsilon^{\sigma\delta\alpha(\mu}\nabla_{\alpha}R^{\nu)\delta}
    +\left(\nabla_{\sigma}\nabla_{\delta}\vartheta\right)^*\! R^{\delta(\mu\nu)\sigma}\,, 
    \label{eq:Ctensor} \\
    \bar{T}_{\mu\nu}^{\vartheta}\equiv
    & \;\left(\nabla_{\mu}\vartheta\right)
    \left(\nabla_{\nu}\vartheta\right)\,. 
    \label{eq:T_theta}
\end{align}
We have here adopted the trace-reversed form of the field equations, using the fact that the C-tensor is traceless, as it will render future calculations simpler.

The dCS action presented above is an effective theory that includes only linear in $\alpha$ and quadratic in curvature corrections to the Einstein-Hilbert action, thus ignoring higher order terms in $\alpha$ and in curvature. Therefore, the resulting field equations are also similarly effective, and their solutions ought to be truncated at leading order in $\alpha$ and considered only for systems (and regimes of spacetime) with small curvatures. Various previous work~\cite{Yagi:2012ya, Alexander:2021ssr} have studied the regime of validity of this effective action and its curvature cutoff. In essence, the effective theory remains valid provided $(\alpha^2/\kappa_g) P^2 \ll 1$, where recall that $P$ has been defined as the Pontryagin density. When this is the case, the higher order in $\alpha$ and in curvature terms neglected in the action above can continue to be ignored. The systems studied in this paper involve the exterior spacetime of remnant BHs with masses in the range $3M_{\odot}<M<10^7M_{\odot}$. For such systems, the theory remains effective, and $\zeta \ll 1$ provided $\sqrt{\alpha}\ll10^7\mathrm{km}$, which will be assumed henceforth; the best current constraints on $\alpha$ come from NICER and advanced LIGO observations, and they require that $\alpha \leq 8.5 \mathrm{km}$ at $90\%$ confidence~\cite{Silva:2020acr}. In this range of $\alpha$ and for these BH masses, the quadratic curvature corrections to the Einstein Hilbert action will remain perturbative, and the higher-order in $\alpha$ and curvature terms will remain controlled relative to the quadratic term included in the dCS action.

%%%%%%%%%%%%%%%%%%%%%%%%%%%%%%%%%%%%%%%%%%%%%%%%%%%%%%%%%%%%
\subsection{Slowly rotating BHs in dCS gravity} 

Solutions to BHs in dCS gravity have been found both numerically~\cite{Delsate:2018ome} and analytically~\cite{Yunes:2009hc, Konno:2009kg, Yagi:2012ya, Maselli:2017kic}. For this work, we consider the analytical solution found in~\cite{Yunes:2009hc,Konno:2009kg}, which were obtained by perturbatively solving the field equations~\eqref{eq:EOM_R} and~\eqref{eq:EOM_theta} to linear order in both the dimensionless spin parameter $\chi$, where $\chi\equiv a/M$ (with the dimensional spin parameter $a = S/M$, where $S$ is the spin angular moment, and $M$ is the BH mass), and the dCS expansion parameter $\zeta$, defined in Eq.~\eqref{eq:zeta}. This analytical solution casts the line element of a  slowly rotating BH in dCS gravity as
\begin{equation} \label{eq:full_line_element}
    ds^2 = ds_{\mathrm{Kerr}}^2 + ds_{\mathrm{dCS}}^2 \,,
\end{equation}
where, following the convention in \cite{Chandrasekhar_1983}, the line element for the Kerr metric in Boyer-Lindquist coordinates $(t,r,\theta,\phi)$ is given by
\begin{equation} \label{eq:kerr_metric}
    \begin{aligned}
        ds_{\mathrm{Kerr}}^{2}=& ~ g^{\textrm{Kerr}}_{\alpha \beta} dx^\alpha dx^\beta \\
        =&-\left(1-\frac{2Mr}{\tilde{\rho}^2}\right)dt^{2}
        -\frac{4Mar\sin^{2}\theta}{\tilde{\rho}^2}dtd\phi +\frac{\tilde{\rho}^2}{\Delta} dr^{2} \\
        &+\tilde{\rho}^2d\theta^{2}+\left(r^{2}+a^{2}+\frac{2Ma^{2}r\sin^{2} \theta}{\tilde{\rho}^2}\right)\sin^{2}\theta d\phi^{2}\,,
    \end{aligned}
\end{equation}
with $\tilde{\rho}^2\equiv r^2+a^2\cos^2\theta$ and $\Delta\equiv r^2-2Mr+a^2$. 
To leading order in $\chi$ and $\zeta$, the dCS modification to the Kerr line element is given by
\begin{equation} \label{eq:metric_dCS}
    ds_{\mathrm{dCS}}^2 = - \zeta \chi \frac{\tilde{G}(r)}{2} \sin^2 \theta dt d\phi \,,
\end{equation}
where 
\begin{equation}
    \tilde{G}(r)=-\frac{5M^5}{4r^4}\left(1+\frac{12M}{7r}+\frac{27M^2}{10r^2}\right) \,.
\end{equation}
The above dCS correction to the line element is of $\mathcal{O}(\zeta^1,\chi^1,\epsilon^0)$, and thus, we can use the tri-variate notation that we will introduce in Sec.~\ref{sec:trivariate_expansion} to write
\begin{equation}
    h_{t\phi}^{(1,1,0)} = - \frac{\tilde{G}(r)}{2} \sin^2 \theta \,.
\end{equation}
The dCS metric of a slowly-rotating BH is then identical to the Kerr metric, except for the $(t,\phi)$ component, which acquires the correction presented above. Similarly, the background scalar field at $\mathcal{O}(\zeta^1,\chi^1,\epsilon^0)$ is given by
\begin{equation} \label{eq:scalar_stationary}
    \vartheta^{(1,1,0)}=\frac{5M^2}{32\sqrt{\pi}r^2}\left(1+\frac{2M}{r}+\frac{18M^2}{5r^2}\right)\cos{\theta}\,,
\end{equation}
where we have absorbed a factor of $\zeta^{1/2}$ into the expansion of $\vartheta$ as explained in more detail in \cite{Li:2022pcy}. 

%%%%%%%%%%%%%%%%%%%%%%%%%%%%%%%%%%%%%%%%%%%%%%%%%%%%%%%%%%%%
%%%%%%%%%%%%%%%%%%%%%%%%%%%%%%%%%%%%%%%%%%%%%%%%%%%%%%%%%%%%	
\section{BH Perturbations in Teukolsky formalism }
\label{sec:BH_perturbation}
	
In this section, we review the modified Teukolsky formalism developed in~\cite{Li:2022pcy}. In this paper, we extend the two-parameter expansion scheme in~\cite{Li:2022pcy} to a three-parameter expansion discussed in Sec.~\ref{sec:trivariate_expansion} to incorporate the slow-rotation approximation, following \cite{Wagle:2021tam, Srivastava:2021imr}.

%%%%%%%%%%%%%%%%%%%%%%%%%%%%%%%%%%%%%%%%%%%%%%%%%%%%%%%%%%%%
\subsection{Modified Teukolsky Equation} 
\label{sec:modified_Teuk_eqn}
	
As discussed previously in Sec.~\ref{sec:introduction}, for studying perturbations of non-rotating BHs, we obtain the perturbed field equations and decompose these into master equations by making use of the metric perturbations~\cite{Regge:PhysRev.108.1063, Zerilli:1971wd, Moncrief:1974am, Maggiore:2018sht}. These metric perturbations are separated into two sectors, depending on their behavior under a parity transformation. For each parity, all the metric degrees of freedom are then packed into one master function: the Regge-Wheeler function for odd parity~\cite{Regge:PhysRev.108.1063} and the Zerilli-Moncrief function for even parity~\cite{Zerilli:1971wd, Moncrief:1974am}. The master equations governing these master functions are decoupled from other dynamical degrees of freedom of the metric fields and are separable into radial and angular equations. 

However, for rotating BHs in GR, due to the lack of spherical symmetry, to obtain the decoupled and separable perturbed field equations, one has to use the Teukolsky equations~\cite{Teukolsky:1973ha,Press:1973zz, Teukolsky:1974yv}, where the fundamental variables to solve for are the Weyl scalars $\Psi_{0,4}$ characterizing curvature perturbations. In this case, the master equations of $\Psi_{0,4}$ are decoupled from other NP quantities and are separable into purely radial and purely angular equations. For a quick review of the NP formalism and the Teukolsky formalism in GR, one can refer to the original papers~\cite{Newman-Penrose,Teukolsky:1973ha, Press:1973zz, Teukolsky:1974yv}, the book~\cite{Chandrasekhar_1983}, or more recent papers that work in the Teukolsky formalism~\cite{Loutrel_Ripley_Giorgi_Pretorius_2020, Li:2022pcy}.
	
In modified gravity, most calculations for rotating BHs have so far been done using metric perturbations and the slow-rotation expansion, e.g.,~\cite{Wagle:2021tam, Srivastava:2021imr} in dCS gravity, \cite{Pierini:2021jxd, Pierini:2022eim} in EdGB theory, and~\cite{Cano:2020cao, Cano_Fransen_Hertog_Maenaut_2021} in higher-derivative gravity. However, these approaches cannot deal with BHs with arbitrary spin, which motivated the development of the modified Teukolsky formalism in~\cite{Li:2022pcy, Hussain:2022ins}. Following the formalism in~\cite{Li:2022pcy, Hussain:2022ins}, one can find separable and decoupled equations for $\Psi_{0,4}$ of BHs with arbitrary spin in a wide class of modified gravity theories, such as in dCS gravity, which can be treated as an EFT extension of GR. In this paper, we will use the modified Teukolsky equations of $\Psi_{0,4}$ in~\cite{Li:2022pcy}. For an alternative approach following~\cite{Wald_1978} by projecting the Einstein equations to the Teukolsky equations, one can refer to~\cite{Hussain:2022ins}.
	
In~\cite{Li:2022pcy}, the authors introduced a two-parameter expansion, in terms of $\zeta$ and $\epsilon$ where
\begin{enumerate}
    \item $\zeta$ is the parameter characterizing the strength of modifications to GR. In the case of dCS gravity, $\zeta$ is given by Eq.~\eqref{eq:zeta}. 
    \item $\epsilon$ is the parameter characterizing the strength of gravitational perturbations, which also appears in GR.
\end{enumerate}
In this way, one can expand all the NP quantities as
\begin{align} \label{eq:expansion_NP}
   \Psi_i
   &=\Psi_i^{(0)}+\epsilon\Psi_i^{(1)} \nonumber\\&=\Psi_i^{(0,0)}+\zeta\Psi_{i}^{(1,0)}+\epsilon\Psi_{i}^{(0,1)}
   +\zeta\epsilon\Psi_{i}^{(1,1)}
\end{align}
and the extra non-metric fields, such as the pseudoscalar field $\vartheta$ in dCS gravity, as
\begin{equation} \label{eq:expansion_scalar}
    \vartheta
    =\vartheta^{(0)}+\epsilon\vartheta^{(1)} 
    =\zeta\vartheta^{(1,0)}+\zeta\epsilon\vartheta^{(1,1)}\,,
\end{equation}
where we have reserved the single superscript notation for only an expansion in $\epsilon$. When using the double-superscript notation, however, the first superscript will also refer to contributions proportional to $\zeta$ to a given power. In contrast, the second superscript will refer to terms proportional to $\epsilon$ to a given power.
	
Using the expansion in Eqs.~\eqref{eq:expansion_NP} and \eqref{eq:expansion_scalar}, the authors in~\cite{Li:2022pcy} found that for a rotating BH described by a matter vacuum, non-Ricci-flat, Petrov type I spacetime that perturbatively deviates from a Petrov type D spacetime in GR, the gravitational wave perturbation $\Psi_0$ satisfies
\begin{align} \label{eq:master_eqn_non_typeD_Psi0}
    H_{0}^{(0,0)}\Psi_0^{(1,1)}
    =\mathcal{S}_{\geo}^{(1,1)}+\mathcal{S}^{(1,1)}\,,
\end{align}
where $H_{0}^{(0,0)}$ is the Teukolsky operator for $\Psi_0$ in GR~\cite{Teukolsky:1973ha}, and the source terms are divided into a ``geometric piece'', 
\begin{align} 
    \mathcal{S}_{\geo}^{(1,1)}=\mathcal{S}_{0,\typeD}^{(1,1)}
    +\mathcal{S}_{0,\nonD}^{(1,1)}+\mathcal{S}_{1,\nonD}^{(1,1)}\,,  \label{eq:S_geo}
\end{align}
with
\begin{subequations} \label{eq:source_non_typeD_Psi0}
\begin{align}
    \label{eq:Sd} & \mathcal{S}_{0,\typeD}^{(1,1)}=-H_0^{(1,0)}\Psi_0^{(0,1)}\,, \\
    \label{eq:nond1} & \mathcal{S}_{0,\nonD}^{(1,1)}=-H_0^{(0,1)}\Psi_0^{(1,0)}\,, \\
    \label{eq:nond2} & \mathcal{S}_{1,\nonD}^{(1,1)}=H_1^{(0,1)}\Psi_1^{(1,0)}\,,
\end{align}    
\end{subequations}
and a ``Ricci piece'', 
\begin{align} \label{eq:S}
    & \begin{aligned} 
        \mathcal{S}^{(1,1)}
        =& \;\mathcal{E}_2^{(0,0)}S_2^{(1,1)}
        +\mathcal{E}_2^{(0,1)}S_2^{(1,0)}
        -\mathcal{E}_1^{(0,0)}S_1^{(1,1)} \\
        & \;-\mathcal{E}_1^{(0,1)}S_1^{(1,0)}\,,
    \end{aligned}
\end{align}
with $S_{1,2}$ given by
\begin{subequations} \label{eq:source_bianchi}
\begin{align}
    \label{eq:source_bianchi_1}
    \begin{split} 
        S_1\equiv& \;\delta_{[-2,-2,1,0]}\Phi_{00}
        -D_{[-2,0,0,-2]}\Phi_{01} \\
        & \;+2\sigma\Phi_{10}-2\kappa\Phi_{11}-\bar{\kappa}\Phi_{02}\,,
    \end{split} \\
    \label{eq:source_bianchi_2}
    \begin{split} 
        S_2\equiv& \;\delta_{[0,-2,2,0]}\Phi_{01}
        -D_{[-2,2,0,-1]}\Phi_{02} \\
        & \;-\bar{\lambda}\Phi_{00}+2\sigma\Phi_{11}-2\kappa\Phi_{12}\,.
    \end{split}
\end{align}
\end{subequations}
The operators $H_{0,1}$, $\mathcal{E}_{0,1}$ are defined as
\begin{equation} \label{eq:H_in_Teuk}
    \begin{aligned}
        & H_0 = \mathcal{E}_2F_2-\mathcal{E}_1 F_1-3\Psi_2\,,\quad
        H_1 = \mathcal{E}_2J_2-\mathcal{E}_1 J_1\,, \\
        & \mathcal{E}_1=E_1-\frac{1}{\Psi_2}\delta\Psi_2 \,,\quad
        \mathcal{E}_2=E_2-\frac{1}{\Psi_2}D\Psi_2\,,
    \end{aligned}
\end{equation}
where $\Psi_2$ is a NP scalar, and we have also defined
\begin{equation} \label{eq:operators_in_Teuk}
\begin{aligned}
    & F_1\equiv\bar{\delta}_{[-4,0,1,0]}\,,\quad
    && F_2\equiv\boldsymbol{\Delta}_{[1,0,-4,0]}\,, \\
    & J_1\equiv D_{[-2,0,-4,0]}\,,\quad
    && J_2\equiv\delta_{[0,-2,0,-4]}\,, \\
    & E_1\equiv\delta_{[-1,-3,1,-1]}\,,\quad
    && E_2\equiv D_{[-3,1,-1,-1]}\,. \\
\end{aligned}
\end{equation}
The operators that appear in the above definitions are defined for convenience to be
\begin{subequations} \label{eq:derivative-notation}
    \begin{align}
        D_{[a,b,c,d]}
        &=D+a\varepsilon+b\bar{\varepsilon}+c\rho+d\bar{\rho}\,, \\
        \boldsymbol{\Delta}_{[a,b,c,d]}
        &=\boldsymbol{\Delta}+a\mu+b\bar{\mu}+c\gamma+d\bar{\gamma}\,, \\
        \delta_{[a,b,c,d]}
        &=\delta+a\bar{\alpha}+b\beta+c\bar{\pi}+d\tau\,, \\
        \bar{\delta}_{[a,b,c,d]}
        &=\bar{\delta}+a\alpha+b\bar{\beta}+c\pi+d\bar{\tau}\,,
    \end{align}
\end{subequations}
where $(D,\boldsymbol{\Delta},\delta,\bar{\delta})$ are the usual NP differential operators (constructed by contracting the tetrad with partial derivatives), while $(\varepsilon,\rho,\mu,\gamma,\alpha,\beta, \pi, \tau)$ are spin coefficients, with the overhead bar denoting complex conjugation, and $(a,b,c,d)$ are certain constants. For a complete derivation of the equations above and the definition of Weyl scalars, spin coefficients, directional derivatives, and Ricci NP scalars, one can refer to \cite{Newman:1961qr, Li:2022pcy}. 

For this work, we are only interested in studying slowly rotating BHs in dCS gravity up to $\mathcal{O}(\chi)$. Such BHs are described by vacuum non-Ricci-flat Petrov type D spacetimes~\cite{Owen:2021eez}. The modified Teukolsky equations for these BHs hold the same form as given in Eq.~\eqref{eq:master_eqn_non_typeD_Psi0} with the source terms $\mathcal{S}_{0,\nonD}^{(1,1)}$ and $\mathcal{S}_{1,\nonD}^{(1,1)}$ in Eqs.~\eqref{eq:nond1}--\eqref{eq:nond2} vanishing.

The equation of $\Psi_4^{(0,1)}$ can be obtained from Eq.~\eqref{eq:master_eqn_non_typeD_Psi0} by the Geroch-Held-Penrose (GHP) transformation \cite{Geroch_Held_Penrose_1973} and is given in \cite{Li:2022pcy}:
\begin{align} \label{eq:master_eqn_non_typeD_Psi4}
    & H_{4}^{(0,0)}\Psi_4^{(1,1)}
    =\mathcal{T}_{\geo}^{(1,1)}+\mathcal{T}^{(1,1)}\,,
\end{align}
where $H_{4}^{(0,0)}$ is the Teukolsky operator in GR for $\Psi_4$, and the ``geometric piece'' of the source terms is defined as
\begin{equation} \label{eq:T_geo}
    \mathcal{T}_{\geo}^{(1,1)}=\mathcal{T}_{4,\typeD}^{(1,1)}
    +\mathcal{T}_{4,\nonD}^{(1,1)}+\mathcal{T}_{3,\nonD}^{(1,1)}\,,
\end{equation}
with
\begin{subequations} \label{eq:source_non_typeD_Psi4}
\begin{align}
    & \mathcal{T}_{4,\typeD}^{(1,1)}=-H_4^{(1,0)}\Psi_4^{(0,1)}\,,  \\
    & \mathcal{T}_{4,\nonD}^{(1,1)}=-H_4^{(0,1)}\Psi_4^{(1,0)}\,, \\
    & \mathcal{T}_{3,\nonD}^{(1,1)}=H_3^{(0,1)}\Psi_3^{(1,0)}\,, 
\end{align} \\
\end{subequations}
whereas the ``Ricci piece'' is defined as
\begin{align}
& \begin{aligned} \label{eq:T}
        \mathcal{T}^{(1,1)}
        =& \;\mathcal{E}_4^{(0,0)}S_4^{(1,1)}+
        \mathcal{E}_4^{(0,1)}S_4^{(1,0)}
        -\mathcal{E}_3^{(0,0)}S_3^{(1,1)} \\
        & \;-\mathcal{E}_3^{(0,1)}S_3^{(1,0)}\,,
\end{aligned} 
\end{align}
with
\begin{subequations} \label{eq:s3s4}
\begin{align}
& \begin{aligned}
    S_3\equiv& \;-\mathbf{\Delta}_{[0,2,2,0]}\Phi_{21}
    +\bar{\delta}_{[2,2,0,-1]}\Phi_{22} \\
    & \;+2\nu\Phi_{11}+\bar{\nu}\Phi_{20}-2\lambda\Phi_{12}\,,
\end{aligned} \\
& \begin{aligned}
    S_4\equiv& \;-\mathbf{\Delta}_{[0,1,2,-2]}\Phi_{20}
    +\bar{\delta}_{[2,0,0,-2]}\Phi_{21} \\
    & \;+2\nu\Phi_{10}-2\lambda\Phi_{11}+\bar{\sigma}\Phi_{22}\,.
\end{aligned}
\end{align}
\end{subequations}
The operators $H_{3,4}$ and $\mathcal{E}_{3,4}$ are defined as 
\begin{equation} \label{eq:H_in_Teuk_ORG}
\begin{aligned}
    & H_4=\mathcal{E}_4F_4-\mathcal{E}_3F_3-3\Psi_2\,,\quad
    H_3=\mathcal{E}_4J_4-\mathcal{E}_3J_3\,, \\
    & \mathcal{E}_3=E_3-\frac{1}{\Psi_2}\bar{\delta}\Psi_2\,,\quad
    \mathcal{E}_4=E_4-\frac{1}{\Psi_2}\mathbf{\Delta}\Psi_2\,,
\end{aligned}
\end{equation}
with
\begin{equation}
\begin{aligned}
    & F_3\equiv\delta_{[0,4,0,-1]}\,,\quad
    && F_4\equiv D_{[4,0,-1,0]}\,, \\
    & J_3\equiv\mathbf{\Delta}_{[4,0,2,0]}\,,\quad
    && J_4\equiv\bar{\delta}_{[2,0,4,0]}\,, \\
    & E_3\equiv\bar{\delta}_{[3,1,1,-1]}\,,\quad 
    && E_4\equiv\mathbf{\Delta}_{[1,1,3,-1]}\,.
\end{aligned}
\end{equation}
Although the formalism above works for BHs with arbitrary spin in dCS gravity, we choose to use the slow-rotation expansion in this paper, so we can check the consistency of our results with prior work using metric perturbations \cite{Wagle:2021tam, Srivastava:2021imr} in our next paper~\cite{dcstyped2}. We implement a slow-rotation expansion of the above equations in Sec.~\ref{sec:trivariate_expansion}. 

%%%%%%%%%%%%%%%%%%%%%%%%%%%%%%%%%%%%%%%%%%%%%%%%%%%%%%%%%%%%
\subsection{Structure of the source terms}

Here, we further discuss the structure of the source terms in Eq.~\eqref{eq:master_eqn_non_typeD_Psi0} presented in~\cite{Li:2022pcy}. In particular, we will focus on the source terms that are non-vanishing for a non-Ricci-flat, Petrov type D BH in dCS gravity, given in Eqs.~\eqref{eq:Sd} and \eqref{eq:S}. The source term in Eq.~\eqref{eq:Sd} only depends on the perturbed Weyl scalar $\Psi_0^{(0,1)}$ in GR and the dCS corrections to the stationary NP quantities at $\mathcal{O}(\zeta^1,\epsilon^0)$. One can solve the Teukolsky equation in GR \cite{Teukolsky:1973ha,Teukolsky:1974yv} to calculate $\Psi_0^{(0,1)}$ directly. The NP quantities at $\mathcal{O}(\zeta^1,\epsilon^0)$ can be computed from the dCS metric in Eq.~\eqref{eq:metric_dCS}, as shown in more detail in Sec.~\ref{sec:NP_quantities_background}.

Due to the non-minimal coupling between the scalar field and the metric, the source term $\mathcal{S}^{(1,1)}$ in Eq.~\eqref{eq:S} depends on both the scalar field perturbations and the metric perturbations. To compute $\mathcal{S}^{(1,1)}$, we first need to calculate the NP Ricci scalars $\Phi_{ij}$ using the stress tensor or the Ricci tensor, i.e.,
\begin{equation} \label{eq:phiij}
    \Phi_{00} = -\frac{1}{2} R_{11} = - \frac{1}{2} R_{\mu\nu}l^\mu l^\nu \,,
\end{equation}  
where $l^\mu$ is one of the NP tetrad basis vectors. Since the background and perturbed scalar field in GR vanish, we have that $\vartheta^{(0,0)}=\vartheta^{(0,1)}=0$. Therefore, the NP Ricci scalars $\Phi_{ij}^{(1,1)}$ can be expressed as a function of the scalar field perturbation $\vartheta^{(1,1)}$ and the metric perturbation $h^{(0,1)}$ as
\begin{equation} \label{eq:phiij_pert}
    \Phi_{ij}^{(1,1)} = 
    {\cal{O}}(\vartheta^{(1,0)}h^{(0,1)})+ {\cal{O}}(\vartheta^{(1,1)}g^{(0,0)}) \,,
\end{equation}
where $g^{(0,0)}$ and  $h^{(0,1)}$ are shorthand for terms that depend on the metric tensor of the GR background and of the metric perturbation due to GWs reconstructed in GR, respectively. From Eqs.~\eqref{eq:S}, \eqref{eq:source_bianchi}, and \eqref{eq:phiij_pert}, we notice that $\mathcal{S}^{(1,1)}$ couples the GWs in GR and the scalar field $\vartheta$, so we need to solve the equations of motions of these non-gravitational fields to find their contributions to the stress tensor and $\mathcal{S}^{(1,1)}$. Morevover, from Eqs.~\eqref{eq:H_in_Teuk}--\eqref{eq:derivative-notation}, we see that $\mathcal{S}^{(1,1)}$ in Eq.~\eqref{eq:S} depends on $\Psi_2^{(0,1)}$, the directional derivatives at $\mathcal{O}(\zeta^0,\epsilon^1)$, and the perturbed spin coefficients at $\mathcal{O}(\zeta^0,\epsilon^1)$, which need to be retrieved from the reconstructed metric perturbation $h_{\mu\nu}^{(0,1)}$ for GR GWs. One can either follow the metric reconstruction approach in \cite{Cohen_Kegeles_1975, Chrzanowski:1975wv, Kegeles_Cohen_1979, Lousto_Whiting_2002, Ori_2003, Whiting_Price_2005, Yunes_Gonzalez_2006, Keidl_Friedman_Wiseman_2007, Keidl_Shah_Friedman_Kim_Price_2010}, the so-called Chrzanowski-Cohen-Kegeles (CCK) procedures, which involves defining an intermediate quantity called the Hertz potential, or the approach in \cite{Loutrel_Ripley_Giorgi_Pretorius_2020, Chandrasekhar_1983}, which solves the remaining NP equation directly. In this paper, we choose to follow the more widely used CCK procedures and apply them to compute the source term $\mathcal{S}^{(1,1)}$.

%%%%%%%%%%%%%%%%%%%%%%%%%%%%%%%%%%%%%%%%%%%%%%%%%%%%%%%%%%%%
\subsection{Slow-rotation expansion} 
\label{sec:trivariate_expansion}
	
When considering a slow-rotation expansion, in addition to the quantities given in Eqs.~\eqref{eq:expansion_NP} and \eqref{eq:expansion_scalar}, one needs to consider an additional expansion in the dimensionless spin parameter $\chi=a/M$. As an extension of Eqs.~\eqref{eq:expansion_NP} and \eqref{eq:expansion_scalar}, all the NP quantities now admit a three-parameter expansion in $\zeta$, $\epsilon$, and $\chi$, where
\begin{equation} \label{eq:NP_expansion_3}
    \Psi=\sum_{l,m,n}\zeta^l\chi^m\epsilon^n
    \Psi^{(l,m,n)}\,,
\end{equation}
as well as the pseudoscalar field,
\begin{equation} \label{eq:scalar_expansion_3}
    \vartheta=\sum_{l,m,n}\zeta^l\chi^m\epsilon^n
    \vartheta^{(l,m,n)}\,.
\end{equation}
In what follows, it will sometimes be convenient to hide the $\chi$ expansion in more compact notation, such as $\psi^{(1,1)} = \psi^{(1,0,1)} + \chi \psi^{(1,1,1)}$. When only a two-parameter expansion is denoted, the $\chi$ expansion will be assumed.
	
In this paper, we will focus only on linear perturbations in $\epsilon$, along with the small-coupling approximation and the slow-rotation approximation, i.e., up to linear order terms in $\zeta$ and $\chi$, respectively. Therefore, our Eqs.~\eqref{eq:NP_expansion_3} and \eqref{eq:scalar_expansion_3} can be expanded as
\begin{subequations} \label{eq:triexpansion}
\begin{align} 
    & \begin{aligned} \label{eq:NP_expansion_leading}
        \Psi
        =& \;\Psi^{(0,0,0)}+\chi\left(\Psi^{(0,1,0)}
        +\zeta \Psi^{(1,1,0)}\right) \\
        & \;+\epsilon\left(\Psi^{(0,0,1)}+\chi\Psi^{(0,1,1)}\right)
        +\zeta\epsilon\left(\Psi^{(1,0,1)}+\chi\Psi^{(1,1,1)}\right)\,,
    \end{aligned} \\
    & \vartheta=\zeta\left(\chi\vartheta^{(1,1,0)}
    +\epsilon\vartheta^{(1,0,1)}+\chi\epsilon\vartheta^{(1,1,1)}\right)
    \label{eq:scalar_expansion_leading}\,.
\end{align} 
\end{subequations}

Equation~\eqref{eq:NP_expansion_leading} groups the corrections to $\Psi^{(0,0,0)}$ into three sets organized by parenthesis. The first set of terms are stationary corrections to the Schwarzschild metric due to the slow-rotation approximation in GR, $\Psi^{(0,1,0)}$, and in dCS gravity, $\Psi^{(1,1,0)}$. The term $\Psi^{(0,1,0)}$ can be retrieved from the slow-rotation expansion of the Kerr metric in Eq.~\eqref{eq:kerr_metric}, and $\Psi^{(1,1,0)}$ can be evaluated with the $\mathcal{O}(\zeta^1,\chi^1,\epsilon^0)$ correction to the metric in Eq.~\eqref{eq:metric_dCS} found by~\cite{Yunes:2009hc}. Since the Pontryagin density vanishes for any spherically symmetric spacetime, and the Schwarzschild metric is the unique stationary spherically symmetric solution to the Einstein equations, there is no correction to the metric at $\mathcal{O}(\zeta^1,\chi^0,\epsilon^0)$ \cite{Yagi:2012ya}. Thus, we have dropped the term $\Psi^{(1,0,0)}$ in Eq.~\eqref{eq:NP_expansion_leading}. For the same reason, $\vartheta^{(1,0,0)}=0$.
	
The second set of terms are the GW perturbations to the Kerr metric in GR up to $\mathcal{O}(\zeta^0,\chi^1,\epsilon^1)$. These terms include perturbed Weyl scalars, NP spin coefficients, and directional derivatives, all of which need to be evaluated in GR but include spin perturbations. To evaluate this type of terms, we need metric reconstruction of GW perturbations at $\mathcal{O}(\zeta^0,\epsilon^1)$, the procedures of which are discussed in detail in Sec.~\ref{sec:metric_reconstruction}.
	
The third set of terms are the one we want to solve for, which are corrections to GW perturbations in dCS gravity. The term $\Psi^{(1,0,1)}$ corresponds to gravitational perturbations sourced by non-rotating BHs in dCS gravity. Since $\vartheta^{(1,0,0)}=0$, $\Psi^{(1,0,1)}$ is purely sourced by the leading contribution to the dynamical pseudoscalar field $\vartheta^{(1,0,1)}$, so only $\mathcal{S}^{(1,0,1)}$ contributes to Eq.~\eqref{eq:master_eqn_non_typeD_Psi0}, and no metric reconstruction is needed. The term $\Psi^{(1,1,1)}$ corresponds to leading-order corrections to gravitational perturbations of slowly rotating BHs in dCS gravity. Unlike the non-rotating case, since both the metric and $\vartheta$ are corrected at $\mathcal{O}(\zeta^1,\chi^1,\epsilon^0)$, $\Psi_{0,4}^{(1,1,1)}$ can either be driven by dynamical GW perturbations in GR or dynamical $\vartheta$. 
	
For the first type of correction, the driving terms can come from $\mathcal{S}_{\geo}^{(1,1)}$ in the form of terms proportional to the product $h^{(1,0)}h^{(0,1)}$. As discussed in Sec.~\ref{sec:modified_Teuk_eqn} and \cite{Li:2022pcy}, this kind of terms is due to the correction to the background geometry, so they are independent of bGR theories. Up to $\mathcal{O}(\zeta^1,\chi^1,\epsilon^1)$, the background spacetime is still Petrov type D \cite{Yagi:2012ya}, so $\mathcal{S}_{0,\nonD}^{(1,1,1)}=\mathcal{S}_{1,\nonD}^{(1,1,1)}=0$ in Eq.~\eqref{eq:S_geo}, and one does not need metric construction to evaluate these terms \cite{Li:2022pcy}. In Sec.~\ref{sec:modified_Teuk_operator}, we will compute $\mathcal{S}_{\geo}^{(1,1)}$ in detail. Besides $\mathcal{S}_{\geo}^{(1,1)}$, there is also contribution from $\mathcal{S}^{(1,1)}$ in the form of terms proportional to the product $\vartheta^{(1,0)}h^{(0,1)}$ due to the effective stress tensor. In this case, metric reconstruction is needed, and we will compute $\mathcal{S}^{(1,1)}$ in Sec.~\ref{sec:corrections_from_stress}.
	
For the second type of correction, the driving terms only come from $\mathcal{S}^{(1,1)}$. Since both $\vartheta^{(1,0,1)}$ and $\vartheta^{(1,1,1)}$ are nonzero, the metric field in these terms needs to be evaluated on the Kerr background, expanded to $\mathcal{O}(\chi)$. To find $\vartheta^{(1,1)}$, one needs to solve Eq.~\eqref{eq:EOM_theta} at $\mathcal{O}(\zeta^1,\epsilon^1)$, i.e.,
\begin{align} \label{eq:EOM_scalar_11}
    \square^{(0,0)}\vartheta^{(1,1)}
    =-\pi^{-\frac{1}{2}}M^2\left[R^*\!R\right]^{(0,1)}
    -\square^{(0,1)}\vartheta^{(1,0)}\,.
\end{align}
In Sec.~\ref{sec:source_scalar}, we will compute the source terms of Eq.~\eqref{eq:EOM_scalar_11}. In Sec.~\ref{sec:corrections_from_stress}, we will compute the source terms driven by $\vartheta^{(1,1)}$ but leave $\vartheta^{(1,1)}$ unevaluated. In our follow-up work \cite{dcstyped2}, we will solve both the modified Teukolsky equation and the scalar field equation jointly to find the QNM shifts. Since metric reconstruction at $\mathcal{O}(\zeta^0,\epsilon^1)$ is required for both the modified Teukolsky equation and the scalar field equation, we present a review of the procedures in Sec.~\ref{sec:metric_reconstruction}. 

%%%%%%%%%%%%%%%%%%%%%%%%%%%%%%%%%%%%%%%%%%%%%%%%%%%%%%%%%%%%
\subsection{NP quantities on background} 
\label{sec:NP_quantities_background}

In this subsection, we will present the background tetrad for a non-Ricci-flat, Petrov type D, slowly rotating, dCS gravity BH spacetime. To obtain the null tetrad for the metric given by the line element in Eq.~\eqref{eq:full_line_element}, one can follow the general procedures prescribed in~\cite{Li:2022pcy} or the standard procedures for finding the Kinnersley tetrad in GR given in~\cite{Chandrasekhar_1983}. In \cite{Sopuerta:2009iy}, such a tetrad was found by following the second approach. Nonetheless, for completeness, let us re-derive the tetrad following the prescription in \cite{Chandrasekhar_1983,Sopuerta:2009iy}. Our result is consistent with the one in \cite{Sopuerta:2009iy}, but with additional tetrad rotations to set $\Psi^{(1,0)}_{0,1,3,4}=0$. 

To begin, we first find the null geodesics in the equatorial plane for a dCS BH to be
\begin{equation} \label{eq:null_geodesics}
    \begin{aligned}
        \frac{dt}{d\tau}= 
        & \left[r^2+a^2-\frac{aG(r)}{2}\right]\frac{E}{\tilde{\Delta}(r)}\,, \\
        \frac{dr}{d\tau}= 
        & \pm\sqrt{\Delta(r)\tilde{\Delta}(r)
        \left(1-\frac{aG(r)}{r^2}\right)}
        \frac{E}{\tilde{\Delta}(r)}\,, \\
        \frac{d\theta}{d\tau}=& ~0\,, \\
        \frac{d\phi}{d\tau}= 
        & \left(a+\frac{G(r)}{2}\right)\frac{E}{\tilde{\Delta}(r)}\,,
    \end{aligned}
\end{equation}
where $E=-{\partial\mathscr{L}}/{\partial t}$ is a constant of motion, $\mathscr{L}$ is the Lagrangian for Kerr in \cite{Chandrasekhar_1983}, $\tilde{\Delta}(r)=\Delta(r)+2aMG(r)/r+G(r)^2/4$, and $G(r)=\zeta\chi\tilde{G}(r)$. Following the procedures outlined in~\cite{Chandrasekhar_1983}, we align the tetrad basis vectors $l^{\mu}$ and $n^{\mu}$ along the outgoing and ingoing null geodesics respectively at the equilateral plane with $E=1$ such that 
\begin{align} 
    \begin{split} \label{eq:Kinnersley_l}
        l^{\mu}= 
        & \;\frac{1}{\tilde{\Delta}(r)}\Bigg(r^2+a^2-\frac{aG(r)}{2}\,,\;
        \sqrt{\Delta(r)\tilde{\Delta}(r)\left(1-\frac{aG(r)}{r^2}\right)}\,, \\ &\;0\,,\;a+\frac{G(r)}{2}\Bigg)\,,
    \end{split} \\
    \begin{split} \label{eq:Kinnersley_n}
        n^{\mu}= 
        & \;N\left(r^2+a^2-\frac{aG(r)}{2}\,,\;
        -\sqrt{\Delta(r)\tilde{\Delta}(r)
        \left(1-\frac{aG(r)}{r^2}\right)}\right.\,, \\ &\;\left.0,\;a+\frac{G(r)}{2}\right)\,,
    \end{split}
\end{align}
where $N$ is the normalization factor introduced to impose $l^{\mu}n_{\mu}=-1$. Since $l^{\mu}$ and $n^{\mu}$ are along null geodesics, $l^{\mu}l_{\mu}=n^{\mu}n_{\mu}=0$ is satisfied automatically. Expanding Eqs.~\eqref{eq:Kinnersley_l} and \eqref{eq:Kinnersley_n} up to $\mathcal{O}(\zeta^1,\chi^1,\epsilon^0)$, we find 
\begin{align}
    & l^{\mu}=\left(\frac{r}{r-r_s}\,,\;1\,,\;0\,,\;\frac{\chi M}{r(r-r_s)}+
    \frac{\zeta\chi\tilde{G}(r)}{2r(r-r_s)}\right)\,,
    \label{eq:Kinnersley_l_expansion} \\
    & n^{\mu}=\tilde{N}(r)\left(\frac{r}{r-r_s}\,,\;-1\,,\;0\,,\;
    \frac{\chi M}{r(r-r_s)}+
    \frac{\zeta\chi\tilde{G}(r)}{2r(r-r_s)}\right)\,,
    \label{eq:Kinnersley_n_expansion}
\end{align}
where $r_s$ is the Schwarzschild radius given by $r_s=2M$, and $\tilde{N}(r)=(r-r_s)/2r$. When $\zeta=0$, $l^{\mu}$ and $n^{\mu}$ reduce to the Kinnersley tetrad of Kerr BHs expanded to $\mathcal{O}(\chi^1)$. The tetrad basis vectors $l^{\mu}$ and $n^{\mu}$ in Eqs.~\eqref{eq:Kinnersley_l_expansion} and \eqref{eq:Kinnersley_n_expansion} are the same as the principal null directions in Eq.~(31) of \cite{Sopuerta:2009iy}.

To obtain the remaining components of the null tetrad, notice that the correction to the Kerr metric due to dCS gravity enters at $\mathcal{O}(\zeta^1,\chi^1,\epsilon^0)$ only in the $t\phi$-component. Therefore, it can be expected that the corrections to the Kinnersley tetrad are only along the $\partial_t$ and $\partial_{\phi}$ directions, which is seen to be true for $l^{\mu}$ and $n^{\mu}$. Thus, at $\mathcal{O}(\zeta^1,\chi^1,\epsilon^0)$, the corrections to the remaining null tetrad components, $m^{\mu}$ and $\bar{m}^{\mu}$, take the form
\begin{equation} \label{eq:deltam}
    m^{\mu (1,1,0)}
    =\left(m_{t}(r,\theta),\,0,\,0,\,m_{\phi}(r,\theta)\right)\,,
\end{equation}
where $\bar{m}^{\mu}$ can be obtained by taking the complex conjugation of Eq.~\eqref{eq:deltam}.
Imposing the remaining orthogonality conditions to $\mathcal{O}(\zeta^1, \chi^1, \epsilon^0)$, we find $m_{t}(r,\theta)=m_{\phi}(r,\theta)=0$. Therefore,
\begin{equation} \label{eq:Kinnersley_m_expansion}
    \begin{aligned}
        m^{\mu}=
        & \;\frac{1}{\sqrt{2}r}\Bigg(i\chi M\sin\theta,\,0,\,
        1-\frac{i\chi M\cos\theta}{r}, \\ & \;i\left(1-\frac{i\chi M\cos\theta}{r}\right)\csc\theta\Bigg)\,.
    \end{aligned}
\end{equation}
Notice that $m^{\mu}$ (and therefore $\bar{m}^{\mu}$) holds the same form as the Kinnersley tetrad of Kerr BH expanded to $\mathcal{O}(\chi)$. Using Eqs.~\eqref{eq:Kinnersley_l_expansion}, \eqref{eq:Kinnersley_n_expansion}, and \eqref{eq:Kinnersley_m_expansion}, to $\mathcal{O}(\zeta^1, \chi^1, \epsilon^0)$, we obtain
\begin{equation}
    \begin{aligned}
        & \Psi_0=\Psi_4=0\,,\;
        \Psi_1=-\frac{3\sqrt{2}i\zeta\chi
            A_1(r)}{32r^9}\sin\theta\,,\\
        & \Psi_2=-\frac{M}{r^3}-\frac{3i\chi}{r^4} \left(M^2-\frac{\zeta A_2(r)}{8r^5}\right)\cos\theta\,,\\
        & \Psi_3=-\frac{3\sqrt{2}i\zeta\chi A_3(r)}{64r^{10}}\sin\theta\,,		
    \end{aligned}
\end{equation}
where $A_i(r)$ are listed in Appendix~\ref{appendix:background_NP_more}.
	
Such a calculation contradicts the claim that BHs in dCS gravity are Petrov type D spacetimes to $\mathcal{O}(\chi)$. We see that the seeming contradiction arises due to the non-vanishing $\Psi_1$ and $\Psi_3$ on the background. However, we can perform tetrad rotations to eliminate $\Psi_{1}^{(1,1,0)}$ and $\Psi_{3}^{(1,1,0)}$, as described in detail in Appendix~\ref{appendix:background_NP_more}. This can be achieved by a type \rom{2} rotation [i.e., Eq.~\eqref{eq:rotate2}] with $b^{(1,1,0)}=-\sqrt{2}iA_1(r)\sin\theta/(32Mr^6)$ and a type \rom{1} rotation [i.e., Eq.~\eqref{eq:rotate1}] with $a^{(1,1,0)}=\sqrt{2}i A_3(r)\sin\theta/(64Mr^{7})$, respectively. Following this, we finally obtain
\begin{align} \label{eq:Psi2_bg}
    \Psi_{i} &= 0 \, \quad \forall ~i \in \{0,1,3,4\} \,, \nonumber\\ 
    \Psi_{2} &=-\frac{M}{r^3}-\frac{3i\chi}{r^4} \left(M^2-\frac{\zeta A_2(r)}{8r^5}\right)\cos\theta\,.
\end{align}
The explicit expression for the rotated tetrad is listed in Eq.~\eqref{eq:principal_tetrad}. Different from \cite{Sopuerta:2009iy}, we will call the tetrad in Eq.~\eqref{eq:principal_tetrad} the ``principal tetrad." Notice that, in \cite{Owen:2021eez}, a Kinnersley-like tetrad for the dCS metric expanded up to $\mathcal{O}(\zeta^1,\chi^2, \epsilon^0)$ was also found. To impose that $\Psi_{0,4}^{(1,0)}$ vanish at $\mathcal{O}(\chi^2)$, when the background spacetime is Petrov type I, Ref.~\cite{Owen:2021eez} added terms at $\mathcal{O}(\zeta^{1/2},\chi^1,\epsilon^0)$ to the tetrad. In this paper, however, we only want to impose $\Psi_{0,1,3,4}^{(1,0,0)}=\Psi_{0,1,3,4}^{(1,1,0)}=0$, and prefer to keep the expansion scheme in Eq.~\eqref{eq:expansion_NP}, so the tetrad in Eq.~\eqref{eq:principal_tetrad} is more suitable for our purposes.
	
We have also listed all the spin coefficients up to $\mathcal{O}(\zeta^1,\chi^1,\epsilon^0)$ in the principal tetrad in Appendix~\ref{appendix:background_NP_more}. For any vacuum Petrov type D spacetimes in GR, the Goldberg-Sachs theorem requires that in the tetrad where $\kappa=\sigma=\lambda=\nu=0$, the Weyl scalars $\Psi_{0,1,3,4}=0$ and vice versa. However, in dCS gravity, since the effective stress tensor is nonzero, the background spacetime is non-Ricci-flat. Thus, $\Psi_{0,1,3,4}^{(1,0)}$ do not necessarily vanish in the tetrad where $\kappa^{(1,0)}=\sigma^{(1,0)}=\lambda^{(1,0)}=\nu^{(1,0)}=0$ and vice versa. For the tetrad in Eqs.~\eqref{eq:Kinnersley_l_expansion}, \eqref{eq:Kinnersley_n_expansion}, and \eqref{eq:Kinnersley_m_expansion}, we found that $\kappa^{(1,0)}=\sigma^{(1,0)}=\lambda^{(1,0)}=\nu^{(1,0)}=0$ while $\Psi_{1,3}^{(1,0)}\neq0$ up to $\mathcal{O}(\zeta^1,\chi^1,\epsilon^0)$. This tetrad is along the principal null directions found in \cite{Sopuerta:2009iy}. Nonetheless, the master equation Eq.~\eqref{eq:master_eqn_non_typeD_Psi0} is more simplified when $\Psi_{0,1,3,4}^{(1,0)}=0$, so we will use the principal tetrad in Eq.~\eqref{eq:principal_tetrad} for the remaining calculations even if the spin coefficients mentioned above do not vanish along the principal tetrad.

%%%%%%%%%%%%%%%%%%%%%%%%%%%%%%%%%%%%%%%%%%%%%%%%%%%%%%%%%%%%
%%%%%%%%%%%%%%%%%%%%%%%%%%%%%%%%%%%%%%%%%%%%%%%%%%%%%%%%%%%%
\vspace{0.1in}
\section{Metric reconstruction}
\label{sec:metric_reconstruction}
    
This section reviews how to reconstruct the perturbed metric and the corresponding NP quantities from solutions to the Teukolsky equation for Kerr BHs in GR. There are two approaches to metric reconstruction in general: the first approach involves systematically solving the Bianchi identities, Ricci identities, and commutation relations \cite{Chandrasekhar_1983, Loutrel_Ripley_Giorgi_Pretorius_2020}, whereas the second approach, or the CCK procedures, utilizes an intermediate Hertz potential to reconstruct the metric \cite{Cohen_Kegeles_1975, Chrzanowski:1975wv, Kegeles_Cohen_1979, Lousto_Whiting_2002, Ori_2003, Whiting_Price_2005, Yunes_Gonzalez_2006, Keidl_Friedman_Wiseman_2007, Keidl_Shah_Friedman_Kim_Price_2010}. In this work, the second approach is employed to perform metric reconstruction.

%%%%%%%%%%%%%%%%%%%%%%%%%%%%%%%%%%%%%%%%%%%%%%%%%%%%%%%%%%%%
\subsection{Metric perturbations}
\label{sec:metric_perturbations}
    
In this subsection, we present the reconstructed metric perturbation $h_{\mu\nu}^{(0,1)}$ for GR GWs. For convenience, in this section, we will drop the superscript $(0,1)$ of $h_{\mu\nu}^{(0,1)}$ and always assume that $h_{\mu\nu}$ is at $\mathcal{O}(\zeta^0,\epsilon^1)$. The CCK procedures can be carried out in two different gauge choices:
\begin{align} \label{eq:IRG-def}
    \textrm{Ingoing radiation gauge (IRG):}
    &\quad h_{\alpha\beta}l^\beta=0\,,\; h=0\,, \\
    \label{eq:ORG-def}
    \textrm{Outgoing radiation gauge (ORG):}
    &\quad h_{\alpha\beta}n^\beta=0\,,\; h=0\,,
\end{align}
where $h$ is the trace of $h_{\alpha\beta}$ with respect to the background metric. The reconstructed metric $h_{\alpha\beta}$ in the IRG and ORG are given in Eqs.~\eqref{eq:metricpert_irg} and \eqref{eq:metricpert_org}, respectively \cite{Keidl_Friedman_Wiseman_2007, Keidl_Shah_Friedman_Kim_Price_2010},
\begin{widetext}
\begin{align} 
    (h_{\alpha\beta})_{\textrm{IRG}}
    =& \left[ l_\alpha l_\beta \left( \bar{\delta}_{[1,3,0,-1]} \bar{\delta}_{[0,4,0,3]}-\lambda D_{[0,4,0,3]}\right)
    +\bar{m}_\alpha\bar{m}_\beta\left( D_{[-1,3,0,-1]}D_{[0,4,0,3]}\right) \right. \nonumber\\ 
    &\left.-l_{(\alpha}\bar{m}_{\beta)}\left(D_{[1,3,1,-1]} \bar{\delta}_{[0,4,0,3]} \right)+\bar{\delta}_{[-1,3,-1,-1]}
    D_{[0,4,0,3]}\right]\bar{\Psi}_\Hertz+\textrm{c.c.} \,, \label{eq:metricpert_irg} \\
    (h_{\alpha\beta})_{\textrm{ORG}}
    =& -\rho^{-4}\left[ n_\alpha n_\beta\left(\delta_{[-3,-1,5,0]} \delta_{[-4,0,1,0]}\right)
    +m_\alpha m_\beta\left(\Delta_{[0,5,1,-3]}\Delta_{[0,1,0,-4]}\right) \right. \nonumber\\ 
    & \left.+n_{(\alpha}m_{\beta)}\left( \delta_{[-3,1,5,1]} \Delta_{[0,1,0,-4]}\right)
    +\Delta_{[-1,5,-1,-3]}\delta_{[-4,0,1,0]} \right]\Psi_\Hertz
    +\textrm{c.c.} \,, \label{eq:metricpert_org}
\end{align}
\end{widetext}
where the notation for the derivatives is given by Eq.~\eqref{eq:derivative-notation}, and $\bar{\Psi}_\Hertz$ is the complex conjugate of the Hertz potential. We have also dropped the superscript $(0,1)$ of the Hertz potential for simplicity. Notice, since we use an opposite signature from \cite{Keidl_Friedman_Wiseman_2007, Keidl_Shah_Friedman_Kim_Price_2010}, our Eqs.~\eqref{eq:metricpert_irg} and \eqref{eq:metricpert_org} have an opposite sign from the results in \cite{Keidl_Friedman_Wiseman_2007, Keidl_Shah_Friedman_Kim_Price_2010}.

%%%%%%%%%%%%%%%%%%%%%%%%%%%%%%%%%%%%%%%%%%%%%%%%%%%%%%%%%%%%
\subsection{Hertz potential}
\label{sec:hertz_potential}

The Hertz potential $\Psi_{\Hertz}$ that appears in Eq.~\eqref{eq:metricpert_irg} in the IRG and Eq.~\eqref{eq:metricpert_org} in the ORG satisfies the Teukolsky equation for $\rho^{-4}\Psi_4^{(0,1)}$ and $\Psi_0^{(0,1)}$, respectively~\cite{Cohen_Kegeles_1975, Chrzanowski:1975wv, Kegeles_Cohen_1979, Campanelli_Lousto_1999, Whiting_Price_2005}. For convenience, let us drop the superscript ${(0,1)}$ of $\Psi_{0,4}^{(1,1)}$ and always assume that $\Psi_{0,4}$ are at $\mathcal{O}(\zeta^0,\epsilon^1)$ in this subsection. Using the perturbed metric in Eq.~\eqref{eq:metricpert_irg}, the relation between the Hertz potential $\Psi_{\Hertz}$ and $\Psi_{0,4}$ can be found by directly evaluating the Riemann tensor or by using the Ricci identities. In the IRG, the perturbed Weyl scalars can then be expressed in terms of the Hertz potential using
\begin{subequations} \label{eq:Hertz_IRG}
    \begin{align}
        & \Psi_{0}=-\frac{1}{2}D^{4}\bar{\Psi}_{\Hertz}\,, \label{eq:Hertz_IRG_Psi0} \\
        & \Psi_{4}
        =-\frac{1}{8}\rho^{4}\left[\mathcal{L}^{\dagger4}\bar{\Psi}_{\Hertz}
        -12M\partial_{t}\Psi_{\Hertz}\right]\,, \label{eq:Hertz_IRG_Psi4}
    \end{align}
\end{subequations}
and in the ORG,
\begin{subequations} \label{eq:Hertz_ORG}
    \begin{align}
	    & \Psi_{4}
	    =-\frac{1}{32}\rho^{4}\Delta^{2}D^{\dagger4}
        \Delta^{2}\bar{\Psi}_{\Hertz}\,, \label{eq:Hertz_ORG_Psi4} \\
	    & \Psi_{0}=-\frac{1}{8}\left[\mathcal{L}^{4}\bar{\Psi}_{\Hertz}
	    +12M\partial_{t}\Psi_{\Hertz}\right]\,, \label{eq:Hertz_ORG_Psi0}
    \end{align}
\end{subequations}
where $\rho=-1/(1-ia\cos{\theta})$ and
\begin{equation}
    \begin{aligned}
        & D=l^{\mu}\partial_{\mu}
        =\frac{r^2+a^2}{\Delta}\partial_t+\partial_r
        +\frac{a}{\Delta}\partial_{\phi}\,, \\
        & D^{\dagger}
        =-\frac{r^2+a^2}{\Delta}\partial_t+\partial_r
        -\frac{a}{\Delta}\partial_{\phi}\,, \\
        & \mathcal{L}_{s}^{\dagger}=ia\sin{\theta}\partial_t-
        \left[\partial_{\theta}-s\cot{\theta}
        -i\csc{\theta}\partial_{\phi}\right]\,, \\
        & \mathcal{L}^{\dagger 4}
        =\mathcal{L}_1^{\dagger}\mathcal{L}_0^{\dagger}
        \mathcal{L}_{-1}^{\dagger}\mathcal{L}_{-2}^{\dagger}\,.
    \end{aligned}
\end{equation}
Note that the $D$ operator was introduced before in Eq.~\eqref{eq:derivative-notation}, but here we provide its expression using the Kinnersely tetrad. We point out that these are the same operators that appear in the Teukolsky-Starobinsky identity~\cite{Teukolsky:1974yv, Chandrasekhar_1983}. Notice, Eqs.~\eqref{eq:Hertz_IRG} and \eqref{eq:Hertz_ORG} follow~\cite{Keidl_Friedman_Wiseman_2007}, which corrected a factor of one-half in earlier papers~\cite{Cohen_Kegeles_1975, Chrzanowski:1975wv, Kegeles_Cohen_1979}. 

Similar to the perturbations of $\Psi_0$ and $\Psi_4$, the Hertz potential can be defined in the coordinate basis $(t,r,\theta,\phi)$ as 
\begin{subequations} \label{eq:Hertz_decompose}
\begin{align}
    \textrm{IRG}:\quad &\bar{\Psi}_{\Hertz}
    ={}_{2}\hat{R}_{\ell m}(r)\,{}_{2}\mathcal{Y}_{\ell m}(\theta,\phi)e^{-i\omega t} \,, \\
    \textrm{ORG}:\quad &\bar{\Psi}_{\Hertz}
    ={}_{-2}\hat{R}_{\ell m}(r)\,{}_{-2}\mathcal{Y}_{\ell m}(\theta,\phi)e^{-i\omega t} \,,
\end{align}
\end{subequations}
where ${}_{\pm2}\mathcal{Y}_{\ell m}(\theta,\phi)={}_{\pm2}S_{\ell m}(\theta)e^{im\phi}$ are spin-weighted spheroidal harmonics of spin weight $\pm2$ solving the angular Teukolsky equation in GR. ${}_{\pm2}\hat{R}_{\ell m}(r)$ are radial functions that can be expressed in terms of the radial Teukolsky functions ${}_{2}R_{\ell m}^{(0,1)}(r)$ and ${}_{-2}R_{\ell m}^{(0,1)}(r)$ of $\Psi_{0}^{(0,1)}$ and $\rho^{-4}\Psi_{0}^{(0,1)}$, respectively, by inverting Eqs.~\eqref{eq:Hertz_IRG_Psi0} and \eqref{eq:Hertz_ORG_Psi4} using the Teukolsky-Starobinsky identity \cite{Ori_2003},
\begin{subequations} \label{eq:hertztoteukall}
\begin{align}
    {}_{2}\hat{R}_{\ell m}(r)
    & =-\frac{2}{\mathfrak{C}}\Delta^2(D^\dagger_{m\omega})^4 \left[\Delta^2\,{}_{2}R_{\ell m}^{(0,1)}(r) \right]\,, \label{eq:hertztoteuk}\\
    {}_{-2}\hat{R}_{\ell m}(r)
    & =-\frac{32}{\mathfrak{C}}(D_{m\omega})^4
    {}_{-2}R_{\ell m}^{(0,1)}(r)\,. \label{eq:hertztoteuk_ORG}
\end{align}
\end{subequations}
Here, the operators $D_{m\omega}$ and $D^\dagger_{m\omega}$ are mode decomposition of $D$ and $D^\dagger$, respectively \cite{Ori_2003}, 
\begin{equation} \label{eq:reducedop}
\begin{aligned}
    & D_{m\omega}
    =\partial_r+i\frac{am-(r^2 +a^2)\omega}{\Delta}\,, \\
    & D^\dagger_{m\omega}
    =\partial_r-i\frac{am-(r^2 +a^2)\omega}{\Delta}\,.
\end{aligned}
\end{equation}
$\mathfrak{C}$ is the mode-dependent Teukolsky-Starobinsky constant \cite{Teukolsky:1974yv, Starobinsky:1973aij, Starobinskil:1974nkd, Ori_2003} 
\begin{align}
    \label{eq:p-value}
    \mathfrak{C}
    =& \;\lambda^2\left(\lambda+2\right)^2-8\omega^2\lambda \left[\tilde{\alpha}^2\left(5\lambda+6\right)-12a^2\right]\nonumber \\ 
    & \;+144\omega^4\tilde{\alpha}^4+144\omega^2M^2 \,,
\end{align}
where $\tilde{\alpha}=a^2-am/\omega$ and $\lambda={}_{s}A_{\ell m}+s+|s|$ with ${}_{s}A_{\ell m}$ being the Teukolsky's angular separation constant~\cite{Teukolsky:1973ha}. For a Schwarzschild BH, ${}_{s}A_{\ell m}=(\ell-s)(\ell+s+1)$. Furthermore, one can notice from Eq.~\eqref{eq:Hertz_IRG} that any $(\ell,m,\omega)$ mode of $\Psi_0^{(0,1)}$ in the IRG generates a mixture of $(\ell,m,\omega)$ and $(\ell,-m,-\bar{\omega})$ modes of $\Psi_4^{(0,1)}$. Thus, it is more convenient to use the ORG when solving the modified Teukolsky equation of $\Psi_4^{(1,1)}$. For a similar reason, when solving the modified Teukolsky equation of $\Psi_0^{(1,1)}$, we will use the IRG.

Substituting the differential operators $D^\dagger_{m\omega}$ and $D_{m\omega}$ in Eq.~\eqref{eq:reducedop} into the expression for the radial part of the Hertz potential in Eq.~\eqref{eq:hertztoteukall}, we have
\begin{subequations} \label{eq:hertztoteukcoord}
    \begin{align}
    {}_{s}\hat{R}_{\ell m}(r)
    =& {}_s f^{\ell m}_1(r,\omega,M) {}_{s}R_{\ell m}^{(0,1)}(r) \nonumber \\ 
    & \quad + {}_s f^{\ell m}_2(r,\omega,M) {}_{s}{R'}_{\ell m}^{(0,1)}(r) \,, \\
    {}_{s}\hat{R}'_{\ell m}(r)
    =&{}_s f^{\ell m}_3(r,\omega,M) {}_{s}R_{\ell m}^{(0,1)}(r) \nonumber \\ 
    & \quad + {}_s f^{\ell m}_4(r,\omega,M) {}_{s}{R'}_{\ell m}^{(0,1)}(r) \,, 
    \end{align}
\end{subequations}
where we have made use of the radial Teukolsky equation to reduce all second- and higher-order derivatives of ${}_{s}{R}_{\ell m}^{(0,1)}(r)$. In Eq.~\eqref{eq:hertztoteukcoord}, the prime denotes the first derivative with respect to the radial coordinate $r$. The functions $f_i$ are spin weight $s$ and mode dependent. These functions are lengthy and non-illuminating, so they have been presented in a separate Mathematica notebook~\cite{Pratikmodteuk}.

%%%%%%%%%%%%%%%%%%%%%%%%%%%
\subsection{Spin-weighted spheroidal harmonics} \label{sec:sphharm}

Spin-weighted spheroidal harmonics that appear in Eq.~\eqref{eq:Hertz_decompose} are solutions to the angular Teukolsky equation in GR~\cite{Teukolsky:1973ha, Teukolsky:1974yv}. In general, these are eigenfunctions of an equation of the form~\cite{Shah:2015sva}
\begin{align} \label{eq:swsh_equation}
    \frac{1}{\sin\theta}&\frac{d}{d\theta}\left(\sin\theta \frac{dz}{d\theta} \right) -\left( \frac{m^2 + s^2 + 2 m s \cos\theta}{\sin^2 \theta} \right. \nonumber \\ & \left.-\gamma^2 \cos^2 \theta+ 2s\gamma \cos\theta - {}_s A^\gamma_{\ell m} \right) z  = 0 \,,
\end{align}
where $s$ represents the spin weight, ${}_s A^\gamma_{\ell m}$ is the eigenvalue of the equation, which has been numerically calculated in the literature~\cite{Teukolsky:1974yv}. Comparing Eq.~\eqref{eq:swsh_equation} with the angular Teukolsky equation in GR \cite{Teukolsky:1973ha}, we see that $\gamma=\chi M\omega$. In the slow-rotation expansion, spin-weighted spheroidal harmonics ${}_s \mathcal{Y}^\gamma_{\ell m}$ can be expanded as~\cite{Shah:2015sva,Teukolsky:1974yv}
\begin{align}\label{eq:StoY}
    {}_s \mathcal{Y}^\gamma_{\ell m}
    =& \;{}_s Y_{\ell m}+\gamma
    \left({}_s b^m_{\ell,\ell+1 }\,{}_sY_{\ell+1\, m}
    +{}_s b^m_{\ell,\ell-1 }\,{}_s Y_{\ell-1\,m}\right)\nonumber \\ 
    & \;+\mathcal{O}(\gamma^2) \,,
\end{align}
where ${}_sY_{\ell m}$ are spin-weighted spherical harmonics with spin weight $s$. The factors ${}_s b_{\ell,\ell\pm1}^m $ in Eq.~\eqref{eq:StoY} hold the form~\cite{Shah:2015sva}
\begin{subequations} \label{eq:bvalues}
\begin{align}
    {}_s b_{\ell,\ell+1}^m 
    & =-\frac{s\sqrt{\left[(\ell+1)^2-m^2\right]
    \left[(\ell+1)^2-s^2\right]}}{(\ell+1)^2\sqrt{(2\ell+1)(2\ell+3)}}\,, \\
    {}_s b_{\ell,\ell-1}^m 
    &=\frac{s\sqrt{\left(\ell^2-m^2\right)
    \left(\ell^2-s^2\right)}}{\ell^2\sqrt{(2\ell-1)(2\ell+1)}}\,.
\end{align}
\end{subequations}
To evaluate spin-weighted spherical harmonics, one can use
\begin{widetext}
    \begin{align} \label{eq:spherical_formlua}
    { }_s Y_{\ell m}(\theta, \phi)
    =(-1)^{\ell+m-s} \sqrt{\frac{(\ell+m)!(\ell-m)!(2 \ell+1)}{4\pi(\ell+s)!(\ell-s) !}}\sin^{2\ell}\left(\frac{\theta}{2}\right)e^{im\phi}
    \times\sum_{q=0}^{\ell-s}(-1)^q\left(
    \begin{array}{c}
    \ell-s \\ q
    \end{array}
    \right)\left(
    \begin{array}{c}
    \ell+s \\ q+s-m
    \end{array}
    \right)\cot^{2q+s-m}\left(\frac{\theta}{2}\right)\,.
\end{align}
\end{widetext}
In the special case that $s=0$, the spin-weighted spherical harmonics become the standard spherical harmonics, i.e.,
\begin{equation}
    {}_0Y_{\ell m}=Y_{\ell m}\,.
\end{equation}
Spin-weighted spherical harmonics and spin-weighted spheroidal harmonics also obey the following orthogonality relations,
\begin{align}
    & \int_{S^2}dS\;{}_s\mathcal{Y}^{\gamma}_{\ell m}\;
    {}_s\bar{\mathcal{Y}}^{\gamma}_{\ell' m'}
    =\delta_{\ell \ell'} \delta_{mm'}\,, 
    \label{eq:ortho_spheroidal} \\
    & \int_{S^2}dS\;{}_sY_{\ell m}\;
    {}_s\bar{Y}_{\ell' m'}
    =\delta_{\ell \ell'} \delta_{mm'}
    \label{eq:ortho_spherical}\,, 
\end{align}
where $dS$ is the solid angle element, and the integration is over the entire 2-sphere. 
At certain places, we might drop the superscript $\gamma$ of ${}_{s}\mathcal{Y}^{\gamma}_{\ell m}$ denoting its eigenvalue for simplicity. 

A spin-weighted spherical harmonic ${}_sY_{\ell m}$ with spin weight $s$ can also be raised to ${}_{s+1}Y_{\ell m}$ of spin weight $s+1$ via the raising operator $\eth$ or lowered to ${}_{s-1}Y_{\ell m}$ of spin weight $s$ via the lowering operator $\bar{\eth}$. The operators $\eth$ and $\bar{\eth}$ are defined to be \cite{Goldberg:1966uu}
\begin{subequations} \label{eq:eth_ethc}
\begin{align}
    & \eth\,{}_{s}\mathcal{F}
    =-\left(\partial_{\theta}-i\csc{\theta}\partial_\phi
    -s\cot{\theta}\right){}_{s}\mathcal{F}\,, \\
    & \bar{\eth}\,{}_{s}\mathcal{F}
    =-\left(\partial_{\theta}+i\csc{\theta}\partial_\phi
    +s\cot{\theta}\right){}_{s}\mathcal{F}\,,
\end{align}
\end{subequations}
where ${}_{s}\mathcal{F}$ is some function of spin weight $s$, such that
\begin{subequations} \label{eq:eth_ethc_on_Y}
\begin{align}
    & \eth\,{}_{s}Y_{\ell m}
    =[(l-s)(l+s+1)]^{1/2}~_{s+1}\!Y_{\ell m}\,, \\
    & \bar{\eth}\,{}_{s}Y_{\ell m}=-[(l+s)(l-s+1)]^{1/2}~_{s-1}\!Y_{\ell m}\,.
\end{align}
\end{subequations}
One can further rewrite the directional derivatives $\delta^{(0,0,0)}$ and $\bar{\delta}^{(0,0,0)}$ on the Schwarzschild background in terms of $\eth$ and $\bar{\eth}$, respectively, i.e.,
\begin{subequations} \label{eq:raising_lowering}
\begin{align}
    & \delta^{(0,0,0)}{}_{s}\mathcal{F}
    =-\frac{1}{\sqrt{2}r}\left(\bar{\eth}-s\cot\theta\right){}_{s}\mathcal{F}\,,
    \label{eq:raising} \\
    & \bar{\delta}^{(0,0,0)}{}_{s}\mathcal{F}
    =-\frac{1}{\sqrt{2}r}\left(\bar{\eth}+s\cot\theta\right){}_{s}\mathcal{F}\,.
    \label{eq:lowering}
\end{align}
\end{subequations}
By expanding $\delta^{(0,0)}$ and $\bar{\delta}^{(0,0)}$ in terms of $\delta^{(0,0,0)}$ and $\bar{\delta}^{(0,0,0)}$ in the slow-rotation limit, one can also replace $\delta^{(0,0)}$ and $\bar{\delta}^{(0,0)}$ with $\eth$ and $\bar{\eth}$. In Secs.~\ref{sec:source_scalar}--\ref{sec:source_Teuk_ORG}, we will use Eq.~\eqref{eq:raising_lowering} to simplify the terms with $\delta^{(0,0)}$ and $\bar{\delta}^{(0,0)}$ acting on spin-weighted spherical harmonics.

%%%%%%%%%%%%%%%%%%%%%%%%%%%%%%%%%%%%%%%%%%%%%%%%%%%%%%%%%%%%
\subsection{Perturbed NP quantities}

As we are working within the NP basis, in addition to the perturbed metric given by Eq.~\eqref{eq:metricpert_irg}, we also require the reconstruction of the perturbed NP quantities, such as the perturbed tetrad, Weyl scalars, and spin coefficients. We adopt the methodology outlined in~\cite{Campanelli_Lousto_1999, Loutrel_Ripley_Giorgi_Pretorius_2020} to perform this reconstruction. The first step involves expressing the perturbed tetrad in terms of the background tetrad. This is accomplished by expanding the perturbed tetrad in terms of the background tetrad, and then utilizing the transformation properties of the tetrad to obtain the perturbed tetrad components in terms of the background tetrad components such that
\begin{equation} \label{eq:tetrad_expansion}
    e^{\mu(0,1)}_{a}=A_{a}{}^{b(0,1)}e^{\mu(0,0)}_{b}\,,
\end{equation}
where $e^{\mu}_{a}$ represents a null tetrad such that
\begin{equation}
    e^{\mu}_{a}=\{l^\mu, n^\mu, m^\mu, \bar{m}^\mu \} \,,
\end{equation}
and $A_{a}{}^{b}$ are coefficients that map the background tetrad to the perturbed tetrad.

As shown in~\cite{Campanelli_Lousto_1999, Loutrel_Ripley_Giorgi_Pretorius_2020}, one can always perform tetrad rotations to set six real parameters of the $A_{a}{}^{b(0,1)}$ coefficients to zero. Then, expanding $h_{\mu\nu}$ in terms of $e^{a(0,1)}_{\mu}$ and $e^{a(0,0)}_{\mu}$ and using the linearized completeness relation, we find
\begin{equation} \label{eq:completeness_perturbed}
\begin{aligned}
    h_{\mu\nu}^{(0,1)}
    =& \;-2\left[l_{(\mu}^{(0,1)}n_{\nu)}^{(0,0)}
    +l_{(\mu}^{(0,0)}n_{\nu)}^{(0,1)}\right. \\
    & \;\left.-m_{(\mu}^{(0,1)}\bar{m}_{\nu)}^{(0,0)}
    -m_{(\mu}^{(0,0)}\bar{m}_{\nu)}^{(0,1)}\right]\,.
\end{aligned}
\end{equation}
Comparing Eq.~\eqref{eq:completeness_perturbed} to Eq.~\eqref{eq:tetrad_expansion}, we find~\cite{Campanelli_Lousto_1999, Loutrel_Ripley_Giorgi_Pretorius_2020},
\begin{equation} \label{eq:perturbed_tetrad}
    \begin{aligned}
        & l^{\mu(0,1)}=\frac{1}{2}h_{ll}^{(0,1)}n^{\mu}\,,\quad \\
        & n^{\mu(0,1)}=\frac{1}{2}h_{nn}^{(0,1)}l^{\mu}
        +h_{ln}^{(0,1)}n_{\mu}\,,\quad \\
        & m^{\mu(0,1)}=h_{nm}^{(0,1)}l^{\mu}
        +h_{lm}^{(0,1)}n^{\mu}
        -\frac{1}{2}h_{m\bar{m}}^{(0,1)}m^{\mu}
        -\frac{1}{2}h_{mm}^{(0,1)}\bar{m}^{\mu}\,,
    \end{aligned}
\end{equation}
where we have dropped the superscripts of $e^{\mu(0,0)}_{a}$ for simplicity. Since we have adopted the sign convention in~\cite{Chandrasekhar_1983}, our signature is opposite to that used in~\cite{Campanelli_Lousto_1999, Loutrel_Ripley_Giorgi_Pretorius_2020}. Therefore, the perturbed tetrad in Eq.~\eqref{eq:perturbed_tetrad} has an opposite sign from the results of~\cite{Campanelli_Lousto_1999, Loutrel_Ripley_Giorgi_Pretorius_2020}, as seen in Eq.~\eqref{eq:completeness_perturbed}. Equation~\eqref{eq:perturbed_tetrad} works for both the IRG and ORG. In the IRG or ORG, we can further set $h_{la}^{(0,1)}=h_{m\bar{m}}^{(0,1)}=0$ or $h_{na}^{(0,1)}=h_{m\bar{m}}^{(0,1)}=0$ in Eq.~\eqref{eq:perturbed_spin_coefs}, respectively, where $a$ is any index in the NP basis.
	
For the spin coefficients, one can linearize the commutation relation following~\cite{Chandrasekhar_1983},
\begin{equation} \label{eq:commutation}
    \left[e_{a}^{\mu},e_{b}^{\mu}\right]
    =\left(\gamma^{c}{}_{ba}-\gamma^{c}{}_{ab}\right)e_{c}^{\mu}
    =C_{ab}{}^{c}e_{c}^{\mu}\,,
\end{equation}
where $\gamma^{a}{}_{bc}$ is the Ricci rotation coefficients. Using the relation between spin coefficients and Ricci rotation coefficients in Eq.~\eqref{eq:spin_coefs}, one can write $C_{ab}{}^{c}$ in terms of spin coefficients, as listed in Eq.~\eqref{eq:C_spin_coefs}~\cite{Newman:1961qr,Chandrasekhar_1983}. From Eq.~\eqref{eq:C_spin_coefs}, one can also solve for spin coefficients in terms of $C_{ab}{}^{c}$, and the results are in Eq.~\eqref{eq:spin_coefs_C}. Expanding Eq.~\eqref{eq:commutation} using Eq.~\eqref{eq:tetrad_expansion}, one finds
\begin{equation} \label{eq:C_expand}
    \begin{aligned}
        C_{ab}{}^{c(0,1)}
        =& \;\partial_{a}A_{b}{}^{c(0,1)}
        -\partial_{b}A_{a}{}^{c(0,1)} \\
        &\;-\left(A_{a}{}^{d(0,1)}C_{bd}{}^{c}
        -A_{b}{}^{d(0,1)}C_{ad}{}^{c}
        +A_{d}{}^{c(0,1)}C_{ab}{}^{d}\right)\,,  
    \end{aligned}
\end{equation}
where the superscripts of $C_{ab}{}^{c(0,0)}$ are dropped for convenience. The coefficients $A_{a}{}^{b(0,1)}$ can be retrieved from Eq.~\eqref{eq:perturbed_tetrad}. The GR structure constants $C_{ab}{}^{c(0,0)}$ are directly given by Eq.~\eqref{eq:C_spin_coefs} and the spin coefficients in GR. With all the quantities in Eq.~\eqref{eq:C_expand}, one can then use Eq.~\eqref{eq:C_expand} and \eqref{eq:spin_coefs_C} to evaluate the spin coefficients at $\mathcal{O}(\zeta^0,\epsilon^1)$. We have listed our result in Eq.~\eqref{eq:perturbed_spin_coefs}, which works for both the ORG and IRG. Our result is consistent with \cite{Loutrel_Ripley_Giorgi_Pretorius_2020} up to the overall minus sign due to different signatures, which corrects some errors in \cite{Campanelli_Lousto_1999}. 
	
To reconstruct Weyl scalars, one can either directly linearize the Riemann tensor and project it onto the NP basis to find Weyl scalars or use the Ricci identities in Eq.~\eqref{eq:Weyl_scalar_Ricci}. For both approaches, we use the perturbed tetrad in Eq.~\eqref{eq:perturbed_tetrad}, and we check that the results are consistent. We also compare our results in the IRG to the equations in \cite{Keidl_Friedman_Wiseman_2007}, which corrected a factor of $1/2$ missed in \cite{Kegeles_Cohen_1979} and are listed in Eq.~\eqref{eq:Weyl_scalar_Hertz}. After expressing everything in terms of the Hertz potential, our results of $\Psi_{0,1,2,4}^{(0,1)}$ in the IRG agree perfectly with Eq.~\eqref{eq:Weyl_scalar_Hertz} but not for $\Psi_{3}^{(0,1)}$. Since $\Psi_{3}^{(0,1)}$ is not invariant either under tetrad rotations or coordinate transformations at $\mathcal{O}(\zeta^0,\epsilon^1)$, this difference indicates that we might have a $\mathcal{O}(\zeta^0,\epsilon^1)$ difference in the choices of coordinate or tetrad. 
	
For coordinate- and tetrad-invariant quantities $\Psi_{0,4}^{(0,1)}$, our results are consistent with \cite{Kegeles_Cohen_1979, Keidl_Friedman_Wiseman_2007}. In addition, since $\Psi_2^{(0,1)}$ is invariant under tetrad rotations but not coordinate transformations at $\mathcal{O}(\zeta^0,\epsilon^1)$ [i.e., Eqs.~\eqref{eq:Weyl_scalars_rotated_01} and \eqref{eq:Weyl_scalars_coord_01}], we are in the same coordinate as \cite{Kegeles_Cohen_1979, Keidl_Friedman_Wiseman_2007}, consistent with that we all use the IRG. Thus, the difference in $\Psi_3^{(0,1)}$ is only due to tetrad choices at $\mathcal{O}(\zeta^0,\epsilon^1)$, where we explicitly follow the convention in \cite{Campanelli_Lousto_1999, Loutrel_Ripley_Giorgi_Pretorius_2020}, but Refs.~\cite{Kegeles_Cohen_1979, Keidl_Friedman_Wiseman_2007} were not explicit about their tetrad choices at $\mathcal{O}(\zeta^0,\epsilon^1)$. More specifically, we find that the tetrad in Eq.~\eqref{eq:perturbed_tetrad} after setting $h_{la}^{(0,1)}=h_{m\bar{m}}^{(0,1)}=0$ differs from the tetrad in \cite{Kegeles_Cohen_1979, Keidl_Friedman_Wiseman_2007} by a type I rotation. In Schwarzschild, this difference in $\Psi_{3}^{(0,1)}$ can be compactly written as
\begin{equation}
    \Psi_3^{(0,1)}=\Psi_{3,\CCK}^{(0,1)}
    +\frac{3}{2}\Psi_2h^{(0,1)}_{n\bar{m}}\,,
\end{equation}
where $\Psi_{3,\CCK}^{(0,1)}$ is the result in \cite{Kegeles_Cohen_1979, Keidl_Friedman_Wiseman_2007}. The results of other Weyl scalars at $\mathcal{O}(\zeta^0,\epsilon^1)$ in Schwarzschild are listed in Eq.~\eqref{eq:Weyl_scalar_Hertz_Schw}. For Kerr, we do not have such a simple correction to $\Psi_3^{(0,1)}$, so we will just use the Ricci identity in Eq.~\eqref{eq:Weyl_scalar_Ricci_Psi3}. Similarly, in the ORG, no previous literature provided results of all the Weyl scalars in terms of $\Psi_{\Hertz}^{(0,1)}$ directly, so we also use the Ricci identity to evaluate them.

When deriving the modified Teukolsky equations, we made the gauge choice that $\Psi_{1,3}^{(0,1)}=0$, but this is not the case for the tetrad in Eq.~\eqref{eq:perturbed_tetrad}, as one can see in Eqs.~\eqref{eq:Weyl_scalar_Ricci}--\eqref{eq:Weyl_scalar_Hertz_Schw}. Thus, to be consistent with the gauge we chose for the modified Teukolsky equations, we need to perform additional type I and type II rotations to remove $\Psi_{1,3}^{(0,1)}$. From Eq.~\eqref{eq:Weyl_scalars_rotated_01}, we find the rotation parameters to be
\begin{equation} \label{eq:rotate_coefs_01}
    a^{(0,1)}=-\frac{\bar{\Psi}_{3}^{(0,1)}}{3\Psi_2}\,,\quad
    b^{(0,1)}=-\frac{\Psi_{1}^{(0,1)}}{3\Psi_2}\,.
\end{equation}
Since $\Psi_{0,2,4}^{(0,1)}=0$ are invariant under tetrad rotations at $\mathcal{O}(\zeta^0,\epsilon^1)$, one can continue using Eqs.~\eqref{eq:Weyl_scalar_Ricci}-\eqref{eq:Weyl_scalar_Hertz_Schw} by just setting $\Psi_{1,3}^{(0,1)}=0$. For spin coefficients, their values after the rotation are listed in Eq.~\eqref{eq:spin_coefs_rotated_01} following \cite{Chandrasekhar_1983}. With these reconstructed quantities, we are now ready to evaluate the source terms in the equation of $\vartheta^{(1,1)}$ in Eq.~\eqref{eq:EOM_scalar_11}.

%%%%%%%%%%%%%%%%%%%%%%%%%%%%%%%%%%%%%%%%%%%%%%%%%%%%%%%%%%%%
%%%%%%%%%%%%%%%%%%%%%%%%%%%%%%%%%%%%%%%%%%%%%%%%%%%%%%%%%%%%
\section{The evolution equation for $\vartheta^{(1,1)}$ in the IRG}
\label{sec:source_scalar}
	
\begin{figure*}
    \centering
    \begin{tikzpicture}
    \node (a) at (0,0) [label= \textrm{Scalar field equation for $\vartheta^{(1,1)}$ in the IRG}]{};
    \draw (-4,0) -- (4,0);
    \draw (-4,0.8) -- (4,0.8);
    \draw (-4,0) -- (-4,0.8);
    \draw (4,0) -- (4,0.8);
    %\draw (0,0.4) ellipse (3.5cm and 0.4cm);
    \draw [->,thick] (0,0) -- (-4,-1);
    \draw [->,thick] (0,0) -- (4,-1);
    \node (b1) at (-4,-1) [label=below: \textrm{Left-hand side of Eq.~\eqref{eq:EOM_scalar_11}}]{};
    \node (b2) at (4,-1) [label=below: \textrm{Right-hand side of Eq.~\eqref{eq:EOM_scalar_11}}]{};
    \draw (-4,-1.37) ellipse (2.5cm and 0.3cm);
    \draw (4,-1.37) ellipse (2.5cm and 0.3cm);
    \draw [->,thick] (-4,-1.7) -- (-4,-2.5);
    \draw [->,thick] (4,-1.7) -- (4,-2.5);
    \node (c1a) at (-4,-2.1) [label=left: \textrm{In coordinate basis}]{};
    \node (c1) at (-4,-2.5) [label=below: \textrm{Equation~\eqref{eq:H_theta_coord}}]{};
    \draw (-4,-2.87) ellipse (2cm and 0.3cm);
    \draw (-4,-4.37) ellipse (2cm and 0.3cm);
    \draw [->,thick] (-4,-3.2) -- (-4,-4);
    \node (c2a) at (-4,-3.6) [label=left: \textrm{Separated radial equation}]{};
    \node (c2) at (-4,-4) [label=below: \textrm{Equation~\eqref{eq:scalar_hom_radial}}]{};
    \node (c3) at (4,-2.45) [label=below: \textrm{Source term $\mathcal{S}_{\vartheta}^{(1,1)}$ rewritten as Eq.~\eqref{eq:source_scalar_11_decomposed}}]{};
    \draw (4,-2.88) ellipse (3.5cm and 0.33cm);
    \draw [->,thick] (4,-3.2) -- (4,-4.5);
    \node (c4a) at (4,-3.7) [label=right:{\textrm{In coordinate basis}}]{};
    \node (c4b) at (4,-4.1) [label=right:{\textrm{using the Hertz potential Eq.~\eqref{eq:Hertz_decompose_slow_rot}}}]{};
    \node (c5) at (4,-5.7) [label={[align=left]\textrm{Expression for $(R^*\!R)^{(0,1)}$ and $\square^{(0,0,1)}\vartheta^{(1,1,0)}$ in } \\ \textrm{ $\{t,r,\theta,\phi\}$ coordinates in Eq.~\eqref{eq:R*R_coord_IRG} and Eq.~\eqref{eq:box_coord_IRG} } }]{};
    \draw (4,-5.13) ellipse (4.4cm and 0.59cm);
    \draw [->,thick] (4,-5.7) -- (4,-6.5);
    \node (c6) at (4,-6.1) [label=right:{\textrm{Using the recipe in Sec.~\ref{sec:elim-ylm}}}]{};
    \node (c7) at (4,-7.6) [label={[align=center]\textrm{Source terms as functions of the radial coordinate} \\ \textrm{in the IRG, $V_{\ell m}^{R}(r)$ and $V_{\ell m}^\Box (r)$,  are given by Eqs.~\eqref{eq:R*R_radial_IRG} and \eqref{eq:box_radial_IRG}}}]{};
    \draw (4,-7.07) ellipse (5.4cm and 0.57cm);
    \node (c8) at (0,-9.6) [label={[align=center] The radial evolution equation of the scalar field perturbation $\vartheta^{(1,1)}$\\  for slowly rotating BHs in dCS gravity in the IRG is given by Eq.~\eqref{eq:scalar_radial_IRG}}]{};
    \draw [->,thick] (4,-7.63) -- (4,-8.4);
    \draw [->,thick] (-4,-4.7) -- (-4,-8.4);
    \draw (-6.5,-8.4) -- (6.5,-8.4);
    \draw (-6.5,-9.5) -- (6.5,-9.5);
    \draw (-6.5,-8.4) -- (-6.5,-9.5);
    \draw (6.5,-8.4) -- (6.5,-9.5);
    \end{tikzpicture}
    \caption{\label{Fig. 1} A schematic flowchart prescribing the steps involved in obtaining a separated radial evolution equation for the scalar field perturbation $\vartheta^{(1,1)}$ for slowly rotating BHs in dCS gravity. This flowchart summarizes the details of the calculations described in Sec.~\ref{sec:source_scalar} and Sec.~\ref{sec:scalar_radial_eqn}. Initial and final outcomes are represented by rectangular boxes, while intermediate results are symbolized by encapsulating bubbles. The directional arrows are meant to seamlessly guide the reader through the logical flow of the calculations. }
\end{figure*}
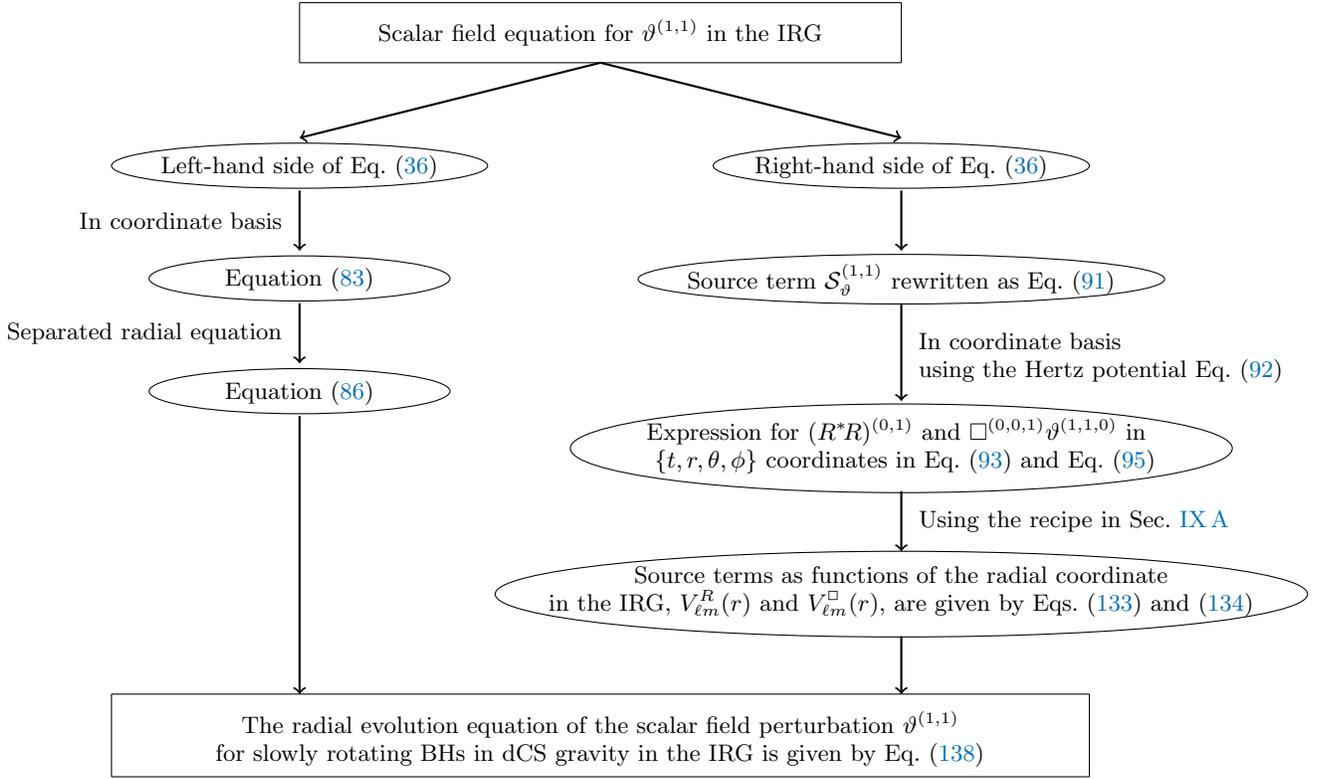

In this section, we project the equation governing $\vartheta^{(1,1)}$ [Eq.~\eqref{eq:EOM_scalar_11}] onto the NP basis using the IRG. For convenience, we define the right-hand side of Eq.~\eqref{eq:EOM_scalar_11} as
\begin{equation} 
\label{eq:source_scalar_11}
    \mathcal{S}_{\vartheta}^{(1,1)}
    \equiv-\pi^{-\frac{1}{2}}M^2\left[R^*\!R\right]^{(0,1)}
    -\square^{(0,1)}\vartheta^{(1,0)}
\end{equation}
so that the evolution equation for $\vartheta^{(1,1)}$ becomes
\begin{align} \label{eq:EOM_scalar_11_New}
    \square^{(0,0)}\vartheta^{(1,1)}
    = \mathcal{S}_{\vartheta}^{(1,1)}\,.
\end{align}
This equation is first expressed in terms of the NP quantities, following which we evaluate its left-hand side using the background NP quantities in Sec.~\ref{sec:NP_quantities_background} and Appendix~\ref{appendix:background_NP_more} and its right-hand side using the reconstructed NP quantities at $\mathcal{O}(\zeta^0,\epsilon^1)$ in Sec.~\ref{sec:metric_reconstruction} and Appendix~\ref{appendix:metric_reconstruction_more}. The same methodology demonstrated in this section is applied to computing the modified Teukolsky equation in Sec.~\ref{sec:source_Teuk}. Figure~\ref{Fig. 1} presents a schematic illustration of the steps involved in calculating a completely separated radial evolution equation for the scalar field perturbation in the IRG.

%%%%%%%%%%%%%%%%%%%%%%%%%%%%%%%%%%%%%%%%%%%%%%%%%%%%%%%%%%%%
\subsection{Projection onto the NP basis} \label{sec:projnp}

In this subsection, we project Eq.~\eqref{eq:EOM_scalar_11} onto the NP basis. This projection involves the projection of two fundamental quantities: the D'Alembertian operator $\square$ and the Pontryagin density $R^*\!R$ onto the NP basis. In particular, our goal is to express these quantities in terms of NP quantities, particularly the below-mentioned quantities in the NP basis, namely
\begin{equation} \label{eq:Ricci_components_NP}
    \begin{aligned}
        {}^{*}R_{abcd}
        =& \;\frac{1}{2}\epsilon_{cd}{}^{ef}R_{abef} \\
        \nabla_{b}\nabla_{a}\vartheta
        =& \;\nabla_{b}\left(\vartheta_{,a}\right)
        =\vartheta_{,ab}-\gamma^{d}{}_{ab}\vartheta_{,d}\,.
    \end{aligned}
\end{equation}
Here, $\eta_{ab}$ is the metric in the NP basis. The notation $f_{,a}$ denotes the directional derivative of $f$ defined by the tetrad basis $e_{a}^{\mu}$. The quantities $\gamma_{abc}$ are Ricci rotation coefficients, which can be expressed in terms of spin coefficients using Eq.~\eqref{eq:Ricci_spin}~\cite{Newman:1961qr}. The tensor $R_{abcd}$ can be expressed in terms of Weyl scalars using Eq.~\eqref{eq:Riemann_Weyl}~\cite{Newman:1961qr}. Therefore, the Pontryagin density and the $\square$ operator can be expressed in the NP basis as
\begin{subequations} \label{eq:pontandbox}
    \begin{align}
        & R^*\!R=8i(3\Psi_2^2-4\Psi_1\Psi_3+\Psi_0\Psi_4-c.c.)\,, 
        \label{eq:R*R_NP}\\
        & \begin{aligned}
        \square
        =& \;-\left[\{D,\boldsymbol{\Delta}\}-\{\delta,\bar{\delta}\}
        +(\mu+\bar{\mu}-\gamma-\bar{\gamma})D\right. \\
        & \;\left.+(\epsilon+\bar{\epsilon}-\rho-\bar{\rho})\boldsymbol{\Delta}
        +(\alpha-\bar{\beta}-\pi+\bar{\tau})\delta\right. \\
        & \;\left.+(\bar{\alpha}-\beta-\bar{\pi}+\tau)
        \bar{\delta}\right]\,.\label{eq:box_NP}
        \end{aligned}
    \end{align}
\end{subequations}
The factor of $i$ in Eq.~\eqref{eq:R*R_NP} arises from the normalization of the Levi-Civita tensor $\epsilon_{abcd}$ in the NP basis. In the literature, such as in~\cite{Carroll:2004st}, the covariant Levi-Civita tensor is typically defined as $\epsilon_{\mu\nu\cdots\gamma}=\sqrt{|g|}\tilde{\epsilon}_{\mu\nu\cdots\gamma}$, where $\tilde{\epsilon}_{\mu\nu\cdots\gamma}$ denotes the Levi-Civita symbol. However, this definition encounters issues when attempting to convert a Levi-Civita tensor from Boyer-Lindquist coordinates to the NP basis, due to the determinant of the Jacobian relating these two bases often being complex. Thus, to convert the tensor density $\tilde{\epsilon}_{\mu\nu\cdots\gamma}$ to a tensor, we instead need to define
\begin{equation} \label{eq:def_levi_civita}
    \epsilon_{\mu\nu\cdots\gamma}
    =\sqrt{-g}\,\tilde{\epsilon}_{\mu\nu\cdots\gamma}\,,
\end{equation}
which has the same normalization factor as the Einstein-Hilbert action. The absolute value in the usual definition is to impose that the Levi-Civita tensor is a real tensor in the Lorentzian signature, but the minus sign of $\sqrt{-g}$ will do the same trick. Since $\eta=1$, we find the normalization factor to be $i$ rather than $1$ from Eq.~\eqref{eq:def_levi_civita}. This is also consistent with that
\begin{equation}
    \begin{aligned}
        & \epsilon_{lnm\bar{m}}
        =\frac{1}{2}\left(\epsilon_{lnm\bar{m}}
        -\overline{\epsilon_{lnm\bar{m}}}\right)\,, 
    \end{aligned}
\end{equation}
which shows that $\epsilon_{lnm\bar{m}}$ is an imaginary number. We have now expressed all the terms in Eq.~\eqref{eq:EOM_scalar_11} in the NP basis.
	
%%%%%%%%%%%%%%%%%%%%%%%%%%%%%%%%%%%%%%%%%%%%%%%%%%%%%%%%%%%%	
\subsection{Left-hand side of Eq.~\eqref{eq:EOM_scalar_11}}
\label{sec:lhs_scalar_eqn}

In this subsection, we compute the operator $\square^{(0,0)}$ acting on $\vartheta^{(1,1)}$ to obtain the homogeneous component of Eq.~\eqref{eq:EOM_scalar_11}. The operator $\square^{(0,0)}$ can be evaluated directly using the Kerr metric presented in Eq.~\eqref{eq:kerr_metric}. Alternatively, one can use Eq.~\eqref{eq:box_NP} and the NP quantities of Kerr, expanded up to $\mathcal{O}(\chi)$, as given by Eqs.~\eqref{eq:principal_tetrad}--\eqref{eq:spin_coefs_background}, and then setting $\zeta$ to zero. We therefore find
\begin{equation}
    \square^{(0,0)}=-\frac{1}{r^2}H_{\vartheta}^{(0,0)}\,,
\end{equation}
where $H_{\vartheta}^{(0,0)}$ is the Teukolsky operator for particles with spin weight $s=0$ in \cite{Teukolsky:1973ha},
\begin{equation} \label{eq:H_theta_coord}
    \begin{aligned}
        H_{\vartheta}^{(0,0)}
        =&\;-r(r-r_s)\partial_{r}^2-2(r-M)\partial_r
        -\frac{\omega^2r^3-4m\chi\omega M^2}{r-r_s} \\
        & \;-\partial_\theta^2-\cot{\theta}\partial_\theta
        +m^2\csc^2{\theta}\,,
    \end{aligned}
\end{equation}
where we have only kept the terms up to $\mathcal{O}(\chi)$ and separated $\vartheta^{(1,1)}$ as
\begin{equation} \label{eq:scalar_separation}
    \vartheta^{(1,1)}=\Theta_{\ell m}(r){}_{0}\mathcal{Y}_{\ell m}(\theta)e^{-i\omega t}\,,
\end{equation}
or in the slow-rotation expansion
\begin{equation} \label{eq:scalar_decompose_slow_rot}
\begin{aligned}
    \vartheta^{(1,1)}
     =& \;\Theta_{\ell m}(r)
     \left[{}_{0}{Y}_{\ell m}(\theta,\phi)+\chi M\omega
     \left({}_{0}b^m_{\ell,\ell+1 }\,{}_{0}Y_{\ell+1\,m}\right.\right. \\
     & \;\left.\left.+\,{}_{0}b^m_{\ell,\ell-1}\,{}_{0}Y_{\ell-1\,m}\right)
     +\mathcal{O}(\chi^2)\right]e^{-i\omega t}\,.
\end{aligned}
\end{equation}
Thus, in the absence of sources, $\Theta_{\ell m}(r)$ satisfies
\begin{align} \label{eq:scalar_hom_radial}
    & \left[r(r-r_s)\partial_{r}^2\right. 
    \left.+2(r-M)\partial_r+\frac{\omega^2r^3-4\chi m M^2 \omega}{r-r_s}\right.\nonumber \\ 
    & \left.-{}_{0}A_{\ell m}\right]\Theta_{\ell m}(r)=0\,,
\end{align}
where ${}_{0}A_{\ell m}$ is the Teukolsky's separation constant for $s=0$ \cite{Teukolsky:1973ha}. We therefore see that the left-hand side of Eq.~\eqref{eq:EOM_scalar_11} is separable in the radial and angular coordinates. Further, in Sec.~\ref{sec:separation_variable}, we show that the complete expression in Eq.~\eqref{eq:EOM_scalar_11} can be separated into radial and angular parts using spin-weighted spheroidal harmonics of spin weight zero.

%%%%%%%%%%%%%%%%%%%%%%%%%%%%%%%%%%%%%%%%%%%%%%%%%%%%%%%%%%%%
\subsection{$\mathcal{S}_{\vartheta}^{(1,1)}$ in terms of $h^{(1,1)}$ and $\vartheta^{(1,1)}$}
	
To systematically calculate $\mathcal{S}_{\vartheta}^{(1,1)}$ in Eq.~\eqref{eq:source_scalar_11}, we can partition it into three parts based on the terms that necessitate metric reconstruction, namely those at $\mathcal{O}(\zeta^0,\epsilon^1)$:
\begin{enumerate}
    \item Weyl scalars at $\mathcal{O}(\zeta^0,\epsilon^1)$: These terms are solely determined by the Pontryagin density $(R^*\!R)^{(0,1)}$. For slowly rotating BHs in dCS gravity, these Weyl scalars in Eq.~\eqref{eq:source_scalar_11} receive contributions from both $h_{\mu\nu}^{(0,0,1)}$ and $h_{\mu\nu}^{(0,1,1)}$. We expand Eq.~\eqref{eq:R*R_NP} up to $\mathcal{O}(\zeta^0,\epsilon^1)$. Since in Petrov type D spacetimes $\Psi_{0,1,3,4}^{(0,0)}=0$, the only terms that survive are proportional to $\Psi_2$, such that
    \begin{equation} \label{eq:R*R_Psi2}
        (R^*\!R)^{(0,1)}=48i\left(\Psi_2^{(0,0)}\Psi_2^{(0,1)}
        -\bar{\Psi}_2^{(0,0)}\bar{\Psi}_2^{(0,1)}\right)\,,
    \end{equation}
    where $\Psi_2^{(0,0)}$ is given by Eq.~\eqref{eq:Psi2_bg} by setting $\zeta=0$, and $\Psi_2^{(0,1)}$ is given by Eq.~\eqref{eq:Psi2_Hertz} using the metric reconstruction procedures.
    
    \item Spin coefficients at $\mathcal{O}(\zeta^0,\epsilon^1)$: This dependence arises from $\square^{(0,1)}\vartheta^{(1,0)}$. Since we are only interested in the terms up to $\mathcal{O}(\zeta^1,\chi^1,\epsilon^1)$ in this work, and $\vartheta^{(1,0,0)}=0$ as explained in Sec.~\ref{sec:trivariate_expansion}, the metric fields in $\square^{(0,1)}\vartheta^{(1,0)}$ only have the contribution from $h_{\mu\nu}^{(0,0,1)}$. Thus, we only need metric reconstruction at $\mathcal{O}(\zeta^0, \chi^0,\epsilon^1)$ for these terms.
    The first two terms in Eq.~\eqref{eq:box_NP} will not contribute directly, although one can find additional spin coefficients by using the commutation relations to combine the anti-commutators. For the rest of the terms, we find at $\mathcal{O}(\zeta^0,\epsilon^1)$,
    \begin{equation} \label{eq:source_scalar_type2}
    \begin{aligned}
        & \;\left[\left(\mu^{(0,1)}+\bar{\mu}^{(0,1)}
        -\gamma^{(0,1)}-\bar{\gamma}^{(0,1)}\right)D  \right. \\
        & \;\left.+\left(\epsilon^{(0,1)}+\bar{\epsilon}^{(0,1)}
        -\rho^{(0,1)}-\bar{\rho}^{(0,1)}\right)\boldsymbol{\Delta} \right. \\
        & \;\left.+\left(\alpha^{(0,1)}-\bar{\beta}^{(0,1)}
        -\pi^{(0,1)}+\bar{\tau}^{(0,1)}\right)\delta \right. \\
        & \;\left.+\left(\bar{\alpha}^{(0,1)}-\beta^{(0,1)}
        -\bar{\pi}^{(0,1)}+\tau^{(0,1)}\right)\bar{\delta}\right]\vartheta^{(1,0)}\,,
    \end{aligned}
    \end{equation}
    where the spin coefficients at $\mathcal{O}(\zeta^0,\epsilon^1)$ are given by Eqs.~\eqref{eq:perturbed_spin_coefs} and \eqref{eq:spin_coefs_rotated_01} using metric reconstruction procedures.
    
    \item Tetrad/directional derivatives are at $\mathcal{O}(\zeta^0,\epsilon^1)$. Similar to the second situation, these types of terms also arise from $\square^{(0,1)}\vartheta^{(1,0)}$ and vanish at $\mathcal{O}(\zeta^1,\chi^0,\epsilon^1)$. Thus, we must only reconstruct the NP quantities at $\mathcal{O}(\zeta^0,\chi^0,\epsilon^1)$. Using the Schwarzschild properties of all the spin coefficients in Eq.~\eqref{eq:spin_coefs_Schw}, we find
    \begin{equation} \label{eq:source_scalar_type3}
    \begin{aligned}
        & \;\left[D^{(0,1)}\boldsymbol{\Delta}
        +D\boldsymbol{\Delta}^{(0,1)}
        +\boldsymbol{\Delta}^{(0,1)}D
        +\boldsymbol{\Delta}D^{(0,1)}\right. \\
        & \left.-\delta^{(0,1)}\bar{\delta}
        -\delta\bar{\delta}^{(0,1)}
        -\bar{\delta}^{(0,1)}\delta
        -\bar{\delta}\delta^{(0,1)}\right.\\
        & \;\left.-2(\gamma-\mu)D^{(0,1)} -2\rho\boldsymbol{\Delta}^{(0,1)}
        +2\alpha\left(\delta^{(0,1)}
        +\bar{\delta}^{(0,1)}\right)\right]\vartheta^{(1,0)}\,,
    \end{aligned}
    \end{equation}
    where the tetrad at $\mathcal{O}(\zeta^0,\epsilon^1)$ is given by Eqs.~\eqref{eq:perturbed_tetrad} and \eqref{eq:tetrad_rotated_01}. Notice, the terms at $\mathcal{O}(\zeta^0,\epsilon^1)$ in Eqs.~\eqref{eq:source_scalar_type2} and \eqref{eq:source_scalar_type3} are all at $\mathcal{O}(\zeta^0,\chi^0,\epsilon^1)$ as $\vartheta^{(1,0)}$ is non-vanishing only at $\mathcal{O}(\zeta^1, \chi^1, \epsilon^0)$, but we choose to hide the expansion in $\chi$ for simplicity.
\end{enumerate}
A similar classification will be used when we compute the source terms in the modified Teukolsky equation for $\Psi_0^{(1,1)}$ and $\Psi_4^{(1,1)}$.
	
One can further combine the second and the third type of source terms and express them as functionals of the metric components in the NP basis (e.g., $h_{nn}^{(0,1)}$, $h_{nm}^{(0,1)}$, $h_{mm}^{(0,1)}$ in the IRG) and the rotation coefficients (e.g., $a^{(0,1)}$ and $b^{(0,1)}$) such that the separation of variables can be more easily carried out in Sec.~\ref{sec:separation_variable}. In this case, we find
\begin{equation} \label{eq:box_hab}
    \begin{aligned}
        & \square^{(0,0,1)}\vartheta^{(1,0,0)}
        =-\square^{(0,1,1)}\vartheta^{(1,0,0)}=0\,, \\
        & \square^{(0,0,1)}\vartheta^{(1,1,0)} \\
        & =\left\{h_{nn}^{(0,0,1)}D^2
        -h_{nm}^{(0,0,1)}\{D,\bar{\delta}\}
        +h_{mm}^{(0,0,1)}\bar{\delta}^2 \right. \\
        & \left.+\left[(D-2\rho)h_{nn}^{(0,0,1)}
        -(\bar{\delta}-2\alpha)h_{nm}^{(0,0,1)}\right]D\right. \\
        & \left.-\left[(D-2\rho)h_{nm}^{(0,0,1)}
        -(\bar{\delta}-2\alpha)h_{mm}^{(0,0,1)}\right]\bar{\delta}
        +c.c.\right\}\vartheta^{(1,1,0)}\,.
    \end{aligned}
\end{equation}
	
Finally, we have
\begin{equation} \label{eq:source_scalar_11_decomposed}
    \begin{aligned}
        & \mathcal{S}_{\vartheta}^{(1,0,1)}
        =-\pi^{-\frac{1}{2}}M^2(R^*\!R)^{(0,0,1)}\,, \\
        & \mathcal{S}_{\vartheta}^{(1,1,1)}
        =-\pi^{-\frac{1}{2}}M^2(R^*\!R)^{(0,1,1)}
        -\square^{(0,0,1)}\vartheta^{(1,1,0)}\,,
    \end{aligned}
\end{equation}
where $(R^*\!R)^{(0,0,1)}$ and $(R^*\!R)^{(0,1,1)}$ are given by Eq.~\eqref{eq:R*R_Psi2}, and $\square^{(0,0,1)}\vartheta^{(1,1,0)}$ is given by Eq.~\eqref{eq:box_hab}. 

%%%%%%%%%%%%%%%%%%%%%%%%%%%%%%%%%%%%%%%%%%%%%%%%%%%%%%%%%%%%
\vspace{0.1in}
\subsection{$\mathcal{S}_{\vartheta}^{(1,1)}$ in the coordinate basis}

We now rewrite $\mathcal{S}_{\vartheta}^{(1,1)}$ [Eqs.~\eqref{eq:source_scalar_11} and \eqref{eq:source_scalar_11_decomposed}] in the coordinate basis $(t,r,\theta,\phi)$ using the perturbed NP quantities found in Sec.~\ref{sec:metric_reconstruction} and Appendix~\ref{appendix:background_NP_more}. From Eqs.~\eqref{eq:source_scalar_11} and \eqref{eq:source_scalar_11_decomposed}, we notice that $\mathcal{S}_{\vartheta}^{(1,1)}$ contains two pieces: the term proportional to $(R^*\!R)^{(0,1)}$ and the term proportional to $\square^{(0,1)}\vartheta^{(1,0)}$. 

For the first piece, according to Eq.~\eqref{eq:R*R_Psi2}, we essentially need to evaluate $\Psi_2^{(0,1)}$ up to $\mathcal{O}(\chi)$. The value of $\Psi_2^{(0,1)}$ in terms of the Hertz potential $\Psi_{\Hertz}^{(0,1)}$ is given by Eq.~\eqref{eq:Psi2_Hertz}, and $\Psi_{\Hertz}^{(0,1)}$ has the expansion in Eq.~\eqref{eq:Hertz_decompose}. Since we use the slow-rotation approximation in this work, we can further decompose spin-weighted spheroidal harmonics in terms of spin-weighted spherical harmonics using Eqs.~\eqref{eq:StoY} and \eqref{eq:bvalues} such that Eq.~\eqref{eq:Hertz_decompose} becomes
\begin{equation} \label{eq:Hertz_decompose_slow_rot}
\begin{aligned}
    \bar{\Psi}_{\Hertz}
     =& \;{}_{2}\hat{R}_{\ell m}(r)
     \left[{}_{2}{Y}_{\ell m}(\theta,\phi)+\chi M\omega
     \left({}_{2}b^m_{\ell,\ell+1 }\,{}_{2}Y_{\ell+1\,m}\right.\right. \\
     & \;\left.\left.+\,{}_{2}b^m_{\ell,\ell-1}\,{}_{2}Y_{\ell-1\,m}\right)
     +\mathcal{O}(\chi^2)\right]e^{-i\omega t} \,,
\end{aligned}
\end{equation}
%
%\begin{widetext}
Now, one can insert into Eqs.~\eqref{eq:Psi2_Hertz} and \eqref{eq:R*R_Psi2} the decomposition in Eq.~\eqref{eq:Hertz_decompose_slow_rot} and the background NP quantities at $\mathcal{O}(\zeta^0,\epsilon^0)$ in Eqs.~\eqref{eq:principal_tetrad}--\eqref{eq:spin_coefs_background} after setting $\zeta=0$. After using Eqs.~\eqref{eq:raising} and \eqref{eq:lowering} to simplify the terms with $\delta^{(0,0)}$ and $\bar{\delta}^{(0,0)}$ acting on ${}_{s}Y_{\ell m}(\theta,\phi)$, we find
\begin{widetext}
\begin{align} \label{eq:R*R_coord_IRG}
    (R^*\!R)^{(0,1)}
    =& \;e^{-i\omega t}\left[\left(g^{\ell m}_1(r,\omega,M){}_{2}\hat{R}_{\ell m}(r)
    +g_2^{\ell m}(r,\omega,M)\,{}_{2}\hat{R}_{\ell m}'(r)\right)
    {}_0Y_{\ell m}(\theta,\phi)\right.\nonumber \\
    & \;\left.+\chi\left(g^{\ell m}_3(r,\omega,M)\,{}_{2}\hat{R}_{\ell m}(r)
    +g_4^{\ell m}(r,\omega,M)\,{}_{2}\hat{R}_{\ell m}'(r)\right)
    \sin\theta\,{}_1Y_{\ell m}(\theta,\phi)\right.\nonumber\\
    & \;\left.+\chi\left(g^{\ell m}_5(r,\omega,M)\,{}_{2}\hat{R}_{\ell m}(r)
    +g_6^{\ell m}(r,\omega,M)\,{}_{2}\hat{R}_{\ell m}'(r)\right)
    \cos\theta\,{}_0Y_{\ell m}(\theta,\phi)\right.\nonumber \\
    & \;\left.+\chi\,{}_{2}b^{m}_{\ell,\ell+1}
    \left(g^{\ell m}_7(r,\omega,M)\,{}_{2}\hat{R}_{\ell m}(r)
    +g_8^{\ell m}(r,\omega,M)\,{}_{2}\hat{R}_{\ell m}'(r)\right)
    {}_0Y_{\ell+1\,m}(\theta,\phi)\right.\nonumber \\
    & \;\left.+\chi\,{}_{2}b^{m}_{\ell,\ell-1}
    \left(g^{\ell m}_9(r,\omega,M)\,{}_{2}\hat{R}_{\ell m}(r)
    +g_{10}^{\ell m}(r,\omega,M)\,{}_{2}\hat{R}_{\ell m}'(r)\right)
    {}_0Y_{\ell-1\,m}(\theta,\phi)\right]+\textrm{c.c.} \,,
\end{align}
\end{widetext}
where $f'(r)$ denotes the derivative of $f$ with respect to the $r$ coordinate for any function $f(r)$. Here, ${}_{2}\hat{R}_{\ell m}(r)$ is the radial function of the Hertz potential for slowly rotating Kerr BHs in GR, which can be computed from the radial function of $\Psi_0^{(0,1)}$ using Eq.~\eqref{eq:hertztoteuk}. One can, in principle, expand ${}_{2}\hat{R}_{\ell m}(r)$ further in $\chi$ and drop additional terms above $\mathcal{O}(\chi)$ in Eq.~\eqref{eq:R*R_coord_IRG}. For simplicity, we choose not to do this additional expansion here but implement it when computing QNMs in \cite{dcstyped2}, where we need to explicitly evaluate the radial functions of $\Psi_{0,4}^{(0,1)}$. We have also used the radial Teukolsky equation to reduce any $n$-th radial derivative of ${}_{2}\hat{R}_{\ell m}(r)$ with $n\geq 1$ to ${}_{2}\hat{R}_{\ell m}$ and ${}_{2}\hat{R}_{\ell m}'(r)$. The explicit forms of $g^{\ell m}_i(r,\omega,M)$ are long and therefore presented through a separate Mathematica notebook as Supplementary Material~\cite{Pratikmodteuk}.

For the second piece, according to Eq.~\eqref{eq:source_scalar_11_decomposed}, there is only a contribution from $\square^{(0,0,1)}\vartheta^{(1,1,0)}$ since the scalar field at $\mathcal{O}(\zeta^1,\chi^0,\epsilon^0)$ vanishes in dCS gravity. Thus, we only need the reconstructed metric at $(\zeta^0,\chi^0,\epsilon^1)$. Using Eqs.~\eqref{eq:Hertz_decompose_slow_rot}, \eqref{eq:spin_coefs_background}, and \eqref{eq:metricpert_irg}, we find
\begin{widetext}
\begin{subequations} \label{eq:metric_pert_hertz}
\begin{align}
    h_{nn}^{(0,0,1)} =&\frac{1}{2r^2}\left(\Lambda_{\ell}\,{}_{2}\hat{R}_{\ell m}(r)\,{}_0
    Y_{\ell m}(\theta,\phi)e^{-i\omega t}+\textrm{c.c.}\right) \,, \\
    h_{nm}^{(0,0,1)} =&\sqrt{\frac{\ell^2+\ell-2}{2}}\frac{1}{r^2 (r-r_s)}
    \left\{\left[4M-r(2+i\omega r)\right]{}_{2}\hat{R}_{\ell m}(r)
    +r(r-r_s)\,{}_{2}\hat{R}_{\ell m}'(r)\right\}{}_1Y_{\ell m}(\theta,\phi)e^{-i\omega t} \,, \\
    h_{mm}^{(0,0,1)} =&\frac{1}{r(r-r_s)^2} 
    \left\{\left[(r-r_s)(\ell^2+\ell-2+7i\omega r)
    -\omega r(ir+2\omega r^2)\right]{}_{2}\hat{R}_{\ell m}(r)\right. \nonumber\\
    & \;\left.+2(r-r_s)\left(M-i\omega r^2\right){}_{2}\hat{R}_{\ell m}'(r)\right\}
    {}_2Y_{\ell m}(\theta,\phi)e^{-i\omega t}\,.
\end{align}
\end{subequations}
%\end{widetext}
Now, we can evaluate all the directional derivatives and spin coefficients at $\mathcal{O}{(\zeta^0,\chi^0,\epsilon^1)}$ using Eqs.~\eqref{eq:perturbed_spin_coefs}, \eqref{eq:tetrad_rotated_01}, and \eqref{eq:spin_coefs_rotated_01}. In the end, using Eq.~\eqref{eq:box_hab}, we find
%\begin{widetext}
    \begin{align} \label{eq:box_coord_IRG}
    \square^{(0,0,1)}\vartheta^{(1,1,0)}
    =& \;e^{-i\omega t}\left[\left(h^{\ell m}_1(r,\omega,M)\,{}_{2}\hat{R}_{\ell m}(r)
    +h^{\ell m}_2(r,\omega,M)\,{}_{2}\hat{R}_{\ell m}'(r)\right)
    \sin\theta\,{}_1Y_{\ell m}(\theta,\phi)
    \right.\nonumber \\ 
    & \;\left.+h^{\ell m}_3(r,\omega,M)
    \left(2\vartheta_R'(r)+r\vartheta_R''(r)\right){}_{2}\hat{R}_{\ell m}(r)
    \cos\theta\,{}_0Y_{\ell m}(\theta,\phi)\right]+\textrm{c.c.}\,,
\end{align}
\end{widetext}
where $\vartheta_R(r)$ is the radial part of the background scalar field in Eq.~\eqref{eq:scalar_stationary}. In Eq.~\eqref{eq:box_coord_IRG}, the reconstructed metric only has contribution at $\mathcal{O}(\chi^0)$, so the radial function ${}_{2}\hat{R}_{\ell m}(r)$ is evaluated on the Schwarzschild background. Since we choose not to expand $\hat{R}(r)$ in $\chi$ here, we do not distinguish ${}_{2}\hat{R}_{\ell m}(r)$ evaluated on the Schwarzschild or slowly rotating Kerr background. Combining Eqs.~\eqref{eq:R*R_coord_IRG} and \eqref{eq:box_coord_IRG}, we have the source terms in the equation of $\mathcal{\vartheta}^{(1,1)}$ up to $\mathcal{O}(\chi)$. The master equation of $\vartheta^{(1,1)}$ in the IRG in the coordinate $\{t,r,\theta,\phi\}$ is presented in Eq.~\eqref{eq:general-teuk} of Sec.~\ref{sec:executive_summary_Teuk}.

In Sec.~\ref{sec:separation_variable}, we will show that Eq.~\eqref{eq:source_scalar_11} up to $\mathcal{O}(\chi)$ can be separated into a radial and an angular equation. In the following section, we apply the same procedures to evaluate the source terms in the modified Teukolsky equation of $\Psi_0^{(1,1)}$ in terms of the reconstructed NP quantities and project the equation into the coordinate basis.

%%%%%%%%%%%%%%%%%%%%%%%%%%%%%%%%%%%%%%%%%%%%%%%%%%%%%%%%%%%%
%%%%%%%%%%%%%%%%%%%%%%%%%%%%%%%%%%%%%%%%%%%%%%%%%%%%%%%%%%%%
\section{The modified Teukolsky equation of $\Psi_0^{(1,1)}$ in the IRG}
\label{sec:source_Teuk}
	
\begin{figure*}
\centering
    \begin{tikzpicture}
    \node (a) at (-1,0) [label= \textrm{Perturbed gravitational field equation for $\Psi_0^{(1,1)}$ in the IRG}]{};
    \draw (-6,0) -- (4,0);
    \draw (-6,0.8) -- (4,0.8);
    \draw (-6,0) -- (-6,0.8);
    \draw (4,0) -- (4,0.8);
    %\draw (-1,0.4) ellipse (4.6cm and 0.4cm);
    \draw [->,thick] (-1,0) -- (-5,-1);
    \draw [->,thick] (-1,0) -- (3,-1);
    \node (b1) at (-5,-1) [label=below: \textrm{Left-hand side of Eq.~\eqref{eq:master_eqn_non_typeD_Psi0}}]{};
    \node (b2) at (3,-1) [label=below: \textrm{Right-hand side of Eq.~\eqref{eq:master_eqn_non_typeD_Psi0}}]{};
    \draw (-5,-1.37) ellipse (2.5cm and 0.3cm);
    \draw (3,-1.37) ellipse (2.5cm and 0.3cm);
    \draw [-,thick] (5,-2.1) -- (1,-2.1);
    \draw [->,thick] (-5,-1.7) -- (-5,-2.5);
    \draw [-,thick] (3,-1.7) -- (3,-2.1);
    \draw [->,thick] (1,-2.1) -- (1,-2.5);
    \draw [->,thick] (5,-2.1) -- (5,-2.5);
    \node (c1a) at (-5,-2.1) [label=left: \textrm{In coordinate basis}]{};
    \node (c1) at (-5,-2.5) [label=below: \textrm{Equation~\eqref{eq:H0_00}}]{};
    \draw (-5,-2.87) ellipse (2cm and 0.3cm);
    \draw (-5,-4.9) ellipse (2.5cm and 0.33cm);
    \draw [->,thick] (-5,-3.2) -- (-5,-4.5);
    \node (c2a) at (-6.4,-4.45) [label={[align=right]\textrm{Separated radial} \\ \textrm{equation}} ]{};
    \node (c2) at (-5,-4.5) [label=below: \textrm{Left-hand side of Eq.~\eqref{eq:psi0_mastereqn_radial}}]{};
    \node (c3a) at (5,-2.45) [label=below: \textrm{Source term $\mathcal{S}^{(1,1)}$}]{};
    \node (c3b) at (1,-2.45) [label=below: \textrm{Source term $\mathcal{S}_{\geo}^{(1,1)}$ }]{};
    \draw (5,-2.88) ellipse (1.7cm and 0.33cm);
    \draw (1,-2.88) ellipse (1.7cm and 0.33cm);
    \draw [-,thick] (5,-3.2) -- (5,-5.5);
    \draw [->,thick] (1,-3.2) -- (1,-4.2);
    \draw [-,thick] (3,-5.5) -- (7,-5.5);
    \draw [->,thick] (3,-5.5) -- (3,-6);
    \draw [->,thick] (7,-5.5) -- (7,-6);
    \node (c4a) at (5,-3.9) [label=right:{\textrm{In coordinate basis}}]{};
    \node (c4b) at (5,-4.3) [label=right:{\textrm{using the Hertz }}]{};
    \node (c4b) at (5,-4.7) [label=right:{\textrm{potential Eq.~\eqref{eq:Hertz_decompose_slow_rot}}}]{};
    \node (c5a) at (3,-7.8) [label={[align=center]\textrm{terms proportional} \\ \textrm{to $\Theta_{\ell m}(r)$}  \\ \textrm{$\mathcal{S}_A^{(1,1)}$ \& $\tilde{\mathcal{S}}_A^{(1,1)}$} }]{};
    \draw (3,-7) ellipse (1.8cm and 0.9cm);
    \node (c5b) at (7,-7.8) [label={[align=center]\textrm{terms proportional} \\ \textrm{to ${}_{2}\hat{R}_{\ell m}(r)$} \\ \textrm{$\mathcal{S}_B^{(1,1)}$ \& $\tilde{\mathcal{S}}_B^{(1,1)}$} }]{};
    \draw (7,-7) ellipse (1.8cm and 0.9cm);
    \draw [->,thick] (7,-7.9) -- (7,-9.2);
    \draw [->,thick] (3,-7.9) -- (3,-9.2);
    \node (c6a) at (3,-10.3) [label={[align=center]\textrm{terms given} \\ \textrm{in Eqs.~\eqref{eq:sA_total_radial}} }]{};
    \draw (3,-9.75) ellipse (1.3cm and 0.5cm);
    \node (c6b) at (7,-10.3) [label={[align=center]\textrm{terms given} \\ \textrm{in Eqs.~\eqref{eq:sB_total_radial}}} ]{};
    \draw (7,-9.75) ellipse (1.3cm and 0.5cm);
    \node (c6c) at (5,-9.05) [label={[align=center]\textrm{Using the recipe in} \\ \textrm{Sec.~\ref{sec:elim-ylm}}} ]{};
    \node (c7) at (1,-5.3) [label={[align=center]\textrm{terms given} \\ \textrm{in Eq.~\eqref{eq:S_geo_radial}} }]{};
    \draw (1,-4.72) ellipse (1.3cm and 0.5cm);
    \draw [->,thick] (1,-5.2) -- (1,-10.8);
    \node (c8) at (0,-12) [label={[align=center] The radial evolution equation of the Weyl scalar perturbation $\Psi_0^{(1,1)}$ for slowly rotating BHs \\ in dCS gravity in the IRG is given by Eq.~\eqref{eq:psi0_mastereqn_radial}}]{};
    \draw [->,thick] (3,-10.25) -- (3,-10.8);
    \draw [->,thick] (7,-10.25) -- (7,-10.8);
    \draw [->,thick] (-5,-5.25) -- (-5,-10.8);
    \draw (-8.8,-12) -- (9,-12);
    \draw (-8.8,-10.8) -- (9,-10.8);
    \draw (-8.8,-10.8) -- (-8.8,-12);
    \draw (9,-10.8) -- (9,-12);
    \end{tikzpicture}
    \caption{\label{Fig. 2} A schematic flowchart prescribing the steps involved in obtaining a separated radial evolution equation for the gravitational field perturbation described by the $\Psi_0^{(1,1)}$ Weyl scalar in the IRG for slowly rotating BHs in dCS gravity. This flowchart summarizes the details of the calculations described in detail in Sec.~\ref{sec:source_Teuk} and Sec.~\ref{sec:psi0_radial_equation}. Initial and final outcomes are represented by rectangular boxes, while intermediate results are symbolized by encapsulating bubbles. The directional arrows are meant to seamlessly guide the reader through the logical flow of the calculations. }
\end{figure*}
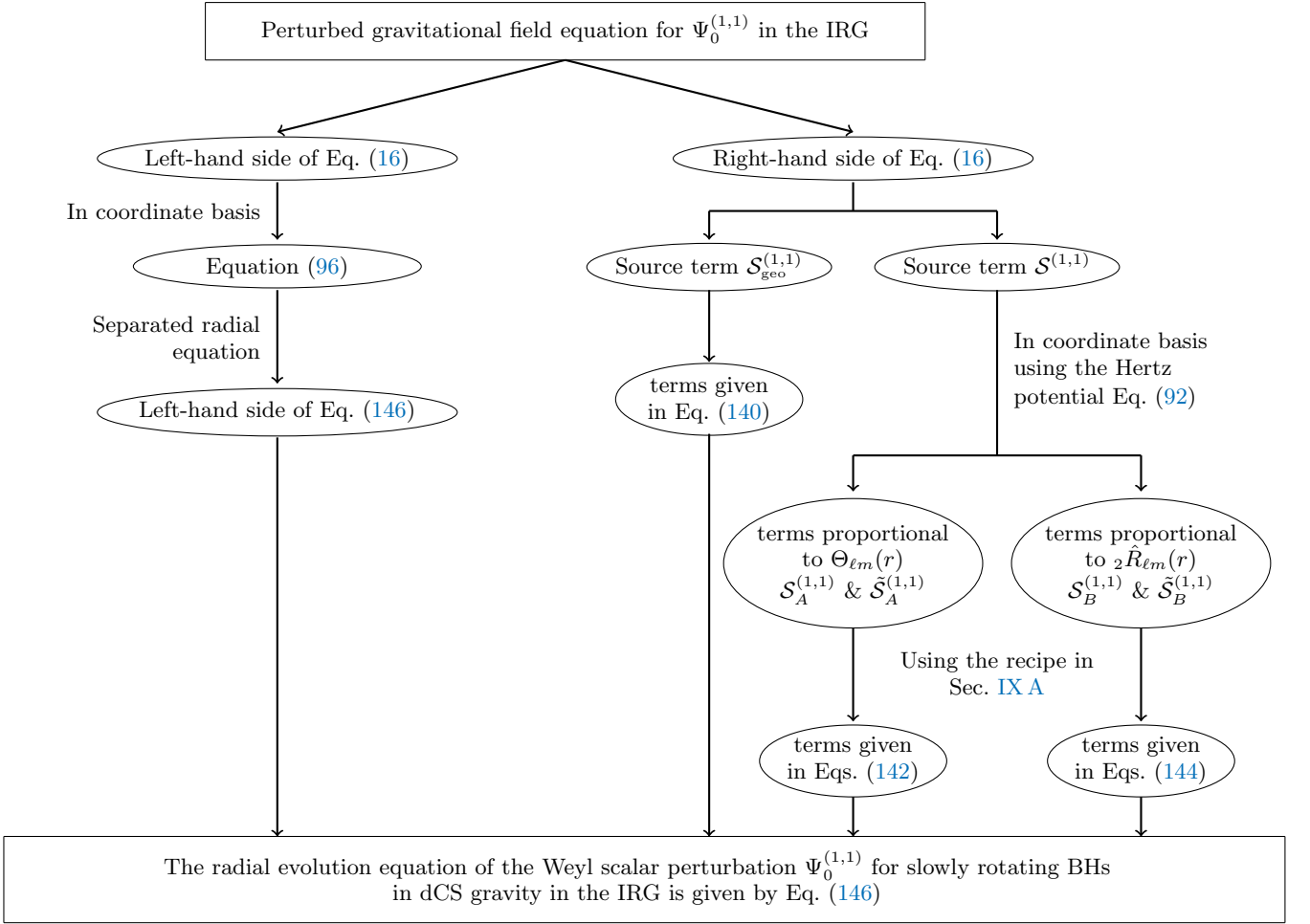

In this section, we evaluate the modified Teukolsky equation of $\Psi_0^{(1,1)}$ in Eq.~\eqref{eq:master_eqn_non_typeD_Psi0} following the similar procedures in Sec.~\ref{sec:source_scalar}. We first evaluate the left-hand side of Eq.~\eqref{eq:master_eqn_non_typeD_Psi0} and the source term $\mathcal{S}_{\geo}^{(1,1)}$ due to the correction to the background geometry using the background NP quantities in Sec.~\ref{sec:NP_quantities_background} and Appendix~\ref{appendix:background_NP_more}. We then project the source term $\mathcal{S}^{(1,1)}$ onto the NP basis and compute its coordinate-based value using the reconstructed NP quantities in the IRG given in Sec.~\ref{sec:metric_reconstruction} and Appendix~\ref{appendix:metric_reconstruction_more}. Figure~\ref{Fig. 2} presents a schematic illustration of the steps involved in calculating a completely separated radial evolution equation for the $\Psi_0^{(1,1)}$ Weyl scalar perturbation in the IRG.

%%%%%%%%%%%%%%%%%%%%%%%%%%%%%%%%%%%%%%%%%%%%%%%%%%%%%%%%%%%%
%%%%%%%%%%%%%%%%%%%%%%%%%%%%%%%%%%%%%%%%%%%%%%%%%%%%%%%%%%%%
\subsection{Right-hand side of Eq.~\eqref{eq:master_eqn_non_typeD_Psi0} and $\mathcal{S}_{\geo}^{(1,1)}$} 
\label{sec:modified_Teuk_operator}

Since slowly rotating BHs in dCS gravity are Petrov type D up to $\mathcal{O}(\zeta^1,\chi^1,\epsilon^0)$, we only need to compute the Teukolsky operator $H_{0}^{(0,0)}$ in GR and its stationary correction $H_{0}^{(1,0)}$ in Eq.~\eqref{eq:Sd}. Thus, we do not need metric reconstruction in this subsection. 
	
Using the Weyl scalars and spin coefficients found in Sec.~\ref{sec:NP_quantities_background} and Eq.~\eqref{eq:H_in_Teuk}, we find that
\begin{subequations} \label{eq:H0_00}
\begin{align} 
    & H_0^{(0,0,0)}
    =\frac{1}{2r^2}H_{0,\mathrm{TK}}^{(0,0,0)}\,,
    \label{eq:H0_000} \\
    & H_0^{(0,1,0)}
    =\frac{1}{2r^2}H_{0,\mathrm{TK}}^{(0,1,0)}\,,
     \label{eq:H0_010}
\end{align}
\end{subequations}
where $H_{0,\mathrm{TK}}^{(0,0,0)}$ and $H_{0,\mathrm{TK}}^{(0,1,0)}$ are $\mathcal{O}(\chi^0)$ and $\mathcal{O}(\chi^1)$ terms of the Teukolsky operator $H_{0,\mathrm{TK}}$ for $\Psi_{0}$ [Eq.~(4.7) in \cite{Teukolsky:1973ha}],
\begin{subequations} \label{eq:H0_00_TK}
\begin{align}
    \begin{split} \label{eq:H0_000_TK} 
        & H_{0,\mathrm{TK}}^{(0,0,0)}
        =-r(r-r_s)\partial^2_{r}-6(r-M)\partial_r
        -\frac{C(r)}{r-r_s} \\ 
        & \;-\partial^2_{\theta}-\cot{\theta}\partial_{\theta}
        +\left(4+m^{2}+4m\cos\theta\right)\csc^2{\theta}-6\,,
    \end{split} \\
    \begin{split} \label{eq:H0_010_TK}
        & H_{0, \mathrm{TK}}^{(0,1,0)}
        =-4M\left[\frac{m(i(r-M)-M\omega r)}{r(r-r_s)}
        -\omega\cos{\theta}\right]\,, \\
    \end{split} \\
    & \label{eq:Cfunc} C(r)
    =4i\omega r(r-3M)+\omega^{2}r^{3}\,,
\end{align}    
\end{subequations}
where we have decomposed the Weyl scalar $\Psi_0^{(1)}$ at $\mathcal{O}(\epsilon)$ as
\begin{equation} \label{eq:psi0_separation}
\begin{aligned}
    \Psi_0^{(1)}
    =& \;\left[{}_{2}R_{\ell m}^{(0,1)}(r)
    +\zeta\,{}_{2}R_{\ell m}^{(1,1)}(r)+\mathcal{O}(\zeta^2)\right] \\
    & \;{}_{2}S_{\ell m}(\theta)e^{-i\omega t+im\phi}\,.
\end{aligned}
\end{equation}
The Teukolsky equation corresponding to Eqs.~\eqref{eq:H0_000_TK} and \eqref{eq:H0_010_TK} is separable with ${}_{2}R_{\ell m}^{(0,1)}(r)$ satisfying
\begin{equation}
\begin{aligned}
    & \left[r(r-r_s)\partial^2_{r}+6(r-M)\partial_r+\frac{C(r)}{r-r_s}\right. \\
    & \left.+\frac{4m\chi M(i(r-M)-M\omega r)}{r(r-r_s)}
    -{}_{2}A_{\ell m}\right]{}_{2}R_{\ell m}^{(0,1)}(r)=0\,.
\end{aligned}
\end{equation}
		
Since there is no correction to the background geometry at $\mathcal{O}(\zeta^1,\chi^0,\epsilon^0)$, $H_0^{(1,0,0)}=0$. For $H_0^{(1,1,0)}$, we find
\begin{equation} \label{eq:H0_110}
    \begin{aligned}
        H_0^{(1,1,0)}
        =& \;\frac{imM^4}{448r^9(r-r_s)}
        \left(C_1(r)+4i\omega r^2C_2(r)\right) \\
        & \;-\frac{iM^4}{16r^9}\cos{\theta}
        \left(C_3(r)-\frac{i\omega r^2C_4(r)}{2}\right) \\
        & \;+\frac{iM^4}{32r^8}\left[(r-r_s)C_4(r)\cos{\theta}\partial_r
        -\frac{C_5(r)}{2r}\sin{\theta}\partial_{\theta}\right]\,,
    \end{aligned}
\end{equation}
where all $C_i(r)$ are listed in Appendix~\ref{appendix:background_NP_more}. The source term $\mathcal{S}_{\geo}^{(1,1)}$ is then given by 
\begin{equation} \label{eq:source_psi0_geo}
    \mathcal{S}_{\geo}^{(1,0,1)}=0\,,\quad
    \mathcal{S}_{\geo}^{(1,1,1)}=H_0^{(1,1,0)}\Psi_{0}^{(0,0,1)}\,.
\end{equation}
$\mathcal{S}_{\geo}^{(1,1,1)}$ can be evaluated in terms of the coordinates using Eqs.~\eqref{eq:psi0_separation} and \eqref{eq:H0_110}. 
	
%%%%%%%%%%%%%%%%%%%%%%%%%%%%%%%%%%%%%%%%%%%%%%%%%%%%%%%%%%%%
\subsection{Source term $\mathcal{S}^{(1,1)}$} 
\label{sec:corrections_from_stress}
	
In this subsection, we evaluate the source term $\mathcal{S}^{(1,1)}$ from the effective stress tensor or the source term of the trace-reversed Einstein equation in Eq.~\eqref{eq:EOM_R}. The first step is to project the Ricci tensor $R_{\mu\nu}$ onto the NP basis such that we can express the NP Ricci scalars $\Phi_{ij}$, where
\begin{equation}
    \Phi_{ij}\sim R_{ab}\sim R_{\mu\nu}e_{a}{}^{\mu}e_{b}{}^{\nu}\,,
\end{equation}
in terms of NP quantities (Weyl scalars, spin coefficients, and tetrad) and the scalar field $\vartheta$. The precise relation between $\Phi_{ij}$ and $R_{ab}$ is given in Eq.~\eqref{eq:Ricci_NP_scalars}. Using Eq.~\eqref{eq:EOM_R}, we find
\begin{equation} \label{eq:R_ab}
    \begin{aligned}
        R_{\mu\nu}=
        & \;-\left(\frac{1}{\kappa_g}\right)^{\frac{1}{2}}M^2
        \Bigg[\left(\nabla^{\sigma}\vartheta\right)
        \epsilon_{\sigma\delta\alpha(\mu}
        \nabla^{\alpha}R_{\nu)}{}^{\delta} \\& \;+\left(\nabla^{\sigma}\nabla^{\delta}
        \vartheta\right)^*\!R_{\delta(\mu\nu)\sigma}\Bigg]
        +\frac{1}{2\kappa_g\zeta}\left(\nabla_{\mu}\vartheta\right)
        \left(\nabla_{\nu}\vartheta\right)\,,
    \end{aligned}
\end{equation}
where we have inserted an additional $\zeta^{-\frac{1}{2}}$ to the term linear in $\vartheta$ and an additional $\zeta^{-1}$ to the term quadratic in $\vartheta$ to compensate the factor of $\zeta^{\frac{1}{2}}$ we have absorbed into the expansion of $\vartheta$ in Eq.~\eqref{eq:expansion_scalar}. Since $\vartheta$ enters at least at $\mathcal{O}(\zeta)$, all the metric fields at the right-hand side of Eq.~\eqref{eq:R_ab} are at $\mathcal{O}(\zeta^0)$, which can be expressed in terms of NP quantities in GR.
\begin{equation} \label{eq:R_ab_components}
    \begin{aligned}
        R_{ab}=& \;\eta^{cd}R_{cadb}\,, \\
        \nabla_{c}R_{ab}
        =& \;R_{ab,c}-\gamma^{d}{}_{ac}R_{db}-\gamma^{d}{}_{bc}R_{ad}\,. \\
    \end{aligned}
\end{equation}
Since we are interested in gravitational perturbations of vacuum spacetime, $R_{\mu\nu}^{(0,0)}=R_{\mu\nu}^{(0,1)}=0$, and all the metric fields in Eq.~\eqref{eq:R_ab} enter at $\mathcal{O}(\zeta^0)$, the first term in Eq.~\eqref{eq:R_ab} vanishes. Evaluating the rest of the terms, we find the seven independent components of $R_{ab}$ in terms of Weyl scalars, spin coefficients, directional derivatives, and the scalar field $\vartheta$ in Eq.~\eqref{eq:R_ab_detail}.
	
We can now evaluate the source terms in the modified Teukolsky equations. Inspecting the source term $\mathcal{S}^{(1,1)}$ in Eq.~\eqref{eq:S}, we can divide it into two parts based on whether $S_{1,2}$ are dynamical, 
\begin{equation} \label{eq:sourceteuk2}
    \begin{aligned}
        & \mathcal{S}^{(1,1)}
        =\mathcal{S}_{I}^{(1,1)}+\mathcal{S}_{II}^{(1,1)}\,, \\
        & \mathcal{S}_{I}^{(1,1)}
        =\mathcal{E}_{2}^{(0,1)}S_2^{(1,0)}	
        -\mathcal{E}_{1}^{(0,1)}S_1^{(1,0)}\,, \\
        & \mathcal{S}_{II}^{(1,1)}
        =\mathcal{E}_{2}^{(0,0)}S_2^{(1,1)}	
        -\mathcal{E}_{1}^{(0,0)}S_1^{(1,1)}\,,
    \end{aligned}
\end{equation}
For $\mathcal{S}_{I}^{(1,1)}$, one can directly evaluate $S_{1,2}^{(1,0)}$ in terms of the stationary scalar field $\vartheta^{(1,0)}$ and the metric in GR using Eqs.~\eqref{eq:source_bianchi}, \eqref{eq:R_ab_detail}, and \eqref{eq:Ricci_NP_scalars}. Then we only need to reconstruct the operators $\mathcal{E}_{1,2}^{(0,1)}$ using our results in Sec.~\ref{sec:metric_reconstruction} and Appendix~\ref{appendix:metric_reconstruction_more}. The results of $S_{1,2}^{(1,0)}$ are provided in Appendix~\ref{appendix:source_terms_detail}.
	
For $\mathcal{S}_{II}^{(1,1)}$, the only pieces involving metric reconstruction are $S_{1,2}^{(1,1)}$. For $S_{1,2}^{(1,1)}$, we can further divide them into two parts based on whether $\Phi_{ij}$ are dynamical,
\begin{equation} \label{eq:S_12AB}
    \begin{aligned}
        S_{1,A}^{(1,1)}
        =& \;\delta_{[-2,-2,1,0]}^{(0,1)}\Phi_{00}^{(1,0)}
        -D_{[-2,0,0,-2]}^{(0,1)}\Phi_{01}^{(1,0)} \\& \;+2\sigma^{(0,1)}\Phi_{10}^{(1,0)}
        -2\kappa^{(0,1)}\Phi_{11}^{(1,0)}
        -\bar{\kappa}^{(0,1)}\Phi_{02}^{(1,0)}\,, \\
        S_{2,A}^{(1,1)}
        =& \;\delta_{[0,-2,2,0]}^{(0,1)}\Phi_{01}^{(1,0)}
        -D_{[-2,2,0,-1]}^{(0,1)}\Phi_{02}^{(1,0)} \\& \;-\bar{\lambda}^{(0,1)}\Phi_{00}^{(1,0)}
        +2\sigma^{(0,1)}\Phi_{11}^{(1,0)}
        -2\kappa^{(0,1)}\Phi_{12}^{(1,0)}\,, \\
        S_{1,B}^{(1,1)}
        =& \;\delta_{[-2,-2,1,0]}^{(0,0)}\Phi_{00}^{(1,1)}
        -D_{[-2,0,0,-2]}^{(0,0)}\Phi_{01}^{(1,1)}\,, \\
        S_{2,B}^{(1,1)}
        =& \;\delta_{[0,-2,2,0]}^{(0,0)}\Phi_{01}^{(1,1)}
        -D_{[-2,2,0,-1]}^{(0,0)}\Phi_{02}^{(1,1)}\,,
    \end{aligned}
\end{equation}
where we used that $\kappa^{(0,0)}=\sigma^{(0,0)}=\lambda^{(0,0)}=0$. Based on this classification, we can then additionally separate $\mathcal{S}_{II}^{(1,1)}$ into two parts
\begin{equation} \label{eq:sourceteuk2a}
    \begin{aligned}
        & \mathcal{S}_{II}^{(1,1)}
        =\mathcal{S}_{IIA}^{(1,1)}+\mathcal{S}_{IIB}^{(1,1)}\,, \\
        & \mathcal{S}_{IIA}^{(1,1)}
        =\mathcal{E}_{2}^{(0,0)}S_{2,A}^{(1,1)}	
        -\mathcal{E}_{1}^{(0,0)}S_{1,A}^{(1,1)}\,, \\
        & \mathcal{S}_{IIB}^{(1,1)}
        =\mathcal{E}_{2}^{(0,0)}S_{2,B}^{(1,1)}	
        -\mathcal{E}_{1}^{(0,0)}S_{1,B}^{(1,1)}\,.
    \end{aligned}
\end{equation}
For $\mathcal{S}_{IIA}^{(1,1)}$, similar to $\mathcal{S}_{I}^{(1,1)}$, we only need to evaluate $\Phi_{ij}^{(1,0)}$ using the background metric and the stationary scalar field $\vartheta^{(1,0)}$. The only quantities need metric reconstruction are these additional $\mathcal{O}(\zeta^0,\epsilon^1)$ operators acting on $\Phi_{ij}^{(1,0)}$, where $\Phi_{ij}^{(1,0)}$ are listed in Appendix~\ref{appendix:source_terms_detail}.
	
The most complicated piece of $\mathcal{S}^{(1,1)}$ is $\mathcal{S}_{IIA}^{(1,1)}$, which needs metric reconstruction for $\Phi_{ij}^{(1,1)}$. Fortunately, from Eq.~\eqref{eq:S_12AB}, we notice that we only need to evaluate $\Phi_{00}^{(1,1)}$, $\Phi_{01}^{(1,1)}$, and $\Phi_{02}^{(1,1)}$. To organize our calculations, we can divide $\Phi_{ij}^{(1,1)}$ into four parts based on which kind of terms need metric reconstruction, similar to what we have done in Sec.~\ref{sec:source_scalar}. Inspecting Eq.~\eqref{eq:R_ab_detail}, we notice that all the terms are some coupling of a Weyl scalar, a scalar field, a spin coefficient, and directional derivatives, which is due to the structure of $R_{\mu\nu}$ in Eq.~\eqref{eq:R_ab}. In this case, we make the following classification:
\begin{enumerate}
    \item Weyl scalars are at $\mathcal{O}(\zeta^0,\epsilon^1)$. In this case, the scalar field $\vartheta$ is at $\mathcal{O}(\zeta^1,\epsilon^0)$. As discussed in Sec.~\ref{sec:trivariate_expansion}, the leading contribution to $\vartheta^{(1,0)}$ is at $\mathcal{O}(\zeta^1,\chi^1,\epsilon^0)$ since non-rotating BHs in dCS gravity are still Schwarzschild. Then since we are interested in $\mathcal{O}(\zeta^1,\chi^1,\epsilon^1)$ corrections, all the spin coefficients and directional derivatives are at $\mathcal{O}(\zeta^0,\chi^0,\epsilon^0)$, the order of the Schwarzschild background.
 
    \item Spin coefficients are at $\mathcal{O}(\zeta^0,\epsilon^1)$. Similar to the first situation, the scalar field $\vartheta$ is at $\mathcal{O}(\zeta^1,\chi^1,\epsilon^0)$, so all the Weyl scalars and directional derivatives are at $\mathcal{O}(\zeta^0,\chi^0,\epsilon^0)$, which are evaluated on the Schwarzschild background in GR.
    
    \item Tetrad/directional derivatives are at $\mathcal{O}(\zeta^0,\epsilon^1)$. Similar to the first two cases, since $\vartheta$ is at $\mathcal{O}(\zeta^1,\chi^1,\epsilon^0)$, so all the Weyl scalars and spin coefficients can be evaluated on the Schwarzschild background.
    
    \item The scalar field $\vartheta$ is at $\mathcal{O}(\zeta^1,\epsilon^1)$, which has contributions from both $\mathcal{O}(\zeta^1,\chi^0, \epsilon^1)$ and $\mathcal{O}(\zeta^1,\chi^1,\epsilon^1)$ terms. Then, all the NP quantities generally need to be evaluated on the Kerr background expanded in the slow-rotation expansion to $\mathcal{O}(\chi^1)$. Since $\vartheta^{(1,1)}$ also requires us to solve the scalar field equation in Eqs.~\eqref{eq:EOM_scalar_11}, we choose not to compute $\vartheta^{(1,1)}$ in this work but only list the source terms in terms of it. We will solve the scalar field equation together with the modified Teukolsky equation in our follow-up work \cite{dcstyped2}.
\end{enumerate}
	
At $\mathcal{O}(\zeta^1,\chi^1,\epsilon^1)$, using the classification above, we can set many terms to 0 since they are evaluated on the Schwarzschild background (i.e., when $\vartheta$ is stationary). Similar to Sec.~\ref{sec:source_scalar}, the results of $\Phi_{00}^{(1,1)}$, $\Phi_{01}^{(1,1)}$, and $\Phi_{02}^{(1,1)}$ up to $\mathcal{O}(\zeta^1,\chi^1,\epsilon^1)$ are expressed in terms of the perturbed Weyl scalars, metric perturbations, and dynamical $\vartheta$ in Appendix~\ref{appendix:source_terms_detail}. Due to the complication of $\mathcal{S}^{(1,1)}$, we will not present the results here but provide its coordinate-based values directly in the next subsection.

%%%%%%%%%%%%%%%%%%%%%%%%%%%%%%%%%%%%%%%%%%%%%%%%%%%%%%%%%%%%
\subsection{$\mathcal{S}^{(1,1)}$ in the coordinate basis} 
\label{sec:S_11_coord}

In this subsection, we evaluate the coordinate-based values of $\mathcal{S}^{(1,1)}$ using the decomposition of $\vartheta^{(1,1)}$ in Eq.~\eqref{eq:scalar_decompose_slow_rot} and the Hertz potential $\Psi_{\Hertz}^{(1,1)}$ in Eq.~\eqref{eq:Hertz_decompose_slow_rot}. Following Sec.~\ref{sec:modified_Teuk_eqn}, we separate $\mathcal{S}^{(1,1)}$ into two parts: the terms coupled to the dynamical scalar field $\vartheta^{(1,1)}$ and the terms coupled to the background scalar field $\vartheta^{(1,0)}$.

For the first part, we find its coordinate-based value to be
\begin{widetext}
\begin{subequations} \label{eq:SAset}
\begin{align} 
    \mathcal{S}_{A}^{(1,1)} 
    =& \;e^{-i\omega t}\left[\left(p_1^{\ell m}(r,\omega,M)\Theta_{\ell m}(r)
    +p_2^{\ell m}(r,\omega,M)\Theta_{\ell m}'(r)
    +p_3^{\ell m}(r,\omega,M)\Theta_{\ell m}''(r)\right)
    {}_2Y_{\ell m}(\theta,\phi)\right. \nonumber\\
    & \;\left.+\chi\left(p_4^{\ell m}(r,\omega,M)\Theta_{\ell m}(r)
    +p_5^{\ell m}(r,\omega,M)\Theta_{\ell m}'(r)
    +p_6^{\ell m}(r,\omega,M)\Theta_{\ell m}''(r)\right)
    \sin\theta\,{}_1Y_{\ell m}(\theta,\phi)\right. \nonumber\\
    & \;\left.+\chi\left(p_7^{\ell m}(r,\omega,M)\Theta_{\ell m}(r)
    +p_8^{\ell m}(r,\omega,M)\Theta_{\ell m}'(r)
    +p_9^{\ell m}(r,\omega,M)\Theta_{\ell m}''(r)\right)
    \cos\theta\,{}_2Y_{\ell m}(\theta,\phi)\right. \nonumber\\
    & \;\left.+\chi\,{}_{0}b^m_{\ell,\ell+1}
    \left(p_{10}^{\ell m}(r,\omega,M)\Theta_{\ell m}(r)
    +p_{11}^{\ell m}(r,\omega,M)\Theta_{\ell m}'(r)
    +p_{12}^{\ell m}(r,\omega,M)\Theta_{\ell m}''(r)\right)
    {}_2Y_{\ell+1\,m}(\theta,\phi)\right. \nonumber\\
    & \;\left.+\chi\,{}_{0}b^m_{\ell,\ell-1}
    \left(p_{13}^{\ell m}(r,\omega,M)\Theta_{\ell m}(r)
    +p_{14}^{\ell m}(r,\omega,M)\Theta_{\ell m}'(r)
    +p_{15}^{\ell m}(r,\omega,M)\Theta_{\ell m}''(r)\right)
    {}_2Y_{\ell-1\,m}(\theta,\phi)\right]\,, 
    \label{eq:sA_coord} \\
    \Tilde{\mathcal{S}}_{A}^{(1,1)} 
    =& \;e^{i\omega t}\left[-\left(\bar{p}_1^{\ell m}(r,\omega,M)\bar{\Theta}_{\ell m}(r)
    +\bar{p}_2^{\ell m}(r,\omega,M)\bar{\Theta}_{\ell m}'(r)
    +\bar{p}_3^{\ell m}(r,\omega,M)\bar{\Theta}_{\ell m}''(r)\right)
    {}_{-2}\bar{Y}_{\ell m}(\theta,\phi)\right. \nonumber\\
    & \;\left.-\chi\left(\bar{p}_4^{\ell m}(r,\omega,M)\bar{\Theta}_{\ell m}(r)
    +\bar{p}_5^{\ell m}(r,\omega,M)\bar{\Theta}_{\ell m}'(r)
    +\bar{p}_6^{\ell m}(r,\omega,M)\bar{\Theta}_{\ell m}''(r)\right)
    \sin\theta\,{}_{-1}\bar{Y}_{\ell m}(\theta,\phi)\right.\nonumber\\
    & \;\left.+\chi\left(\bar{p}_7^{\ell m}(r,\omega,M)\bar{\Theta}_{\ell m}(r)
    +\bar{p}_{8}^{\ell m}(r,\omega,M)\bar{\Theta}_{\ell m}'(r)
    +\bar{p}_{9}^{\ell m}(r,\omega,M)\bar{\Theta}_{\ell m}''(r)\right)
    \cos\theta\,{}_{-2}\bar{Y}_{\ell m}(\theta,\phi)\right. \nonumber\\
    & \;\left.-\chi\,{}_{0}b^m_{\ell,\ell+1}
    \left(\bar{p}_{10}^{\ell m}(r,\omega,M)\bar{\Theta}_{\ell m}(r)
    +\bar{p}_{11}^{\ell m}(r,\omega,M)\bar{\Theta}_{\ell m}'(r)
    +\bar{p}_{12}^{\ell m}(r,\omega,M)\bar{\Theta}_{\ell m}''(r)\right)
    {}_{-2}\bar{Y}_{\ell+1\,m}(\theta,\phi)\right.\nonumber\\
    & \;\left.-\chi\,{}_{0}b^m_{\ell,\ell+1}
    \left(\bar{p}_{13}^{\ell m}(r,\omega,M)\bar{\Theta}_{\ell m}(r)
    +\bar{p}_{14}^{\ell m}(r,\omega,M)\bar{\Theta}_{\ell m}'(r)
    +\bar{p}_{15}^{\ell m}(r,\omega,M)\bar{\Theta}_{\ell m}''(r)\right)
    {}_{-2}\bar{Y}_{\ell-1\,m}(\theta,\phi)\right]\,. 
    \label{eq:sAc_coord}
\end{align}
\end{subequations}
\end{widetext}
In Sec.~\ref{sec:psi0_radial_equation}, after getting the radial part of the equation of $\vartheta^{(1,1)}$, we will further express $\Theta_R''(r)$ in terms of $\Theta_{\ell m}(r)$, $\Theta_{\ell m}'(r)$, ${}_{2}\hat{R}_{\ell m}(r)$, and ${}_{2}\hat{R}_{\ell m}'(r)$.

Similarly, we find the second part to take the form
\begin{widetext}
\begin{subequations}\label{eq:SBset}
\begin{align} 
    \mathcal{S}_{B}^{(1,1)} 
    =& \;\chi e^{-i\omega t}\left[\left(q_1^{\ell m}(r,\omega,M)\,{}_{2}\hat{R}_{\ell m}(r)
    +q_2^{\ell m}(r,\omega,M)\,{}_{2}\hat{R}_{\ell m}'(r)\right)
    \sin\theta\,{}_1Y_{\ell m}(\theta,\phi)\right. \nonumber\\
    & \;\left.+\left(q_3^{\ell m}(r,\omega,M)\,{}_{2}\hat{R}_{\ell m}(r)
    +q_4^{\ell m}(r,\omega,M)\,{}_{2}\hat{R}_{\ell m}'(r)\right)
    \cos\theta\,{}_2Y_{\ell m}(\theta,\phi)\right. \nonumber\\
    & \;\left.+\left(q_5^{\ell m}(r,\omega,M)\,{}_{2}\hat{R}_{\ell m}(r)
    +q_6^{\ell m}(r,\omega,M)\,{}_{2}\hat{R}_{\ell m}'(r)\right)
    \sin\theta\,{}_3Y_{\ell m}(\theta,\phi)\right]\,, 
    \label{eq:sB_coord} \\ 
    \tilde{\mathcal{S}}_{B}^{(1,1)} 
    =& \;\chi e^{i\omega t}\left[\left(\tilde{q}_1^{\ell m}(r,\omega,M)\,{}_{2}\bar{\hat{R}}_{\ell m}(r)
    +\tilde{q}_2^{\ell m}(r,\omega,M)\,{}_{2}\bar{\hat{R}}_{\ell m}'(r)\right)\right.
    \sin\theta\,{}_{-1}\bar{Y}_{\ell m}(\theta,\phi) \nonumber\\
    & \;\left.+\tilde{q}_3^{\ell m}(r,\omega,M)\,
    {}_{2}\bar{\hat{R}}_{\ell m}(r)
    \cos\theta\,{}_{-2}\bar{Y}_{\ell m}(\theta,\phi)\right]\,, 
    \label{eq:sBc_coord}
\end{align}
\end{subequations}
\end{widetext}
where ${}_{2}\hat{R}_{\ell m}(r)$ is the radial function of the Hertz potential given by Eq.~\eqref{eq:hertztoteuk}, and ${}_{2}\bar{\hat{R}}_{\ell m}(r)$ is the complex conjugate of ${}_{2}\hat{R}_{\ell m}(r)$. The radial functions $q_i^{\ell m}(r,\omega, M)$ and $\tilde{q}_i^{\ell m}(r,\omega,M)$ are presented in a Mathematica notebook as Supplementary Material~\cite{Pratikmodteuk}. Note that we have used the radial Teukolsky equation to eliminate any beyond-first-order derivatives of the Hertz potential to obtain a simplified expression in Eqs.~\eqref{eq:sB_coord} and \eqref{eq:sBc_coord}. The master equation of $\Psi_0^{(1,1)}$ in the IRG in the coordinate $\{t,r,\theta,\phi\}$ is presented in Eq.~\eqref{eq:general-teuk} of Sec.~\ref{sec:executive_summary_Teuk}.

%%%%%%%%%%%%%%%%%%%%%%%%%%%%%%%%%%%%%%%%%%%%%%%%%%%%%%%%%%%%
%%%%%%%%%%%%%%%%%%%%%%%%%%%%%%%%%%%%%%%%%%%%%%%%%%%%%%%%%%%%
\section{The evolution equation for $\vartheta^{(1,1)}$ and the modified Teukolsky equation for $\Psi_4^{(1,1)}$ in the ORG}
\label{sec:source_Teuk_ORG}

In this section, we evaluate the equations governing the evolution of the perturbed scalar field $\vartheta^{(1,1)}$ and the perturbed Weyl scalar $\Psi_4^{(1,1)}$. Although one can in principle evaluate the evolution equation of $\Psi_4^{(1,1)}$in the IRG, as briefly discussed in Sec.~\ref{sec:hertz_potential}, the evaluation is more convenient in the ORG. We follow a set of steps similar to those in the IRG for $\vartheta^{(1,1)}$ in Sec.~\ref{sec:source_scalar} and for $\Psi_0^{(1,1)}$ in Sec.~\ref{sec:source_Teuk}. Below, we present the master equations of $\vartheta^{(1,1)}$ and $\Psi_4^{(1,1)}$ in the ORG.

%%%%%%%%%%%%%%%%%%%%%%%%%%%%%%%%%%%%%%%%%%%%%%%%%%%%%%%%%%%%
\vspace{0.11in}
\subsection{The equation of $\vartheta^{(1,1)}$}

The scalar field perturbations are governed by Eq.~\eqref{eq:EOM_scalar_11}. We now represent the right-hand side of Eq.~\eqref{eq:EOM_scalar_11} as
\begin{equation} \label{eq:sourceorg_scalar}
     \mathcal{T}_{\vartheta}^{(1,1)}
    \equiv-\pi^{-\frac{1}{2}}M^2\left[R^*\!R\right]^{(0,1)}
    -\square^{(0,1)}\vartheta^{(1,0)}\,.
\end{equation}
Projecting the Pontryagin density onto the NP basis leads to the same set of equations as described in Sec.~\ref{sec:projnp} since our choice of gauge does not affect the quantities shown in Eqs.~\eqref{eq:pontandbox}. 

The master equation for the scalar field perturbations in the ORG are same as that shown in the IRG
\begin{align} \label{eq:scalareq_org}
     H_{\vartheta}^{(0,0)} \vartheta^{(1,1)} = \mathcal{T}_{\vartheta}^{(1,1)} \,,
\end{align}
with $H_{\vartheta}^{(0,0)}$ and $\vartheta^{(1,1)}$ both given in Eqs.~\eqref{eq:H_theta_coord} and \eqref{eq:scalar_separation} respectively, whereas $\mathcal{T}_{\vartheta}^{(1,1)}$ is given in Eq.\eqref{eq:sourceorg_scalar}. The left-hand side of Eq.~\eqref{eq:scalareq_org} in the ORG remains unchanged from the IRG since the operator acting on the scalar field perturbations is evaluated on the background. On the other hand, since the source term in Eq.~\eqref{eq:scalareq_org} depends on perturbed quantities, the value of these quantities is gauge dependent. In the ORG, the Pontryagin density given in Eq.~\eqref{eq:R*R_Psi2} holds the following form in the coordinate basis

\begin{widetext}
    \begin{align} \label{eq:R*R_coord_ORG}
    (R^*\!R)^{(0,1)}
    =& \;e^{-i\omega t}\left[\left(\textgoth{g}^{\ell m}_1(r,\omega,M){}_{-2}\hat{R}_{\ell m}(r)
    +\textgoth{g}_2^{\ell m}(r,\omega,M)\,{}_{-2}\hat{R}_{\ell m}'(r)\right)
    {}_0Y_{\ell m}(\theta,\phi)\right.\nonumber \\ 
    & \;\left.+\chi\left(\textgoth{g}^{\ell m}_3(r,\omega,M)\,{}_{-2}\hat{R}_{\ell m}(r)
    +\textgoth{g}_4^{\ell m}(r,\omega,M)\,{}_{-2}\hat{R}_{\ell m}'(r)\right)
    \sin\theta\,{}_{-1}Y_{\ell m}(\theta,\phi)\right.\nonumber\\
    & \;\left.+\chi\left(\textgoth{g}^{\ell m}_5(r,\omega,M)\,{}_{-2}\hat{R}_{\ell m}(r)
    +\textgoth{g}_6^{\ell m}(r,\omega,M)\,{}_{-2}\hat{R}_{\ell m}'(r)\right)
    \cos\theta\,{}_0Y_{\ell m}(\theta,\phi)\right.\nonumber \\
    & \;\left.+\chi\,{}_{-2}b^{m}_{\ell,\ell+1}
    \left(\textgoth{g}^{\ell m}_7(r,\omega,M)\,{}_{-2}\hat{R}_{\ell m}(r)
    +\textgoth{g}_8^{\ell m}(r,\omega,M)\,{}_{-2}\hat{R}_{\ell m}'(r)\right)
    {}_0Y_{\ell+1 m}(\theta,\phi)\right.\nonumber \\
    & \;\left.+\chi\,{}_{-2}b^{m}_{\ell,\ell-1}
    \left(\textgoth{g}^{\ell m}_9(r,\omega,M)\,{}_{-2}\hat{R}_{\ell m}(r)
    +\textgoth{g}_{10}^{\ell m}(r,\omega,M)\,{}_{-2}\hat{R}_{\ell m}'(r)\right)
    {}_0Y_{\ell-1 m}(\theta,\phi)\right]+\textrm{c.c.} \,,
\end{align}
%\end{widetext}
where functions $\textgoth{g}_i(r,\omega,M)$ are presented in a separate Mathematica notebook as Supplementary Material~\cite{Pratikmodteuk}, and ${}_{-2}\hat{R}_{\ell m}'(r)$ is the radial function of the Hertz potential for a slowly rotating Kerr BHs in GR computed from the radial function of the $\Psi_4^{(0,1)}$ using Eq.~\eqref{eq:hertztoteuk_ORG}. 

%\begin{widetext}
To evaluate the remaining part of the source term $\mathcal{T}_{\vartheta}^{(1,1)}$, we use the perturbed spin coefficients given in Eq.~\eqref{eq:spin_coefs_C} and metric perturbations given in Eq.~\eqref{eq:metricpert_org}. We obtain
\begin{align} \label{eq:box_coord_ORG}
    \square^{(0,0,1)}\vartheta^{(1,1,0)}
    =& \;e^{-i\omega t}\left[\left(
    \textgoth{h}^{\ell m}_1(r,\omega,M)\,{}_{-2}\hat{R}_{\ell m}(r)
    +\textgoth{h}^{\ell m}_2(r,\omega,M)\,{}_{-2}\hat{R}_{\ell m}'(r)\right)\sin\theta \,{}_{-1}Y_{\ell m}(\theta,\phi)
    \right.\nonumber \\ 
    & \;\left.+\textgoth{h}^{\ell m}_3(r,\omega,M)
    \left(2\vartheta_R'(r)+r\vartheta_R''(r)\right){}_{-2}\hat{R}_{\ell m}(r)
    \cos\theta\,{}_0Y_{\ell m}(\theta,\phi)\right]+\textrm{c.c.}\,,
\end{align}
\end{widetext}
where the functions $\textgoth{h}_i(r)$ are presented in a separate Mathematica notebook as Supplementary Material~\cite{Pratikmodteuk}, and $\vartheta_R(r)$ is the radial part of the background scalar field given in Eq.~\eqref{eq:scalar_stationary}. Similar to the case evaluated in the IRG in Sec.~\ref{sec:source_scalar}, the radial function ${}_{-2}\hat{R}_{\ell m}(r)$ is evaluated on the Schwarzschild background. Combining Eqs.~\eqref{eq:box_coord_ORG} and \eqref{eq:R*R_coord_ORG} gives us the complete source term $\mathcal{T}_{\vartheta}^{(1,1)}$. The master equation of $\vartheta^{(1,1)}$ in the ORG in the coordinate $\{t,r,\theta,\phi\}$ is presented in Eq.~\eqref{eq:general-teuk} of Sec.~\ref{sec:executive_summary_Teuk}.

%%%%%%%%%%%%%%%%%%%%%%%%%%%%%%%%%%%%%%%%%%%%%%%%%%%%%%%%%%%%
\subsection{The equation of $\Psi_4^{(1,1)}$}

In this subsection, we present the modified Teukolsky equation for the Weyl scalar perturbation $\Psi_4^{(1,1)}$ given in Eq.~\eqref{eq:master_eqn_non_typeD_Psi4} in the coordinate basis. Following steps similar to Sec.~\ref{sec:source_Teuk}, we separate the source terms into two categories: $\mathcal{T}_{\geo}^{(1,1)}$ and $\mathcal{T}^{(1,1)}$ whose forms in the NP basis have been given in Eqs.~\eqref{eq:T_geo}--\eqref{eq:T}.

%%%%%%%%%%%%%%%%%%%%%%%%%%%%%%%%%%%%%%%%%%%%%%%%%%%%%%%%%%%%
\subsubsection{Homogeneous part and $\mathcal{T}_{\geo}^{(1,1)}$}

Similar to Sec.~\ref{sec:modified_Teuk_operator}, by using the Weyl scalars and spin coefficients in Sec.~\ref{sec:NP_quantities_background} and Appendix~\ref{appendix:background_NP_more} along with Eq.~\eqref{eq:H_in_Teuk_ORG}, we find
\begin{subequations} \label{eq:H4_00}
\begin{align} 
    & \mathcal{H}_4^{(0,0,0)}
    =\frac{1}{2r^6}H_{4,\mathrm{TK}}^{(0,0,0)}\,,
    \label{eq:H4_000} \\
    & \mathcal{H}_4^{(0,1,0)}
    =\frac{1}{2r^6}H_{4,\mathrm{TK}}^{(0,1,0)}\,,
     \label{eq:H4_010}
\end{align}
\end{subequations}
where we define
\begin{equation}
    \left(H_4^{(0,0,0)}+\chi H_4^{(0,1,0)}\right)\Psi_4^{(1)}
    \equiv\left(\mathcal{H}_4^{(0,0,0)}+\chi\mathcal{H}_4^{(0,1,0)}\right)\psi_4^{(1)}
\end{equation}
by extracting a factor of $\rho^4$ from the Weyl scalar $\Psi_4^{(1)}$ following \cite{Teukolsky:1973ha}, i.e.,
\begin{equation} \label{eq:psi4_separation}
\begin{aligned}
    \Psi_4^{(1)}
    \equiv \rho^4\psi_4^{(1)}
    =& \;\rho^{4}\left[{}_{-2}R_{\ell m}^{(0,1)}(r)
    +\zeta\,{}_{-2}R_{\ell m}^{(1,1)}(r)+\mathcal{O}(\zeta^2)\right] \\
    & \;{}_{-2}S_{\ell m}(\theta)e^{-i\omega t+im\phi}\,.
\end{aligned}
\end{equation}
The operators $H_{4,\mathrm{TK}}^{(0,0,0)}$ and $H_{4,\mathrm{TK}}^{(0,1,0)}$ are $\mathcal{O}(\chi^0)$ and $\mathcal{O}(\chi^1)$ terms of the Teukolsky operator $H_{4,\mathrm{TK}}$ for $\psi_{4}$ [Eq.~(4.7) in \cite{Teukolsky:1973ha}], respectively,
\begin{subequations} \label{eq:H4_00_TK}
\begin{align}
    \begin{split} \label{eq:H4_000_TK} 
        & H_{4,\mathrm{TK}}^{(0,0,0)}
        = -r(r-r_s)\partial^2_{r}+2(r-M)\partial_r
        -\frac{D(r)}{r-r_s} \\ 
        & \;-\partial^2_{\theta}-\cot{\theta}\partial_{\theta}
        +\left(-2 \cot\theta+m \csc\theta\right)^2-2\,,
    \end{split} \\
    \begin{split} \label{eq:H4_010_TK}
        & H_{4, \mathrm{TK}}^{(0,1,0)}
        =4M\left[\frac{m(i(r-M)+M\omega r)}{r(r-r_s)}
        -\omega\cos{\theta}\right]\,,
    \end{split} \\
    & D(r)=-4i\omega r(r-3M)+\omega^{2}r^{3} \nonumber\,.
\end{align}
\end{subequations}
The Teukolsky equation corresponding to Eqs.~\eqref{eq:H4_000_TK} and \eqref{eq:H4_010_TK} is separable with ${}_{-2}R_{\ell m}^{(0,1)}(r)$ satisfying
\begin{equation}
\begin{aligned}
    & \left[r(r-r_s)\partial^2_{r}-2(r-M)\partial_r+\frac{D(r)}{r-r_s}\right. \\
    & \left.-\frac{4m\chi M(i(r-M)+M\omega r)}{r(r-r_s)}
    -{}_{-2}A_{\ell m}\right]{}_{-2}R_{\ell m}^{(0,1)}(r)=0\,.
\end{aligned}
\end{equation}

For the same reason in Sec.~\ref{sec:modified_Teuk_operator}, we do not need metric reconstruction to compute $\mathcal{T}_{\geo}^{(1,1)}$ since the background spacetime is still Petrov type D. The source terms $\mathcal{T}_{\geo}^{(1,1)}$ hold the form
\begin{equation} \label{eq:source_psi4_geo}
    \mathcal{T}_{\geo}^{(1,0,1)} = 0 \,, \quad 
    \mathcal{T}_{\geo}^{(1,1,1)}
    =H_4^{(1,1,0)}\Psi_4^{(0,0,1)}
    =\mathcal{H}_4^{(1,1,0)}\psi_4^{(0,0,1)}
\end{equation}
with
\begin{equation} \label{eq:H4_110}
    \begin{aligned}
        \mathcal{H}_4^{(1,1,0)}
        =& \;\frac{-i m M^4}{448r^{13}(r-r_s)}
        \left(D_1(r)-4i\omega r^2D_2(r)\right) \\
        & \;+\frac{iM^4}{16r^{13}}\cos{\theta}
        \left(D_3(r)-\frac{i\omega r^2D_4(r)}{2}\right) \\
        & \;+\frac{iM^4}{32r^{12}}\left[(r-r_s)D_4(r)\cos{\theta}\partial_r
        -\frac{D_5(r)}{2r}\sin{\theta}\partial_{\theta}\right]\,,
    \end{aligned}
\end{equation}
where we have absorbed the factor of $\rho^4$ into $\mathcal{H}_4^{(1,1,0)}$, and the functions $D_i(r)$ are presented in Appendix~\ref{appendix:background_NP_more}.

%%%%%%%%%%%%%%%%%%%%%%%%%%%%%%%%%%%%%%%%%%%%%%%%%%%%%%%%%%%%
\subsubsection{$\mathcal{T}^{(1,1)}$}

Using the expression for the metric perturbation in the ORG given in Eq.~\eqref{eq:metricpert_org}, one can evaluate the perturbed spin coefficients and perturbed Weyl scalars at $\mathcal{O}(\zeta^0,\epsilon^1)$. These can then be used to evaluate the following source terms, which can be divided into two parts based on whether $S_{3,4}$ are dynamical.

\begin{align}
    \mathcal{T}^{(1,1)} &= \mathcal{T}_{I}^{(1,1)} + \mathcal{T}_{II}^{(1,1)} \,, \nonumber \\
    \mathcal{T}_{I}^{(1,1)} &= \mathcal{E}_4^{(0,1)}S_4^{(1,0)} - \mathcal{E}_3^{(0,1)}S_3^{(1,0)} \,, \nonumber \\
    \mathcal{T}_{II}^{(1,1)} &= \mathcal{E}_4^{(0,0)}S_4^{(1,1)} - \mathcal{E}_3^{(0,0)}S_3^{(1,1)} \,.
\end{align}
Analogous to Sec.~\ref{sec:source_Teuk}, $\mathcal{T}_{I}^{(1,1)}$ consists of terms dependent on the stationary scalar field, the background, and the perturbed metric in GR.

Similarly, we can further divide $\mathcal{T}_{II}^{(1,1)}$ into two categories based on whether terms are proportional to $\Phi_{ij}^{(1,0)}$ or $\Phi_{ij}^{(1,1)}$, which we denote as $\mathcal{T}_{IIA}^{(1,1)}$ and $\mathcal{T}_{IIB}^{(1,1)}$ respectively.
\begin{equation} \label{eq:sourceteuk2a_ORG}
    \begin{aligned}
        & \mathcal{T}_{II}^{(1,1)}
        =\mathcal{T}_{IIA}^{(1,1)}+\mathcal{T}_{IIB}^{(1,1)}\,, \\
        & \mathcal{T}_{IIA}^{(1,1)}
        =\mathcal{E}_{4}^{(0,0)}S_{4,A}^{(1,1)}	
        -\mathcal{E}_{3}^{(0,0)}S_{3,A}^{(1,1)}\,, \\
        & \mathcal{T}_{IIB}^{(1,1)}
        =\mathcal{E}_{4}^{(0,0)}S_{4,B}^{(1,1)}	
        -\mathcal{E}_{3}^{(0,0)}S_{3,B}^{(1,1)}\,.
    \end{aligned}
\end{equation}
\vspace{0.1cm}
Here $S_{3,B}^{(1,1)}$ and $S_{4,B}^{(1,1)}$ have terms proportional to $\Phi_{ij}^{(1,1)}$, whereas $S_{3,A}^{(1,1)}$ and $S_{4,A}^{(1,1)}$ have terms proportional to $\Phi_{ij}^{(1,0)}$. 

Expressing these source terms in the coordinate basis, the terms proportional to the scalar field perturbation are given by
\begin{widetext}
\begin{subequations} \label{eq:ta_coord}
\begin{align} 
    \mathcal{T}_{A}^{(1,1)} 
    =& \;e^{-i\omega t}\left[\left(\textgoth{p}_1^{\ell m}(r,\omega,M)\Theta_{\ell m}(r)
    +\textgoth{p}_2^{\ell m}(r,\omega,M)\Theta_{\ell m}'(r)
    +\textgoth{p}_3^{\ell m}(r,\omega,M)\Theta_{\ell m}''(r)\right)
    {}_{-2}Y_{\ell m}(\theta,\phi)\right. \nonumber\\
    & \;\left.+\chi\left(\textgoth{p}_4^{\ell m}(r,\omega,M)\Theta_{\ell m}(r)
    +\textgoth{p}_5^{\ell m}(r,\omega,M)\Theta_{\ell m}'(r)
    +\textgoth{p}_6^{\ell m}(r,\omega,M)\Theta_{\ell m}''(r)\right)
    \sin\theta\,{}_{-1}Y_{\ell m}(\theta,\phi)\right. \nonumber\\
    & \;\left.+\chi\left(\textgoth{p}_7^{\ell m}(r,\omega,M)\Theta_{\ell m}(r)
    +\textgoth{p}_8^{\ell m}(r,\omega,M)\Theta_{\ell m}'(r)
    +\textgoth{p}_9^{\ell m}(r,\omega,M)\Theta_{\ell m}''(r)\right)
    \cos\theta\,{}_{-2}Y_{\ell m}(\theta,\phi)\right. \nonumber\\
    & \;\left.+\chi\,{}_{0}b^m_{\ell,\ell+1}
    \left(\textgoth{p}_{10}^{\ell m}(r,\omega,M)\Theta_{\ell m}(r)
    +\textgoth{p}_{11}^{\ell m}(r,\omega,M)\Theta_{\ell m}'(r)
    +\textgoth{p}_{12}^{\ell m}(r,\omega,M)\Theta_{\ell m}''(r)\right)
    {}_{-2}Y_{\ell+1\,m}(\theta,\phi)\right. \nonumber\\
    & \;\left.+\chi\,{}_{0}b^m_{\ell,\ell-1}
    \left(\textgoth{p}_{13}^{\ell m}(r,\omega,M)\Theta_{\ell m}(r)
    +\textgoth{p}_{14}^{\ell m}(r,\omega,M)\Theta_{\ell m}'(r)
    +\textgoth{p}_{15}^{\ell m}(r,\omega,M)\Theta_{\ell m}''(r)\right)
    {}_{-2}Y_{\ell-1\,m}(\theta,\phi)\right]\,, 
    \label{eq:tA_coord_org} \\
    \tilde{\mathcal{T}}_{A}^{(1,1)} 
    =& \;e^{i\omega t}\left[-\left(\bar{\textgoth{p}}_1^{\ell m}(r,\omega,M)\bar{\Theta}_{\ell m}(r)
    +\bar{\textgoth{p}}_2^{\ell m}(r,\omega,M)\bar{\Theta}_{\ell m}'(r)
    +\bar{\textgoth{p}}_3^{\ell m}(r,\omega,M)\bar{\Theta}_{\ell m}''(r)\right)
    {}_{2}\bar{Y}_{\ell m}(\theta,\phi)\right. \nonumber\\
    & \;\left.-\chi\left(\bar{\textgoth{p}}_4^{\ell m}(r,\omega,M)\bar{\Theta}_{\ell m}(r)
    +\bar{\textgoth{p}}_5^{\ell m}(r,\omega,M)\bar{\Theta}_{\ell m}'(r)
    +\bar{\textgoth{p}}_6^{\ell m}(r,\omega,M)\bar{\Theta}_{\ell m}''(r)\right)
    \sin\theta\,{}_{1}\bar{Y}_{\ell m}(\theta,\phi)\right.\nonumber\\
    & \;\left.+\chi\left(\bar{\textgoth{p}}_7^{\ell m}(r,\omega,M)\bar{\Theta}_{\ell m}(r)
    +\bar{\textgoth{p}}_{8}^{\ell m}(r,\omega,M)\bar{\Theta}_{\ell m}'(r)
    +\bar{\textgoth{p}}_{9}^{\ell m}(r,\omega,M)\bar{\Theta}_{\ell m}''(r)\right)
    \cos\theta\,{}_{2}\bar{Y}_{\ell m}(\theta,\phi)\right. \nonumber\\
    & \;\left.-\chi\,{}_{0}b^m_{\ell,\ell+1}
    \left(\bar{\textgoth{p}}_{10}^{\ell m}(r,\omega,M)\bar{\Theta}_{\ell m}(r)
    +\bar{\textgoth{p}}_{11}^{\ell m}(r,\omega,M)\bar{\Theta}_{\ell m}'(r)
    +\bar{\textgoth{p}}_{12}^{\ell m}(r,\omega,M)\bar{\Theta}_{\ell m}''(r)\right)
    {}_{2}\bar{Y}_{\ell+1 m}(\theta,\phi)\right.\nonumber\\
    & \;\left.-\chi\,{}_{0}b^m_{\ell,\ell-1}
    \left(\bar{\textgoth{p}}_{13}^{\ell m}(r,\omega,M)\bar{\Theta}_{\ell m}(r)
    +\bar{\textgoth{p}}_{14}^{\ell m}(r,\omega,M)\bar{\Theta}_{\ell m}'(r)
    +\bar{\textgoth{p}}_{15}^{\ell m}(r,\omega,M)\bar{\Theta}_{\ell m}''(r)\right)
    {}_{2}\bar{Y}_{\ell-1 m}(\theta,\phi)\right]\,. 
    \label{eq:tAc_coord_org}
\end{align}
\end{subequations}
%\end{widetext}
The source terms proportional to the background scalar field can be expressed in the coordinate basis as
%\begin{widetext}
\begin{subequations} \label{eq:tb_coord}
\begin{align} 
    \mathcal{T}_{B}^{(1,1)} 
    =& \;\chi e^{-i\omega t}\left[\left(\textgoth{q}_1^{\ell m}(r,\omega,M)\,{}_{-2}\hat{R}_{\ell m}(r)
    +\textgoth{q}_2^{\ell m}(r,\omega,M)\,{}_{-2}\hat{R}_{\ell m}'(r)\right)
    \sin\theta\,{}_{-1}Y_{\ell m}(\theta,\phi)\right. \nonumber\\
    & \;\left.+\left(\textgoth{q}_3^{\ell m}(r,\omega,M)\,{}_{-2}\hat{R}_{\ell m}(r)
    +\textgoth{q}_4^{\ell m}(r,\omega,M)\,{}_{-2}\hat{R}_{\ell m}'(r)\right)
    \cos\theta\,{}_{-2}Y_{\ell m}(\theta,\phi)\right. \nonumber\\
    & \;\left.+\left(\textgoth{q}_5^{\ell m}(r,\omega,M)\,{}_{-2}\hat{R}_{\ell m}(r)
    +\textgoth{q}_6^{\ell m}(r,\omega,M)\,{}_{-2}\hat{R}_{\ell m}'(r)\right)
    \sin\theta\,{}_{-3}Y_{\ell m}(\theta,\phi)\right]\,, 
    \label{eq:tB_coord_org} \\
    \tilde{\mathcal{T}}_{B}^{(1,1)} 
    =& \;\chi e^{i\omega t}\left[\left(\tilde{\textgoth{q}}_1^{\ell m}(r,\omega,M)\,{}_{-2}\bar{\hat{R}}_{\ell m}(r)
    +\tilde{\textgoth{q}}_2^{\ell m}(r,\omega,M)\,{}_{-2}\bar{\hat{R}}_{\ell m}'(r)\right)\right.
    \sin\theta\,{}_{1}\bar{Y}_{\ell m}(\theta,\phi) \nonumber\\
    & \;\left.+\tilde{\textgoth{q}}_3^{\ell m}(r,\omega,M)\,
    {}_{-2}\bar{\hat{R}}_{\ell m}(r)
    \cos\theta\,{}_{2}\bar{Y}_{\ell m}(\theta,\phi)\right]\,, 
    \label{eq:tBc_coord_org}
\end{align}
\end{subequations}
\end{widetext}
where ${}_{-2}\hat{R}_{\ell m}'(r)$ is the radial function of the Hertz potential given by Eq.~\eqref{eq:hertztoteuk_ORG}. The functions $\textgoth{q}_i^{\ell m}(r,\omega,M)$ and $\tilde{\textgoth{q}}^{ \ell m}_i(r,\omega,M)$ (not to be confused with $\bar{\textgoth{q}}^{\ell m}_i(r,\omega,M)$) are functions presented in a Mathematica notebook as Supplementary Material~\cite{Pratikmodteuk}. The master equation of $\Psi_4^{(1,1)}$ in the ORG in the coordinate $\{t,r,\theta,\phi\}$ is presented in Eq.~\eqref{eq:general-teuk} of Sec.~\ref{sec:executive_summary_Teuk}.

%%%%%%%%%%%%%%%%%%%%%%%%%%%%%%%%%%%%%%%%%%%%%%%%%%%%%%%%%%%%
\section{Executive Summary of all Master Equations}
\label{sec:executive_summary_Teuk}
\vspace{0.11in}

This section presents an executive summary of the main results of this paper, whose derivation was presented in Secs.~\ref{sec:source_scalar}, \ref{sec:source_Teuk} and \ref{sec:source_Teuk_ORG}. 
Through Secs.~\ref{sec:source_scalar}, \ref{sec:source_Teuk} and \ref{sec:source_Teuk_ORG}, we used the tetrad defined in Sec.~\ref{sec:NP_quantities_background} to rewrite Eqs.~\eqref{eq:master_eqn_non_typeD_Psi0} and~\eqref{eq:EOM_scalar_11} using the IRG and Eqs.~\eqref{eq:master_eqn_non_typeD_Psi4} and~\eqref{eq:EOM_scalar_11} using the ORG. In this section, we summarize these results and condense them into a single master equation for convenience. In later sections, we will present and apply a procedure to decouple the master equations for the scalar field perturbation $\vartheta^{(1,1)}$ and the Weyl scalar perturbations $\Psi_{0,4}^{(1,1)}$ (or $\Psi_4^{(1,1)}$) into a set of coupled radial ordinary differential equations in the IRG (or ORG).

With this in mind, the master equations for $\vartheta^{(1,1)}$ [Eq.~\eqref{eq:EOM_scalar_11}], $\Psi_0^{(1,1)}$ [Eq.~\eqref{eq:master_eqn_non_typeD_Psi0}], and $\Psi_4^{(1,1)}$ [Eq.~\eqref{eq:master_eqn_non_typeD_Psi4}] can \textit{all} be expressed as
\begin{widetext}
\begin{equation}\label{eq:general-teuk}
\begin{aligned}
    & H\psi=\frac{r^3}{r-2M}\frac{\partial^2\psi}{\partial t^2}
    -\frac{4\chi M^2}{2 M-r}\frac{\partial^2\psi}{\partial t\partial\phi}
    -\csc^2\theta\frac{\partial^2\psi}{\partial\phi^2}
    -r(r-2M)\frac{\partial^2 \psi}{\partial r^2}
    -2(s+1)(r-M) \frac{\partial\psi}{\partial r}
    -\frac{1}{\sin\theta}\frac{\partial}{\partial\theta}
    \left(\sin\theta \frac{\partial\psi}{\partial\theta}\right) \\
    &+ s\left[\chi M\left(\frac{1}{2M-r}-\frac{1}{r}\right)
    -2i\cot\theta\csc\theta\right]\frac{\partial\psi}{\partial\phi}
    +2s\left[\frac{r(3 M-r)}{2M-r}+i\chi M\cos\theta\right]
    \frac{\partial\psi}{\partial t}+\left(s^2\cot^2\theta-s\right)\psi
    =\xi(r)\textgoth{S}\,,
\end{aligned}
\end{equation}
\end{widetext}
where $H$ represents the Teukolsky operator for a spin $s$ field in GR given in \cite{Teukolsky:1973ha}. Recall that $\chi$ is the dimensionless spin parameter, and $M$ is the BH mass. In Table~\ref{tab:psi_table}, we present the field quantities $\psi$ which satisfy this equation and the source terms, $\xi(r)$ and $\textgoth{S}$ on the right-hand side of Eq.~\eqref{eq:general-teuk}, dependent on the gauge and the spin weight $s$ of these fields. Observe that, clearly, the differential operator $H$ in Eq.~\eqref{eq:general-teuk} acting on the field quantity $\psi$ is exactly the same as the one derived by Teukolsky in~\cite{Teukolsky:1973ha} in GR for Kerr BH perturbations [c.f. Eq.~(4.7) therein], expanded to leading order in spin. In addition, from Table~\ref{tab:psi_table} and the source terms in Eqs.~\eqref{eq:source_psi0_geo}, \eqref{eq:SAset}, \eqref{eq:SBset}, \eqref{eq:source_psi4_geo}, \eqref{eq:ta_coord}, and \eqref{eq:tb_coord}, we notice that the $(l,m)$ and $(l,m)$ modes of $\Psi_{0,4}^{(1,1)}$ need to be solved jointly, as we will discuss in more detail in Sec.~\ref{sec:separation_variable}.

\begin{table*}[] 
    \centering
    \begin{tabular}{c|c|c|c||c|c|}
        \hline
        \multirow{3}{*}{$\psi$}
        & \multirow{3}{*}{$\xi(r)$}
        & \multicolumn{2}{c||}{Ingoing radiation gauge} 
        & \multicolumn{2}{c|}{Outgoing radiation gauge} \\
        \cline{3-6}
        & &\begin{tabular}{c} 
            s \\ (spin weight) \end{tabular} 
        & \begin{tabular}{c} 
            \textgoth{S} \\ (equations)
        \end{tabular} 
        & \begin{tabular}{c} 
            s \\ (spin weight) \end{tabular} 
        & \begin{tabular}{c}
            \textgoth{S} \\ (equations)
        \end{tabular} \\
        \hline
        \multirow{2}{*}{$\vartheta^{(1,1)}$} & \multirow{2}{*}{$-r^2$} & \multirow{2}{*}{$0$} & $-\pi^{-\frac{1}{2}}M^2(R^*\!R)^{(0,1)}-\chi\square^{(0,0,1)}\vartheta^{(1,1,0)}$ & \multirow{2}{*}{$0$} & $-\pi^{-\frac{1}{2}}M^2(R^*\!R)^{(0,1)}-\chi\square^{(0,0,1)}\vartheta^{(1,1,0)}$ \\
         & & & Eqs.~\eqref{eq:R*R_coord_IRG} \& \eqref{eq:box_coord_IRG} &  & Eqs.~\eqref{eq:R*R_coord_ORG} \& \eqref{eq:box_coord_ORG} \\
         \hline
        \multirow{2}{*}{$\Psi_0^{(1,1)}$} & \multirow{2}{*}{$2r^2$} & \multirow{2}{*}{$+2$} & $\mathcal{S}_{\geo}^{(1,1)}+\mathcal{S}^{(1,1)}_{A}+\tilde{\mathcal{S}}^{(1,1)}_{A}+\mathcal{S}_{B}^{(1,1)}(r) +\tilde{\mathcal{S}}^{(1,1)}_{B}$ & \multirow{2}{*}{$-$} & \multirow{2}{*}{$-$} \\
        & & & Eqs.~\eqref{eq:source_psi0_geo}, \eqref{eq:SAset}, \& \eqref{eq:SBset} &  &  \\
        \hline
        \multirow{2}{*}{$\rho^{-4}\Psi_4^{(1,1)}$} & \multirow{2}{*}{$2r^6$} & \multirow{2}{*}{$-$} & \multirow{2}{*}{$-$} & \multirow{2}{*}{$-2$} & $\mathcal{T}_{\geo}^{(1,1)}+\mathcal{T}^{(1,1)}_{A}+\tilde{\mathcal{T}}^{(1,1)}_{A}+\mathcal{T}_{B}^{(1,1)}(r) +\tilde{\mathcal{T}}^{(1,1)}_{B}$ \\
        & & &  &  & Eqs.~\eqref{eq:source_psi4_geo}, \eqref{eq:ta_coord}, \& \eqref{eq:tb_coord} \\
        \hline
    \end{tabular}
    \caption{In this table, we present the quantities $\Psi$, the spin weight $s$ and the source terms $\textgoth{S}$ that appear in Eq.~\eqref{eq:general-teuk}.}
    \label{tab:psi_table}
\end{table*}
\vspace{-0.08in}

%%%%%%%%%%%%%%%%%%%%%%%%%%%%%%%%%%%%%%%%%%%%%%%%%%%%%%%%%%%%	
%%%%%%%%%%%%%%%%%%%%%%%%%%%%%%%%%%%%%%%%%%%%%%%%%%%%%%%%%%%%
\section{Separation of variables and extraction of the radial master equation}
\label{sec:separation_variable}
%\vspace{0.12in}
	
In this section, we extract the radial parts of the master equations for $\vartheta^{(1,1)}$ [Eq.~\eqref{eq:EOM_scalar_11}] (both in the IRG and ORG), for $\Psi_0^{(1,1)}$ [Eq.~\eqref{eq:master_eqn_non_typeD_Psi0}], and for $\Psi_4^{(1,1)}$ [Eq.~\eqref{eq:master_eqn_non_typeD_Psi4}]. Let us first present our procedures for eliminating the angular dependence in these equations and then apply them to specific cases.

%%%%%%%%%%%%%%%%%%%%%%%%%%%%%%%%%%%%%%%%%%%%%%%%%%%%%%%%%%%%
\subsection{Eiminating the angular dependence} 
\label{sec:elim-ylm}

To eliminate the angular dependence of these master equations, we utilize the properties of spin-weighted spheroidal harmonics in Sec.~\ref{sec:sphharm} and go through the following procedures:
\begin{enumerate}
    \item From Secs.~\ref{sec:source_scalar}--\ref{sec:source_Teuk_ORG}, we first observe that the master equations of $\vartheta^{(1,1)}$ and $\Psi_{0,4}^{(1,1)}$ after decomposition into spin-weighted spheroidal harmonics [i.e., Eqs.~\eqref{eq:scalar_separation}, \eqref{eq:psi0_separation}, and \eqref{eq:psi4_separation}] all take the form
    \begin{widetext}
    \begin{subequations} \label{eq:master_eqn_scheme}
    \begin{align}
        & {}_{s}H_{\ell m}\left[{}_{s}\psi_{\ell m}(r)
        {}_{s}S_{\ell m}(\theta)\right]e^{-i\omega_{\ell m} t+im\phi}
        =\sum_{k}{}_{s}P_{\ell m}^{k}(r){}_{s}f_{\ell m}^{k}(\theta)e^{-i\omega_{\ell m} t+im\phi}
        +{}_{s}Q_{\ell m}^{k}(r){}_{s}\bar{f}_{\ell m}^{k}(\theta)
        e^{i\bar{\omega}_{\ell m}t-im\phi}\,, 
        \label{eq:master_eqn_scheme_lm} \\
        & {}_{s}H_{\ell\,-m}\left[{}_{s}\psi_{\ell\,-m}(r)
        {}_{s}S_{\ell\,-m}(\theta)\right]e^{-i\omega_{\ell\,-m} t-im\phi}
        =\sum_{k}{}_{s}P_{\ell\,-m}^{k}(r){}_{s}f_{\ell\,-m}^{k}(\theta)
        e^{-i\omega_{\ell\,-m}t-im\phi}
        +{}_{s}Q_{\ell\,-m}^{k}(r){}_{s}\bar{f}_{\ell\,-m}^{k}(\theta)
        e^{i\bar{\omega}_{\ell\,-m}t+im\phi}\,, 
        \label{eq:master_eqn_scheme_l-m}
    \end{align}
    \end{subequations}   
    \end{widetext}
    where ${}_{s}H_{\ell m}$ is the $(\ell,m)$ mode of the Teukolsky operator in GR for particles of spin $0$ [i.e., $H_{\vartheta}^{(0,0)}$ in Eq.~\eqref{eq:H_theta_coord}], spin $2$ [i.e., $H_0^{(0,0)}$ in Eqs.~\eqref{eq:H0_00} and \eqref{eq:H0_00_TK}], and spin $-2$ [i.e., $\mathcal{H}_4^{(0,0)}$ in Eqs.~\eqref{eq:H4_00} and \eqref{eq:H4_00_TK}]. The radial function ${}_{s}\psi_{\ell m}(r)$ is the radial part of $\vartheta^{(1,1)}$ [i.e., $\Theta_{\ell m}(r)$], $\Psi_0^{(1,1)}$ [i.e., ${}_{2}R_{\ell m}^{(1,1)}(r)$], or $\rho^{-4}\Psi_{4}^{(1,1)}$ [i.e., ${}_{-2}R_{\ell m}^{(1,1)}(r)$] to be solved for. The angular function ${}_{s}S_{\ell m}(\theta)$ is the $\theta$-dependent part of spin-weighted spheroidal harmonics ${}_{s}\mathcal{Y}(\theta,\phi)$. The radial functions ${}_{s}P_{\ell m}^{k}(r)$, ${}_{s}Q_{\ell m}^{k}(r)$ and angular functions ${}_{s}f_{\ell m}^{k}(\theta)$ can be extracted from the source terms found in Secs.~\ref{sec:source_scalar}--\ref{sec:source_Teuk_ORG}. In the equation for $\vartheta^{(1,1)}$, we can observe from Eqs.~\eqref{eq:R*R_coord_IRG}, \eqref{eq:box_coord_IRG}, \eqref{eq:R*R_coord_ORG}, and \eqref{eq:box_coord_ORG} that ${}_{0}P_{\ell m}^{k}(r)={}_{0}\bar{Q}_{\ell m}^{k}(r)$ since the scalar field is real, while there is no such a constraint for $\Psi_{0,4}^{(1,1)}$ since they are complex in general.

    \item  Equation~\eqref{eq:master_eqn_scheme} assumes that a single $(\ell,m)$ mode [Eq.~\eqref{eq:master_eqn_scheme_lm}] or a single $(\ell,-m)$ mode [Eq.~\eqref{eq:master_eqn_scheme_l-m}] can solve the modified Teukolsky equation. However, this is in general not true since the source term in Eq.~\eqref{eq:master_eqn_scheme_lm} is a mixture of modes proportional to $e^{-i\omega_{\ell m}t}$ and $e^{i\bar{\omega}_{\ell m}t}$, and similarly for Eq.~\eqref{eq:master_eqn_scheme_l-m}. On the other hand, in GR, one has the well-known symmetry \cite{Leaver:1985ax}
    \begin{equation}
        \omega^{(0)}_{\ell m}=-\bar{\omega}^{(0)}_{\ell -m}\,,
    \end{equation}
    so both Eqs.~\eqref{eq:master_eqn_scheme_lm} and \eqref{eq:master_eqn_scheme_l-m} contain source terms proportional to $e^{-i\omega_{\ell m}t}$ and $e^{-i\omega_{\ell -m}t}$. Thus, one has to consider Eqs.~\eqref{eq:master_eqn_scheme_lm} and \eqref{eq:master_eqn_scheme_l-m} jointly or solve the linear combination , i.e.,
    \begin{widetext}
    \begin{subequations} \label{eq:solution_ansatz}
    \begin{align}
        & \Psi_{\ell m}^{(0,1)}
        ={}_{s}R_{\ell m}^{(0,1)}(r){}_{s}S_{\ell m}(\theta)
        e^{-i\omega_{\ell m} t+im\phi}
        +\eta_{\ell m}\;{}_{s}R_{\ell -m}^{(0,1)}(r){}_{s}S_{\ell\,-m}(\theta)
        e^{i\omega_{\ell m}t-im\phi}\,, \\
        & \Psi_{\ell m}^{(1,1)}
        ={}_{s}\psi_{\ell m}(r){}_{s}S_{\ell m}(\theta)
        e^{-i\omega_{\ell m} t+im\phi}
        +\eta_{\ell m}\;{}_{s}\psi_{\ell -m}(r){}_{s}S_{\ell\,-m}(\theta)
        e^{i\omega_{\ell m}t-im\phi}\,,
    \end{align} 
    \end{subequations}   
    \end{widetext}
    where we have absorbed an overall factor into the normalization of ${}_{s}R_{\ell m}^{(0,1)}(r)$ and ${}_{s}\psi_{\ell m}(r)$. Furthermore, we have also taken $\omega^{(0)}_{\ell m}=-\bar{\omega}^{(0)}_{\ell -m}$ not just in GR but also in dCS gravity due to the structure of Eq~\eqref{eq:master_eqn_scheme}. As discussed and shown in more detail in \cite{paritybreaking}, one can solve for both the complex constant $\eta_{\ell m}$ and the QNM frequencies $\omega_{\ell m}$ using the eigenvalue perturbation method method in \cite{Zimmerman:2014aha, Mark_Yang_Zimmerman_Chen_2015, Hussain:2022ins}. The combination $(\eta_{\ell m},\omega_{\ell m})$ have two independent solutions \cite{Hussain:2022ins, Cano:2023tmv, paritybreaking}, resulting in the breaking of isospectrality. In this case, by plugging the ansatz in Eq.~\eqref{eq:solution_ansatz} into the scalar field equation or the modified Teukolsky equations, using Eq.~\eqref{eq:master_eqn_scheme}, and matching the phase of the terms, we find
    \begin{widetext}
    \begin{subequations} \label{eq:master_eqn_scheme2}
    \begin{align}
        {}_{s}H_{\ell m}\left[{}_{s}\psi_{\ell m}(r)
        {}_{s}S_{\ell m}(\theta)\right]
        & =\sum_{k}{}_{s}P_{\ell m}^{k}(r){}_{s}f_{\ell m}^{k}(\theta)
        +\bar{\eta}_{\ell m}\,{}_{s}Q_{\ell\,-m}^{k}(r){}_{s}\bar{f}_{\ell\,-m}^{k}(\theta)\,, 
        \label{eq:master_eqn_scheme2_lm} \\
        {}_{s}H_{\ell\,-m}\left[{}_{s}\psi_{\ell\,-m}(r)
        {}_{s}S_{\ell\,-m}(\theta)\right]
        & =\sum_{k}{}_{s}P_{\ell\,-m}^{k}(r){}_{s}f_{\ell\,-m}^{k}(\theta)
        +
        \frac{1}{\eta_{\ell m}}{}_{s}Q_{\ell\,m}^{k}(r){}_{s}\bar{f}_{\ell\,m}^{k}(\theta)\,, \label{eq:master_eqn_scheme2_l-m}
    \end{align}
    \end{subequations}   
    \end{widetext}
    where we have divided a factor of $\eta_{\ell m}$ in Eq.~\eqref{eq:master_eqn_scheme2_l-m}, and ${}_{s}\psi_{\ell\pm m}(r)$ are radial solutions tied to $(\eta_{\ell m},\omega_{\ell m})$. Similar procedures for generic modified gravity theories can be found in \cite{paritybreaking}.  
    
    \item For the evolution equation for $\vartheta^{(1,1)}$, we observe that ${}_{0}f_{\ell m}^{k}(\theta)$ consists of the following terms 
    \begin{itemize}
        \item ${}_0Y_{\ell m}(\theta,\phi)$ and ${}_{0}Y_{\ell\pm1\,m}(\theta,\phi)$,
        \item $\cos\theta\,{}_0Y_{\ell m}(\theta,\phi)$, and
        \item $\sin\theta\,{}_{\pm1}Y_{\ell m}(\theta,\phi)$.
    \end{itemize}
    For the master equation for $\Psi_{0,4}^{(1,1)}$, ${}_{\pm 2}f_{\ell m}^{k}(\theta)$ contains
    \begin{itemize}
        \item ${}_{\pm2}Y_{\ell m}(\theta,\phi)$ and ${}_{\pm2}Y_{\ell\pm1\,m}(\theta,\phi)$,
        \item $\cos\theta\,{}_{\pm2}Y_{\ell m}(\theta,\phi)$,
        \item $\sin\theta\,{}_{\pm1}Y_{\ell m}(\theta,\phi)$ and $\sin\theta{}_{\pm3}Y_{\ell m}(\theta,\phi)$.
    \end{itemize}
    Notice that ${}_{s}f_{\ell m}^{k}(\theta)$ are angular functions in the modified Teukolsky equation for the particle of spin weight $s$ and mode $(\ell,m)$. The subscript $s$ and subscripts $(\ell,m)$ do not indicate the mode number of the angular function itself. For example, ${}_{0}f_{\ell m}^{k}(\theta)$ contains terms proportional to $\sin\theta\,{}_{\pm 1}Y_{\ell m}(\theta,\phi)$.
    
    \item As shown in Sec.~\ref{sec:lhs_scalar_eqn}, the homogeneous part of Eq.~\eqref{eq:EOM_scalar_11} for $\vartheta^{(1,1)}$ is separable in $r$ and $\theta$ if one decomposes $\vartheta^{(1,1)}$ into ${}_{0}\mathcal{Y}_{\ell m}(\theta,\phi)$. Thus, to extract the radial part of Eq.~\eqref{eq:EOM_scalar_11}, we multiply Eq.~\eqref{eq:source_scalar_11} by ${}_{0}\bar{\mathcal{Y}}_{\ell m}(\theta,\phi)$ and integrate it over the 2-sphere, utilizing the orthogonality properties of spin-weighted spheroidal harmonics in Eq.~\eqref{eq:ortho_spheroidal}. 
    
    \item Similarly, as shown in Sec.~\ref{sec:modified_Teuk_operator}, the homogeneous part of the modified Teukolsky equation for $\Psi_{0}^{(1,1)}$ and $\rho^{-4}\Psi_{4}^{(1,1)}$ (i.e., $H_{0}^{(0,0)}$ and $\mathcal{H}_{4}^{(0,0)}$) are separable in $r$ and $\theta$ if one decomposes $\Psi_0^{(1,1)}$ and $\rho^{-4}\Psi_4^{(1,1)}$ into ${}_2\mathcal{Y}_{\ell m}(\theta,\phi)$ and ${}_{-2}\mathcal{Y}_{\ell m}(\theta,\phi)$, respectively. Thus, to extract the radial part of Eq.~\eqref{eq:master_eqn_non_typeD_Psi0} and its GHP transformation, we multiply $\mathcal{S}^{(1,1)}$ and $\mathcal{T}^{(1,1)}$ by ${}_{2}\bar{\mathcal{Y}}_{\ell m}(\theta,\phi)$ and ${}_{-2}\bar{\mathcal{Y}}_{\ell m}(\theta,\phi)$, respectively, and integrate them over the 2-sphere.

    \item Since we use the slow-rotation approximation in this work, when computing the integrals involving ${}_{s}\mathcal{Y}_{\ell m}(\theta,\phi)$, one can further expand ${}_{s}\mathcal{Y}_{\ell m}(\theta,\phi)$ in terms of ${}_{s}Y_{\ell m}(\theta,\phi)$ using Eq.~\eqref{eq:StoY}. Thus, there are only spin-weighted spherical harmonics in these integrals. 
    
    \item After the angular integration, the angular functions ${}_{s}f_{\ell m}(\theta)e^{im\phi}$ in Step 3 become coefficients of the form
    \begin{subequations} \label{eq:angular_coeffs}
    \begin{align} 
        & \Lambda^{\ell_1\ell_2 m}_{s_1 s_2}
        \equiv\int_{S^2}dS\;{}_{s_1}Y_{\ell_1 m}
        \;{}_{s_2}\bar{Y}_{\ell_2 m}\,,\\
        & \Lambda^{\ell_1\ell_2 m}_{s_1 s_2 c}
        \equiv\int_{S^2}dS\;\cos{\theta}
        \,{}_{s_1}Y_{\ell_1 m}
        \;{}_{s_2}\bar{Y}_{\ell_2 m}\,,\\
        & \Lambda^{\ell_1\ell_2 m}_{s_1 s_2 s}
        \equiv\int_{S^2}dS\;\sin{\theta}
        \,{}_{s_1}Y_{\ell_1 m}
        \;{}_{s_2}\bar{Y}_{\ell_2 m}\,,
    \end{align}  
    \end{subequations}
    and the ${}_{s}\bar{f}_{\ell\,-m}(\theta)e^{im\phi}$ angular functions become coefficients of the form
    \begin{subequations} \label{eq:angular_coeffs_conj}
    \begin{align} 
        & \Lambda^{\dagger\ell_1\ell_2 -m}_{s_1 s_2}
        \equiv\int_{S^2}dS\;{}_{s_1}\bar{Y}_{\ell_1m}\;{}_{s_2}\bar{Y}_{\ell_2 -m}\,,\\
        & \Lambda^{\dagger\ell_1\ell_2 -m}_{s_1 s_2 c}
        \equiv\int_{S^2}dS\;\cos{\theta}
        \,{}_{s_1}\bar{Y}_{\ell_1m}
        \;{}_{s_2}\bar{Y}_{\ell_2 -m}\,,\\
        & \Lambda^{\dagger\ell_1\ell_2 -m}_{s_1 s_2 s}
        \equiv\int_{S^2}dS\;\sin{\theta}
        \,{}_{s_1}\bar{Y}_{\ell_1m}
        \;{}_{s_2}\bar{Y}_{\ell_2 -m}\,.
    \end{align}  
    \end{subequations}
    Since spin-weighted spherical harmonics are not orthogonal across different spins over the 2-sphere, one has to calculate these coefficients in general directly. Besides evaluating the integrals in Eqs.~\eqref{eq:angular_coeffs} and \eqref{eq:angular_coeffs_conj} for different $(s_1,\ell_1,m)$ and $(s_2,\ell_2,m)$ every time, there are also other approaches. One approach is to use the series-sum representation of ${}_{s}Y_{\ell m}(\theta,\phi)$ in Eq.~\eqref{eq:spherical_formlua}, as discussed in Appendix~\ref{appendix:angular_projection} with the results stored in a Mathematica notebook in \cite{Pratikmodteuk}. Now, the master equations of $\vartheta^{(1,1)}$ and $\Psi_{0,4}^{(1,1)}$ become completely radial, i.e.,
    \begin{widetext}
    \begin{subequations} \label{eq:master_eqn_scheme3}
    \begin{align}
        {}_{s}\tilde{H}_{\ell m}\left[{}_{s}\psi_{\ell m}(r)\right]
        -{}_{s}A_{\ell m}
        & =\sum_{k}{}_{s}\textgoth{f}_{\ell m}^{k}\,{}_{s}P_{\ell m}^{k}(r)
        +\bar{\eta}_{\ell m}\,{}_{s}\bar{\textgoth{f}}_{\ell -m}^{k}\,{}_{s}Q_{\ell\,-m}^{k}(r)\,, 
        \label{eq:master_eqn_scheme3_lm} \\
        {}_{s}\tilde{H}_{\ell\,-m}\left[{}_{s}\psi_{\ell\,-m}(r)\right]
        -{}_{s}A_{\ell -m}
        & =\sum_{k}{}_{s}\textgoth{f}_{\ell -m}^{k}\,P_{\ell\,-m}^{k}(r)
        +
        \frac{1}{\eta_{\ell m}}{}_{s}\bar{\textgoth{f}}_{\ell m}^{k}\,{}_{s}Q_{\ell\,m}^{k}(r)\,, \label{eq:master_eqn_scheme3_l-m}
    \end{align}
    \end{subequations}   
    \end{widetext}
    where ${}_{s}\tilde{H}_{\ell m}$ is the radial Teukolsky operator for a spin $s$ field in GR, and ${}_{s}A_{\ell m}$ is the separation constant in the Teukolsky equation for a spin $s$ field in GR, both of which can be found in \cite{Teukolsky:1973ha}. The coefficient ${}_{s}\textgoth{f}_{\ell m}^{k}$ comes from the integral of ${}_{s}S_{\ell m}(\theta)$ and ${}_{s}f_{\ell m}^{k}(\theta)$ over the $2$-sphere, and similarly for its complex conjugate ${}_{s}\bar{\textgoth{f}}_{\ell m}^{k}$. ${}_{s}\textgoth{f}_{\ell m}^{k}$ and ${}_{s}\bar{\textgoth{f}}_{\ell m}^{k}$ will be given by Eqs.~\eqref{eq:angular_coeffs} and \eqref{eq:angular_coeffs_conj}, respectively.
\end{enumerate}

%%%%%%%%%%%%%%%%%%%%%%%%%%%%%%%%%%%%%%%%%%%%%%%%%%%%%%%%%%%%
\subsection{Radial part of the equation of $\vartheta^{(1,1)}$}
\label{sec:scalar_radial_eqn}

In this subsection, we present the radial part of Eq.~\eqref{eq:EOM_scalar_11} in both the IRG and ORG found by following the procedures in Sec.~\ref{sec:elim-ylm}. In the IRG, we find the radial parts of terms proportional to $e^{-i\omega t}$ in Eqs.~\eqref{eq:R*R_coord_IRG} and \eqref{eq:box_coord_IRG}, respectively, to be
\begin{widetext}
\begin{align} 
    V^{R}_{\ell m}(r)
    =& \;\left(g^{\ell m}_1(r,\omega,M)\,{}_{2}\hat{R}_{\ell m}(r)
    +g_2^{\ell m}(r,\omega,M)\,{}_{2}\hat{R}'_{\ell m}(r)\right)
    +\chi\left(g^{\ell m}_3(r,\omega,M)\,{}_{2}\hat{R}_{\ell m}(r)
    +g_4^{\ell m}(r,\omega,M)\,{}_{2}\hat{R}_{\ell m}'(r)\right)
    \Lambda^{\ell\ell m}_{10s}\,, \label{eq:R*R_radial_IRG}\\
    V^{\square}_{\ell m}(r)
    =& \;\chi\left(h^{\ell m}_1(r,\omega,M)\,{}_{2}\hat{R}_{\ell m}(r)
    +h^{\ell m}_2(r,\omega,M)\,{}_{2}\hat{R}_{\ell m}'(r)\right)\Lambda^{\ell\ell m}_{10s}\,, \label{eq:box_radial_IRG} 
\end{align}
where the terms proportional to $b^{m}_{\ell,\ell\pm1}$ or $\cos{\theta}\,{}_{0}Y_{\ell m}(\theta,\phi)$ in Eq.~\eqref{eq:R*R_coord_IRG} are at $\mathcal{O}(\chi^2)$ after the angular integration. In the ORG, we find
\begin{align} 
    U^{R}_{\ell m}(r)
    =& \;\left(\textgoth{g}^{\ell m}_1(r,\omega,M)\,{}_{-2}\hat{R}_{\ell m}(r)
    +\textgoth{g}_2^{\ell m}(r,\omega,M)\,{}_{-2}\hat{R}'_{\ell m}(r)\right) 
    +\chi\left(\textgoth{g}^{\ell m}_3(r,\omega,M)\,{}_{-2}\hat{R}_{\ell m}(r)
    +\textgoth{g}_4^{\ell m}(r,\omega,M)\,{}_{-2}\hat{R}_{\ell m}'(r)\right)
    \Lambda^{\ell\ell m}_{-10s}\,, \label{eq:R*R_radial_ORG}\\
    U^{\square}_{\ell m}(r)
    =& \;\chi\left(\textgoth{h}^{\ell m}_1(r,\omega,M)\,{}_{-2}\hat{R}_{\ell m}(r)
    +\textgoth{h}^{\ell m}_2(r,\omega,M)\,{}_{-2}\hat{R}_{\ell m}'(r)\right)
    \Lambda^{\ell\ell m}_{-10s}\,,
    \label{eq:box_radial_ORG}
\end{align}
\end{widetext}
where ${}_{s}\hat{R}_{\ell m}(r)$ is the radial part of the Hertz potential given in Eq.~\eqref{eq:Hertz_decompose}, $\Lambda^{\ell\ell m}_{10s}$ and $\Lambda^{\ell\ell m}_{-10s}$ are given by Eq.~\eqref{eq:angular_coeffs}, and the prime denotes a derivative with respect to the radial coordinate $r$. The functions $\left\{g_i^{\ell m}(r,\omega,M),h^{\ell m}_j(r,\omega,M),\textgoth{g}^{\ell m}_i(r,\omega,M),\textgoth{h}^{\ell m}_j(r,\omega,M)\right\}$, where $i \in \left[1,4 \right]$ and $j \in \left[1,2 \right]$, are the same functions in Eqs.~\eqref{eq:R*R_coord_IRG}, \eqref{eq:box_coord_IRG}, \eqref{eq:R*R_coord_ORG} and \eqref{eq:box_coord_ORG} and presented in a separate Mathematica notebook~\cite{Pratikmodteuk}. 

Using Eq.~\eqref{eq:hertztoteukcoord}, we can replace the radial Hertz potential ${}_{s}\hat{R}_{\ell m}(r)$ and its derivative in Eqs.~\eqref{eq:R*R_radial_IRG}--\eqref{eq:box_radial_ORG} with ${}_{s}R_{\ell m}^{(0,1)}(r)$ and ${}_{s}{R'}_{\ell m}^{(0,1)}(r)$. Notice that the form of the equations remains similar with ${}_{s}\hat{R}_{\ell m}(r)$ now replaced by ${}_{s}R_{\ell m}^{(0,1)}(r)$ and the prefactors now new functions of $\{r,\omega,M\}$. For instance, in the first parenthesis of Eq.~\eqref{eq:R*R_radial_IRG}, one finds that the prefactor of ${}_{2}R_{\ell m}^{(0,1)}(r)$ is
\begin{equation}
    g^{\ell m}_1(r,\omega,M){}_{2}f^{\ell m}_1(r,\omega,M)
    +g^{\ell m}_2(r,\omega,M){}_{2}f^{\ell m}_3(r,\omega,M)\,.
\end{equation}
Each of the functions that would appear in $V^{R}_{\ell m}(r)$, $V^{\square}_{\ell m}(r)$, $U^{R}_{\ell m}(r)$, and $U^{\square}_{\ell m}(r)$ are separately presented in the supplementary Mathematica notebook due to their lengthy nature~\cite{Pratikmodteuk}. 

Combining Eq.~\eqref{eq:scalar_hom_radial} with Eqs.~\eqref{eq:R*R_radial_IRG} and \eqref{eq:box_radial_IRG} [or Eqs.~\eqref{eq:R*R_radial_ORG} and \eqref{eq:box_radial_ORG}], we now have a completely radial equation that describes the evolution of the scalar field perturbations,
\begin{widetext}
\begin{align}
    \textrm{IRG:}\quad &\left[r(r-r_s)\partial_{r}^2\right. 
    \left.+2(r-M)\partial_r+\frac{\omega^2r^3-4\chi m M^2 \omega}{r-r_s}
    -{}_{0}A_{\ell m}\right]\Theta_{\ell m}(r) \nonumber\\
    &=-\pi^{-\frac{1}{2}}M^2r^2\left(V^{R}_{\ell m}(r)
    +\bar{\eta}_{\ell m}V^{\dagger R}_{\ell\,-m}(r)\right)
    -r^2\left(V^{\square}_{\ell m}(r)
    +\bar{\eta}_{\ell m}V^{\dagger \square}_{\ell\,-m}(r)\right)\,, 
    \label{eq:scalar_radial_IRG} \\
    \textrm{ORG:}\quad &\left[r(r-r_s)\partial_{r}^2\right. 
    \left.+2(r-M)\partial_r+\frac{\omega^2r^3-4\chi m M^2 \omega}{r-r_s}
    -{}_{0}A_{\ell m}\right]\Theta_{\ell m}(r) \nonumber\\
    &=-\pi^{-\frac{1}{2}}M^2r^2\left(U^{R}_{\ell m}(r)
    +\bar{\eta}_{\ell m}U^{\dagger R}_{\ell\,-m}(r)\right)
    -r^2\left(U^{\square}_{\ell m}(r)
    +\bar{\eta}_{\ell m}U^{\dagger \square}_{\ell\,-m}(r)\right)\,.
    \label{eq:scalar_radial_ORG} 
\end{align}    
\end{widetext}
Recall that $r_s$ is the Schwarzschild radius, $M$ is the mass of the BH, $\chi$ is the dimensionless spin parameter such that $\chi = a/M$ with $a$ being the spin, ${}_{0}A_{\ell m}$ is the separation constant for a spin-0 field~\cite{Teukolsky:1973ha}, and $\left\{V^{R}_{\ell m}(r), V^{\square}_{\ell m}(r), U^{R}_{\ell m}(r), U^{\square}_{\ell m}(r)\right\}$ are radial functions given in Eqs.~\eqref{eq:R*R_radial_IRG}--\eqref{eq:box_radial_ORG}. The constant $\bar{\eta}_{\ell m}$ is the relative coefficient between the $(\ell,m)$ and $(\ell,-m)$ modes of $\Psi_{0,4}^{(1,1)}$ in Eq.~\eqref{eq:solution_ansatz}, of which only certain values can solve Eqs.~\eqref{eq:master_eqn_scheme2} and \eqref{eq:master_eqn_scheme3} consistently. To obtain this coefficient, one has to solve Eq.~\eqref{eq:psi0_mastereqn_radial} for the $(\ell,m)$ and $(\ell,-m)$ modes of $\Psi_0^{(1,1)}$ [or Eq.~\eqref{eq:psi4_mastereqn_radial} for $\Psi_4^{(1,1)}$ in the ORG] jointly. In \cite{paritybreaking}, it was shown that one can turn Eq.~\eqref{eq:psi0_mastereqn_radial} [or Eq.~\eqref{eq:psi4_mastereqn_radial}] into an eigenvalue problem, following \cite{Zimmerman:2014aha, Mark_Yang_Zimmerman_Chen_2015, Hussain:2022ins}, such that the solutions of $\bar{\eta}_{\ell m}$ correspond to the eigenvectors of the system, and the QNM frequencies $\omega_{\ell m}$ are eigenvalues.

In the above equation, $V^{\dagger R}_{\ell\,-m}$ refers to taking the complex conjugate of all the radial functions in $V^{R}_{\ell\,-m}$ but replacing $\{\Lambda^{\ell_1\ell_2m}_{s_1s_2}, \Lambda^{\ell_1\ell_2m}_{s_1s_2c}, \Lambda^{\ell_1\ell_2m}_{s_1s_2s}\}$ with $\{\Lambda^{\dagger\ell_1\ell_2m}_{s_1s_2},\Lambda^{\dagger\ell_1\ell_2m}_{s_1s_2c},\Lambda^{\dagger\ell_1\ell_2m}_{s_1s_2s}\}$, and similarly for $V^{\dagger\square}_{\ell\,-m}$, $U^{\dagger R}_{\ell\,-m}$, and $U^{\dagger\square}_{\ell\,-m}$.
Equations~\eqref{eq:scalar_radial_IRG} and \eqref{eq:scalar_radial_ORG} can now be solved for to obtain the scalar-led QNM frequencies. Notice that there is a coupling between the scalar field perturbations and the gravitational perturbations in GR, which appear in the form of the Hertz potential radial function ${}_{\pm 2}\hat{R}_{\ell m}(r)$ in the IRG or ORG, respectively.

%%%%%%%%%%%%%%%%%%%%%%%%%%%%%%%%%%%%%%%%%%%%%%%%%%%%%%%%%%%%
\subsection{Radial part of the equation of $\Psi_0^{(1,1)}$}
\label{sec:psi0_radial_equation}

In this subsection, we present the radial part of the modified Teukolsky equation for $\Psi_0^{(1,1)}$. Just like in the case of the $\vartheta$ field, the left-hand side of the modified Teukolsky equation Eq.~\eqref{eq:master_eqn_non_typeD_Psi0} is the same as the Teukolsky equation in GR and is separable under the decomposition in Eq.~\eqref{eq:psi0_separation}, with its radial part given by Eqs.~\eqref{eq:H0_00} and \eqref{eq:H0_00_TK}. To extract the radial part of the right-hand side, we follow the recipe provided in Sec.~\ref{sec:elim-ylm} to eliminate all angular dependence. 

First, integrating $H_0^{(1,0)}\Psi_0^{(0,1)}$, where $H_0^{(1,0)}$ is given by Eq.~\eqref{eq:H0_110}, with ${}_{2}\bar{\mathcal{Y}}_{\ell m}(\theta,\phi)$, we find the radial part $\mathcal{S}^{\geo}_{\ell m}(r)$ of $\mathcal{S}_{\geo}^{(1,1)}$ to be
\begin{widetext}
\begin{equation} \label{eq:S_geo_radial}
\begin{aligned}
    \mathcal{S}^{\geo}_{\ell m}(r)
    =& \;\frac{i\chi mM^4}{448r^9(r-r_s)}
    \left(C_1(r)+4i\omega r^2C_2(r)\right){}_{2}R_{\ell m}^{(0,1)}(r) \\
    & \;-\frac{i\chi M^4}{16r^9}\left[C_3(r)-C_4(r)
    \left(\frac{i\omega r^2}{2}+\frac{r(r-r_s)}{2}\partial_r\right)\right]
    {}_{2}R_{\ell m}^{(0,1)}(r)\Lambda^{\ell\ell m}_{22c} \\ 
    & \;-\frac{i\chi M^4}{128r^9}C_5(r)\,{}_{2}R_{\ell m}^{(0,1)}(r)
    \left(\sqrt{(\ell+2)(\ell-1)}\Lambda^{\ell\ell m}_{12s}
    -\sqrt{(\ell+3)(\ell-2)}\Lambda^{\ell\ell m}_{32s}\right)\,,
\end{aligned}
\end{equation}
\end{widetext}
where we have used Eqs.~\eqref{eq:eth_ethc} and \eqref{eq:eth_ethc_on_Y} to replace $\partial_{\theta}\left({}_{2}Y_{\ell m}(\theta,\phi)\right)$ with ${}_{1}Y_{\ell m}(\theta,\phi)$ and ${}_{3}Y_{\ell m}(\theta,\phi)$. In Eq.~\eqref{eq:S_geo_radial}, recall once more that $r_s$ is the Schwarzschild radius, $M$ is the mass of the BH, $\chi$ is the dimensionless spin parameter, $\Lambda^{\ell\ell m}_{22c}$ and $\Lambda^{\ell\ell m}_{32s}$ are given by Eqs.~\eqref{eq:angular_coeffs}, and the functions $C_i$ with $i \in \left[1,5 \right]$ are presented in Eqs.~\eqref{eq:Ceq}.

Next, due to the structure of Eq.~\eqref{eq:master_eqn_scheme2}, multiplying Eq.~\eqref{eq:sA_coord} by ${}_{2}\bar{\mathcal{Y}}_{\ell m}(\theta,\phi)$ and Eq.~\eqref{eq:sAc_coord} by ${}_{2}\bar{\mathcal{Y}}_{\ell -m}(\theta,\phi)$ and integrating over the 2-sphere, we find
\begin{widetext}
\begin{subequations}
\begin{align} 
    \mathcal{S}^{A}_{\ell m}(r)
    =& \;\left(p_1^{\ell m}(r,\omega,M)\Theta_{\ell m}(r)
    +p_2^{\ell m}(r,\omega,M)\Theta_{\ell m}'(r)
    +p_3^{\ell m}(r,\omega,M)\Theta_{\ell m}''(r)\right) \nonumber\\
    & \;+\chi\left(p_4^{\ell m}(r,\omega,M)\Theta_{\ell m}(r)
    +p_5^{\ell m}(r,\omega,M)\Theta_{\ell m}'(r)
    +p_6^{\ell m}(r,\omega,M)\Theta_{\ell m}''(r)\right)
    \Lambda^{\ell\ell m}_{12s} \nonumber\\
    & \;+\chi\left(p_7^{\ell m}(r,\omega,M)\Theta_{\ell m}(r)
    +p_8^{\ell m}(r,\omega,M)\Theta_{\ell m}'(r)
    +p_9^{\ell m}(r,\omega,M)\Theta_{\ell m}''(r)\right)
    \Lambda^{\ell\ell m}_{22c}\,, \\
    \tilde{\mathcal{S}}^{A}_{\ell m}(r)
    =& \;-\left(\bar{p}_1^{\ell m}(r,\omega,M)\bar{\Theta}_{\ell m}(r)
    +\bar{p}_2^{\ell m}(r,\omega,M)\bar{\Theta}_{\ell m}'(r)
    +\bar{p}_3^{\ell m}(r,\omega,M)\bar{\Theta}_{\ell m}''(r)\right)(-1)^m \nonumber\\
    & \;-\chi\left(\bar{p}_4^{\ell m}(r,\omega,M)\bar{\Theta}_{\ell m}(r)
    +\bar{p}_5^{\ell m}(r,\omega,M)\bar{\Theta}_{\ell m}'(r)
    +\bar{p}_6^{\ell m}(r,\omega,M)\bar{\Theta}_{\ell m}''(r)\right)
    \Lambda^{\dagger\ell\ell -m}_{-12s} \nonumber\\
    & \;+\chi\left(\bar{p}_7^{\ell m}(r,\omega,M)\bar{\Theta}_{\ell m}(r)
    +\bar{p}_{8}^{\ell m}(r,\omega,M)\bar{\Theta}_{\ell m}'(r)
    +\bar{p}_{9}^{\ell m}(r,\omega,M)\bar{\Theta}_{\ell m}''(r)\right)
    \Lambda^{\dagger\ell\ell -m}_{-22c}\,.
    \label{eq:sAc_radial_old}
\end{align}
\end{subequations}
\end{widetext}
Here, $\mathcal{S}^{A}_{\ell m}(r)$ and 
%$\mathcal{S}^{\dagger A}_{\ell m}(r)$ 
$\tilde{\mathcal{S}}^{A}_{\ell m}(r)$ denote the radial part of the $(\ell,m)$ mode of $\mathcal{S}_{A}^{(1,1)}(r)$ and $\tilde{\mathcal{S}}_{A}^{(1,1)}(r)$, respectively. Notice that both $\mathcal{S}^{A}_{\ell m}(r)$ and $\tilde{\mathcal{S}}^{A}_{\ell\,-m}(r)$ contribute to the $(\ell, m)$ mode of the radial modified Teukolsky equation in Eq.~\eqref{eq:master_eqn_scheme2_lm}. The coefficient $(-1)^m$ in Eq.~\eqref{eq:sAc_radial_old} comes from that $\Lambda^{\dagger\ell\ell m}_{-s\,s}=(-1)^{m+s}$ since ${}_{-s}\bar{Y}_{\ell\,-m}(\theta,\phi)=(-1)^{m+s}{}_{s}Y_{\ell m}(\theta,\phi)$. The terms proportional to ${}_{0}b^m_{\ell,\ell\pm 1}$ in Eqs.~\eqref{eq:sA_coord} and \eqref{eq:sAc_coord} are at $\mathcal{O}(\chi^2)$ after the angular integration. We can further use Eq.~\eqref{eq:scalar_radial_IRG} to rewrite $\Theta_{\ell m}''(r)$ in term of $\Theta_{\ell m}(r)$, $\Theta_{\ell m}'(r)$, ${}_{2}\hat{R}_{\ell m}(r)$, and ${}_{2}\hat{R}_{\ell m}'(r)$ such that
\begin{widetext}
\begin{subequations} \label{eq:sA_total_radial}
\begin{align} 
    \mathcal{S}^{A}_{\ell m}(r)
    =& \;\left(k_1^{\ell m}(r,\omega,M)\Theta_{\ell m}(r)
    +k_2^{\ell m}(r,\omega,M)\Theta_{\ell m}'(r)
    +k_3^{\ell m}(r,\omega,M)\,{}_{2}\hat{R}_{\ell m}(r)
    +k_4^{\ell m}(r,\omega,M)\,{}_{2}\hat{R}_{\ell m}'(r)\right) \nonumber\\
    & \;+\chi\left(k_5^{\ell m}(r,\omega,M)\Theta_{\ell m}(r)
    +k_6^{\ell m}(r,\omega,M)\Theta_{\ell m}'(r)
    +k_7^{\ell m}(r,\omega,M)\,{}_{2}\hat{R}_{\ell m}(r)
    +k_8^{\ell m}(r,\omega,M)\,{}_{2}\hat{R}_{\ell m}'(r)\right)
    \Lambda^{\ell\ell m}_{12s} \nonumber\\
    & \;+\chi\left(k_9^{\ell m}(r,\omega,M)\Theta_{\ell m}(r)
    +k_{10}^{\ell m}(r,\omega,M)\Theta_{\ell m}'(r)
    +k_{11}^{\ell m}(r,\omega,M)\,{}_{2}\hat{R}_{\ell m}(r)
    +k_{12}^{\ell m}(r,\omega,M)\,{}_{2}\hat{R}_{\ell m}'(r)\right)
    \Lambda^{\ell\ell m}_{22c}\,, \label{eq:sA_radial} \\
    % \mathcal{S}^{\dagger A}_{\ell m}(r)
    \tilde{\mathcal{S}}^{ A}_{\ell m}(r)
    =& \;-\left(\bar{k}_1^{\ell m}(r,\omega,M)\bar{\Theta}_{\ell m}(r)
    +\bar{k}_2^{\ell m}(r,\omega,M)\bar{\Theta}_{\ell m}'(r)
    +\bar{k}_3^{\ell m}(r,\omega,M)\,{}_{2}\bar{\hat{R}}_{\ell m}(r)
    +\bar{k}_4^{\ell m}(r,\omega,M)\,{}_{2}\bar{\hat{R}}_{\ell m}'(r)\right)(-1)^m \nonumber\\
    & \;-\chi\left(\bar{k}_5^{\ell m}(r,\omega,M)\bar{\Theta}_{\ell m}(r)
    +\bar{k}_6^{\ell m}(r,\omega,M)\bar{\Theta}_{\ell m}'(r)
    +\bar{k}_7^{\ell m}(r,\omega,M)\,{}_{2}\bar{\hat{R}}_{\ell m}(r)
    +\bar{k}_8^{\ell m}(r,\omega,M)\,{}_{2}\bar{\hat{R}}_{\ell m}'(r)\right)
    \Lambda^{\dagger\ell\ell -m}_{-12s} \nonumber\\
    & \;+\chi\left(\bar{k}_9^{\ell m}(r,\omega,M)\bar{\Theta}_{\ell m}(r)
    +\bar{k}_{10}^{\ell m}(r,\omega,M)\bar{\Theta}_{\ell m}'(r)
    +\bar{k}_{11}^{\ell m}(r,\omega,M)\,{}_{2}\bar{\hat{R}}_{\ell m}(r)
    +\bar{k}_{12}^{\ell m}(r,\omega,M)\,{}_{2}\bar{\hat{R}}_{\ell m}'(r)\right)
    \Lambda^{\dagger\ell\ell -m}_{-22c}\,, \label{eq:sAc_radial}
\end{align}
\end{subequations}
where some of the radial functions $k_i^{\ell m}(r,\omega,M)$ are 
\begin{subequations}
\begin{align}
    k_1^{\ell m}(r,\omega,M) =& \;-\frac{1}{r^7 (r-2 M)^2}6 i \sqrt{\pi } \sqrt{\ell (\ell+1)} \sqrt{\ell^2+\ell-2} M^3 \left[2 M r (-{}_{2}A_{\ell m}+m \chi  (r \omega -2 i)-i r \omega -4) \right. \nonumber \\ & \;+\left. r^2 \left({}_{2}A_{\ell m}-2
    r^2 \omega ^2+2 i r \omega +2\right)+M^2 (8+2 m \chi  (2 r \omega +3 i))\right] \,, \\
    k_2^{\ell m}(r,\omega,M) =& \;\frac{12 \sqrt{\pi } \sqrt{\ell (\ell+1)} \sqrt{\ell^2+\ell-2} M^3[M (m \chi -3 i)+r (-r \omega +2 i)]}{r^6 (r-2 M)} \,,
\end{align}
\end{subequations}
\end{widetext}
while the remaining functions $k_i^{\ell m}(r,\omega,M)$ for $i \in \left[3,12 \right]$ are provided in \cite{Pratikmodteuk}. Recall that $\bar{k}_i^{\ell m}(r,\omega,M)$ are the complex conjugates of $k_i^{\ell m}(r,\omega,M)$.

Similarly, projecting Eqs.~\eqref{eq:sB_coord} and \eqref{eq:sBc_coord} into the radial direction, we find
\begin{widetext}
\begin{subequations} \label{eq:sB_total_radial}
\begin{align} 
    \mathcal{S}^{B}_{\ell m}(r) 
    =& \;\chi\left[\left(q_1^{\ell m}(r,\omega,M)\,{}_{2}\hat{R}_{\ell m}(r)
    +q_2^{\ell m}(r,\omega,M)\,{}_{2}\hat{R}_{\ell m}'(r)\right)
    \Lambda^{\ell\ell m}_{12s} \right. \nonumber\\
    & \;+\left.\left(q_3^{\ell m}(r,\omega,M)\,{}_{2}\hat{R}_{\ell m}(r)
    +q_4^{\ell m}(r,\omega,M)\,{}_{2}\hat{R}_{\ell m}'(r)\right)
    \Lambda^{\ell\ell m}_{22c} \right.\nonumber\\
    & \;+\left.\left(q_5^{\ell m}(r,\omega,M)\,{}_{2}\hat{R}_{\ell m}(r)
    +q_6^{\ell m}(r,\omega,M)\,{}_{2}\hat{R}_{\ell m}'(r)\right)
    \Lambda^{\ell\ell m}_{32s}\right]\,, \label{eq:sB_radial} \\
    \tilde{\mathcal{S}}^{ B}_{\ell m}(r)
    =& \;\chi\left[\left(\tilde{q}_1^{\ell m}(r,\omega,M)\bar{\hat{R}}_{\ell m}(r)
    +\tilde{q}_2^{\ell m}(r,\omega,M)\,{}_{2}\bar{\hat{R}}_{\ell m}'(r)\right)
    \Lambda^{\dagger\ell\ell -m}_{-12s}\right. \nonumber\\
    & \;\left.+\tilde{q}_3^{\ell m}(r,\omega,M)\,{}_{2}\bar{\hat{R}}_{\ell m}(r)
    \Lambda^{\dagger\ell\ell -m}_{-22c}\right]\,. \label{eq:sBc_radial}
\end{align}    
\end{subequations}
\end{widetext}
The functions ${}_{s}\hat{R}_{\ell m}(r)$ and ${}_{s}\bar{\hat{R}}_{\ell m}(r)$ are the radial parts of the Hertz potential [i.e., Eqs.~\eqref{eq:Hertz_decompose}] and its complex conjugate, respectively. Prime denotes a derivative with respect to the radial coordinate $r$. The coefficients $\left\{\Lambda^{\ell\ell m}_{12s},\Lambda^{\ell\ell m}_{22c},\Lambda^{\ell\ell m}_{32s}\right\}$ and $\left\{\Lambda^{\dagger\ell\ell -m}_{-12s},\Lambda^{\dagger\ell\ell -m}_{-22c}\right\}$ are given by Eqs.~\eqref{eq:angular_coeffs} and \eqref{eq:angular_coeffs_conj}, respectively. Due to the complicated functional form of $q_i^{\ell m}(r,\omega,M)$ with $i\in\left[1,6\right]$, we have presented them in a separate Mathematica notebook~\cite{Pratikmodteuk}. The radial functions $\tilde{q}_i^{\ell m}(r,\omega,M)$ are given by
\begin{widetext}
\begin{subequations}
\begin{align}
    \tilde{q}_1^{\ell m}(r,\omega,M) =& \frac{15 i \ell (\ell+1) \sqrt{\ell^2+\ell-2} M^4 \left(18 M^2+5 M r+r^2\right) (6 M+r (-3+i r \omega ))}{4 r^{12} (r-2 M)} \,, \\
    \tilde{q}_2^{\ell m}(r,\omega,M) =& \frac{15 i \ell (\ell+1) \sqrt{\ell^2+\ell-2} M^4 \left(18 M^2+5 M r+r^2\right)}{4 r^{11}} \,, \\
    \tilde{q}_3^{\ell m}(r,\omega,M) =& -\frac{15 i (\ell-1) \ell (\ell+1) (\ell+2) M^4 \left(54 M^2+10 M r+r^2\right)}{16 r^{12}} \,.
\end{align} 
\end{subequations}

Combining Eqs.~\eqref{eq:master_eqn_non_typeD_Psi0}, \eqref{eq:H0_00}, \eqref{eq:H0_00_TK}, \eqref{eq:S_geo_radial}, \eqref{eq:sA_total_radial}, and \eqref{eq:sB_total_radial}, the modified master equation for the radial part of the $\Psi_0^{(1,1)}$ Weyl scalar is
\begin{equation} \label{eq:psi0_mastereqn_radial}
\begin{aligned}
    & \left[r(r-r_s)\partial^2_{r}+6(r-M)\partial_r+\frac{C(r)}{r-r_s}+\frac{4m\chi M(i(r-M)-M\omega r)}{r(r-r_s)}
    -{}_{2}A_{\ell m}\right]{}_{2}R_{\ell m}^{(1,1)}(r) \\
    & =-2r^2\left[\mathcal{S}^{\geo}_{\ell m}(r)+\left(\mathcal{S}_{\ell m}^{A}(r)+\bar{\eta}_{\ell m}
    \tilde{\mathcal{S}}^{ A}_{\ell\,-m}(r)\right)+
    \left(\mathcal{S}_{\ell m}^{B}(r)
    +\bar{\eta}_{\ell m}\tilde{\mathcal{S}}^{ B}_{\ell\,-m}(r)\right)\right]\,,
\end{aligned}
\end{equation}  
\end{widetext}
where $C(r)$ is given by Eq.~\eqref{eq:Cfunc}, $\mathcal{S}^{\geo}_{\ell m}(r)$ is given in Eq.~\eqref{eq:S_geo_radial}, $\mathcal{S}_{\ell m}^{A}(r)$ and $\tilde{\mathcal{S}}^{A}_{\ell\,-m}(r)$ are given by Eqs.~\eqref{eq:sA_total_radial}, whereas $\mathcal{S}_{\ell m}^{B}(r)$ and $\tilde{\mathcal{S}}^{B}_{\ell\,-m}(r)$ are given by Eqs.~\eqref{eq:sB_total_radial}. Notice that one needs to solve the $(\ell,m)$ and $(\ell,-m)$ modes of Eq.~\eqref{eq:master_eqn_non_typeD_Psi0} jointly, from which one can then obtain the coefficient $\bar{\eta}_{\ell m}$ between these two modes defined in Eq.~\eqref{eq:solution_ansatz} and the QNM frequnecy $\omega_{\ell m}$\footnote{The coefficient $\bar{\eta}_{\ell-m}$ and the QNM frequency $\omega_{\ell -m}$ are redundant with $\bar{\eta}_{\ell m}$ and $\omega_{\ell m}$, respectively, since we solve the $(\ell,m)$ and $(\ell,-m)$ modes jointly. More specifically, from Eq.~\eqref{eq:master_eqn_scheme2} and a more detailed discussion in \cite{paritybreaking}, one can find that $\bar{\eta}_{\ell -m}=1/(\bar{\eta}_{\ell m})$ when $\bar{\eta}_{\ell m}\neq 0$ and $\omega_{\ell-m}=-\bar{\omega}_{\ell m}$.}, as we will work out in \cite{dcstyped2} following the procedures in \cite{paritybreaking}.

Using Eq.~\eqref{eq:hertztoteukcoord}, the radial part ${}_{2}\hat{R}_{\ell m}(r)$ of the Hertz potential $\Psi_{\Hertz}$ in Eqs.~\eqref{eq:sA_total_radial} and \eqref{eq:sB_total_radial} can be further expressed as functions of the radial Teukolsky function ${}_{2}{R}_{\ell m}^{(0,1)}(r)$ for the perturbed $\Psi_0$ in GR~\cite{Teukolsky:1973ha}, as discussed in Sec.~\ref{sec:scalar_radial_eqn}. All necessary functions have been provided in a supplementary Mathematica notebook due to their lengthy nature~\cite{Pratikmodteuk}. One can readily use existing wavefunction ansatz in the literature to evaluate the radial Teukolsky function~\cite{Leaver:1985ax, Leaver:1986gd}. 

Notice that the modified Teukolsky equation for the Weyl scalar perturbation $\Psi_0^{(1,1)}$ exhibits coupling to the scalar field perturbation at $\mathcal{O}(\zeta^1,\epsilon^1)$, but no such coupling is seen for the scalar field perturbation at the same order, unlike the case involving metric perturbations~\cite{Wagle:2021tam}. 

%%%%%%%%%%%%%%%%%%%%%%%%%%%%%%%%%%%%%%%%%%%%%%%%%%%%%%%%%%%%
\subsection{Radial part of the equations of $\Psi_4^{(1,1)}$}

In this subsection, we present the radial part of the modified Teukolsky equation for the Weyl scalar perturbation $\Psi_4^{(1,1)}$ for a slowly rotating BH in dCS gravity. Similar to the case studied in above Sec.~\ref{sec:psi0_radial_equation}, the left-hand side of the modified Teukolsky equation for $\Psi_4^{(1,1)}$ in Eq.~\eqref{eq:master_eqn_non_typeD_Psi4} holds the same form as the left-hand side of the Teukolsky equation for $\Psi_4^{(0,1)}$~\cite{Teukolsky:1973ha}. First, multiply $H_4^{(1,0)}\Psi_4^{(0,1)}=\mathcal{H}_4^{(1,0)}\psi_4^{(0,1)}$ by ${}_{-2}\bar{\mathcal{Y}}_{\ell m}(\theta,\phi)$, with $\mathcal{H}_4^{(1,0)}$ given in Eq.~\eqref{eq:H4_110}, and integrate over the 2-sphere
\begin{widetext}
\begin{align} \label{eq:T_geo_radial}
    \mathcal{T}_{\ell m}^{\geo}(r)
    =& \;\frac{-i \chi m M^4}{448r^{13}(r-r_s)}
    \left(D_1(r)-4i\omega r^2D_2(r)\right) {}_{-2}R_{\ell m}^{(0,1)}(r) \nonumber\\
    & \;+\frac{i\chi M^4}{16r^{13}}
    \left[D_3(r)-D_4(r)\left( \frac{i\omega r^2}{2} - \frac{r(r-r_s)}{2}\partial_r\right) \right] {}_{-2}R_{\ell m}^{(0,1)}(r) \Lambda^{\ell\ell m}_{-2-2c} \nonumber\\
    & \;+\frac{i \chi M^4}{128r^{13}}D_5(r)\,{}_{-2}R_{\ell m}^{(0,1)}(r)
    \left(\sqrt{(\ell+2)(\ell-1)}\Lambda^{\ell\ell m}_{-1-2s}
    -\sqrt{(\ell+3)(\ell-2)}\Lambda^{\ell\ell m}_{-3-2s}\right)\,,
\end{align}
where ${}_{-2}R_{\ell m}^{(0,1)}(r)$ is the radial function of $\rho^{-4}\Psi_4^{(0,1)}$ presented in Eq.~\eqref{eq:psi4_separation}, $\left\{\Lambda^{\ell\ell m}_{-1-2s},\Lambda^{\ell\ell m}_{-3-2s},\Lambda^{\ell\ell m}_{-2-2c}\right\}$ are given in Eqs.~\eqref{eq:angular_coeffs}, and $D_i(r)$ for $i \in \left[1,5\right]$ are presented in Eqs.~\eqref{eq:Deq}.

Next, we multiply Eq.~\eqref{eq:tA_coord_org} by ${}_{-2}\bar{\mathcal{Y}}_{\ell m}(\theta,\phi)$ and Eq.~\eqref{eq:tAc_coord_org} by ${}_{-2}\bar{\mathcal{Y}}_{\ell -m}(\theta,\phi)$ and integrate over the 2-sphere. We also make use of Eq.~\eqref{eq:scalar_radial_ORG} to decompose the $\Theta_{\ell m}''(r)$ dependence in terms of $\Theta_{\ell m}(r)$, $\Theta_{\ell m}'(r)$, ${}_{-2}\hat{R}_{\ell m}(r)$, and ${}_{-2}\hat{R}_{\ell m}'(r)$ such that
\begin{subequations} \label{eq:tA_total_radial}
\begin{align} 
    \mathcal{T}^{A}_{\ell m}
    =& \;\left(\textgoth{k}_1^{\ell m}(r,\omega,M)\Theta_{\ell m}(r)
    +\textgoth{k}_2^{\ell m}(r,\omega,M)\Theta_{\ell m}'(r)
    +\textgoth{k}_3^{\ell m}(r,\omega,M){}_{-2}\hat{R}_{\ell m}(r)
    +\textgoth{k}_4^{\ell m}(r,\omega,M){}_{-2}\hat{R}'_{\ell m}(r)\right)
    \nonumber\\
    & \;\left.+\chi\left(\textgoth{k}_5^{\ell m}(r,\omega,M)\Theta_{\ell m}(r)
    +\textgoth{k}_6^{\ell m}(r,\omega,M)\Theta_{\ell m}'(r)
    +\textgoth{k}_7^{\ell m}(r,\omega,M){}_{-2}\hat{R}_{\ell m}(r)
    +\textgoth{k}_8^{\ell m}(r,\omega,M){}_{-2}\hat{R}'_{\ell m}(r)\right)
    \Lambda^{\ell\ell m}_{-1-2s}\right. \nonumber\\
    & \;+\chi\left(\textgoth{k}_9^{\ell m}(r,\omega,M)\Theta_{\ell m}(r)
    +\textgoth{k}_{10}^{\ell m}(r,\omega,M)\Theta_{\ell m}'(r)
    +\textgoth{k}_{11}^{\ell m}(r,\omega,M){}_{-2}\hat{R}_{\ell m}(r)
    +\textgoth{k}_{12}^{\ell m}(r,\omega,M){}_{-2}\hat{R}'_{\ell m}(r)\right)
    \Lambda^{\ell\ell m}_{-2-2c}\,, 
    \label{eq:tA_radial_org} \\
    \tilde{\mathcal{T}}^{ A}_{\ell m}
    =& \;-\left(\bar{\textgoth{k}}_1^{\ell m}(r,\omega,M)\bar{\Theta}_{\ell m}(r)
    +\bar{\textgoth{k}}_2^{\ell m}(r,\omega,M)\bar{\Theta}_{\ell m}'(r)
    +\bar{\textgoth{k}}_3^{\ell m}(r,\omega,M){}_{-2}\bar{\hat{R}}_{\ell m}(r)
    +\bar{\textgoth{k}}_4^{\ell m}(r,\omega,M){}_{-2}\bar{\hat{R}}'_{\ell m}(r)\right)(-1)^m
    \nonumber\\
    & \;\left.-\chi\left(\bar{\textgoth{k}}_5^{\ell m}(r,\omega,M)\bar{\Theta}_{\ell m}(r)
    +\bar{\textgoth{k}}_6^{\ell m}(r,\omega,M)\bar{\Theta}_{\ell m}'(r)
    +\bar{\textgoth{k}}_7^{\ell m}(r,\omega,M){}_{-2}\bar{\hat{R}}_{\ell m}(r)
    +\bar{\textgoth{k}}_8^{\ell m}(r,\omega,M){}_{-2}\bar{\hat{R}}'_{\ell m}(r)\right)
    \Lambda^{\dagger \ell\ell -m}_{1-2s}\right. \nonumber\\
    & \;+\chi\left(\bar{\textgoth{k}}_9^{\ell m}(r,\omega,M)\bar{\Theta}_{\ell m}(r)
    +\bar{\textgoth{k}}_{10}^{\ell m}(r,\omega,M)\bar{\Theta}_{\ell m}'(r)
    +\bar{\textgoth{k}}_{11}^{\ell m}(r,\omega,M){}_{-2}\bar{\hat{R}}_{\ell m}(r)
    +\bar{\textgoth{k}}_{12}^{\ell m}(r,\omega,M){}_{-2}\bar{\hat{R}}'_{\ell m}(r)\right)
    \Lambda^{\dagger \ell\ell -m}_{2-2c}\,, 
    \label{eq:tAc_radial_org}
\end{align}
\end{subequations}
where we recall that $\Theta_{\ell m}(r)$ is the radial part of the scalar field perturbation, ${}_{-2}\hat{R}_{\ell m}(r)$ is the radial part of the Hertz potential in the ORG, prime denotes a derivative with respect to the radial coordinate $r$, and an overhead bar denotes complex conjugation. The constants $\Lambda$ and $\Lambda^\dagger$ are given by Eqs.~\eqref{eq:angular_coeffs} and Eqs.~\eqref{eq:angular_coeffs_conj}, respectively, with the relevant subscripts and superscripts. The functions $\textgoth{k}_i^{\ell m}(r,\omega,M)$ and $\bar{\textgoth{k}}_i^{\ell m}(r,\omega,M)$ are given in a Mathematica notebook as supplementary material~\cite{Pratikmodteuk}. Similarly, the source terms in Eq.~\eqref{eq:tb_coord} can be decomposed into a radial equation as
\begin{subequations} \label{eq:tB_total_radial}
\begin{align} 
    \mathcal{T}^{B}_{\ell m} 
    =& \;\chi \left[\left(\textgoth{q}_1^{\ell m}(r,\omega,M)\,{}_{-2}\hat{R}_{\ell m}(r)
    +\textgoth{q}_2^{\ell m}(r,\omega,M)\,{}_{-2}\hat{R}_{\ell m}'(r)\right)
    \Lambda^{\ell\ell m}_{-1-2s} \right. \nonumber\\
    & \;\left.+\left(\textgoth{q}_3^{\ell m}(r,\omega,M)\,{}_{-2}\hat{R}_{\ell m}(r)
    +\textgoth{q}_4^{\ell m}(r,\omega,M)\,{}_{-2}\hat{R}_{\ell m}'(r)\right)
    \Lambda^{\ell\ell m}_{-2-2c} \right. \nonumber\\
    & \;\left.+\left(\textgoth{q}_5^{\ell m}(r,\omega,M)\,{}_{-2}\hat{R}_{\ell m}(r)
    +\textgoth{q}_6^{\ell m}(r,\omega,M)\,{}_{-2}\hat{R}_{\ell m}'(r)\right)
    \Lambda^{\ell\ell m}_{-3-2s}\right]\,, 
    \label{eq:tB_radial_org} \\
    \tilde{\mathcal{T}}^{ B}_{\ell m}
    =& \;\chi \left[\left(\textgoth{q}_1^{\dagger\ell m}(r,\omega,M)\,{}_{-2}\bar{\hat{R}}_{\ell m}(r)
    +\textgoth{q}_2^{\dagger\ell m}(r,\omega,M)\,{}_{-2}\bar{\hat{R}}_{\ell m}'(r)\right)\right.
    \Lambda^{\dagger \ell\ell -m}_{1-2s} \nonumber\\
    & \;\left.+\textgoth{q}_3^{\dagger\ell m}(r,\omega,M)\,
    {}_{-2}\bar{\hat{R}}_{\ell m}(r)
    \Lambda^{\dagger \ell\ell -m}_{2-2c}\right]\,. 
    \label{eq:tBc_radial_org}
\end{align}
\end{subequations}

Using Eq.~\eqref{eq:hertztoteukcoord}, the radial function ${}_{-2}\hat{R}_{\ell m}(r)$ of the Hertz potential $\Psi_{\Hertz}$ in Eqs.~\eqref{eq:tA_total_radial} and ~\eqref{eq:tB_total_radial} can be expressed in terms of the radial Teukolsky function ${}_{-2}R_{\ell m}^{(0,1)}$ of $\rho^{-4}\Psi_4^{(0,1)}$ in GR, as described in Sec.~\ref{sec:scalar_radial_eqn}. All necessary functions have been provided in a supplementary Mathematica notebook~\cite{Pratikmodteuk}.  
Combining Eqs.~\eqref{eq:master_eqn_non_typeD_Psi4}, \eqref{eq:H4_00}, \eqref{eq:H4_00_TK}, and \eqref{eq:T_geo_radial}--\eqref{eq:tB_total_radial}, we find
\begin{equation} \label{eq:psi4_mastereqn_radial}
\begin{aligned}
    & \left[r(r-r_s)\partial^2_{r}-2(r-M)\partial_r+\frac{D(r)}{r-r_s}-\frac{4m\chi M(i(r-M)+M\omega r)}{r(r-r_s)}
    -{}_{-2}A_{\ell m}\right]{}_{-2}R_{\ell m}^{(1,1)}(r) \\
    & =-2r^6\left[\mathcal{T}^{\geo}_{\ell m}(r)+\left(\mathcal{T}_{\ell m}^{A}(r)+\bar{\eta}_{\ell m}
    \tilde{\mathcal{T}}^{ A}_{\ell\,-m}(r)\right)+
    \left(\mathcal{T}_{\ell m}^{B}(r)+\bar{\eta}_{\ell m}\tilde{\mathcal{T}}^{B}_{\ell\,-m}(r)\right)\right]\,.
\end{aligned}
\end{equation}  
\end{widetext}
As before, one has to solve the $(\ell,m)$ and $(\ell,-m)$ modes of Eq.~\eqref{eq:psi4_mastereqn_radial} jointly to obtain the coefficient $\bar{\eta}_{\ell m}$ in Eq.~\eqref{eq:solution_ansatz} and the QNM frequency $\omega_{\ell m}$.
This analysis shows that the modified Teukolsky equations for the Weyl scalar perturbations $\Psi_{0,4}^{(1,1)}$  and the scalar field perturbation $\vartheta^{(1,1)}$ can be separated into a radial and angular piece. The radial piece can then be integrated numerically to obtain the QNM frequencies.

%%%%%%%%%%%%%%%%%%%%%%%%%%%%%%%%%%%%%%%%%%%%%%%%%%%%%%%%%%%%
%%%%%%%%%%%%%%%%%%%%%%%%%%%%%%%%%%%%%%%%%%%%%%%%%%%%%%%%%%%%
\section{Discussion} 
\label{sec:discuss}

In this paper, we have employed the modified Teukolsky formalism in~\cite{Li:2022pcy} to investigate the perturbations of slowly rotating BHs in dCS gravity at leading order in spin, where the BH spacetime is non-Ricci-flat, but remains of Petrov type D. To incorporate the slow-rotation approximation, we first extended the two-parameter expansion in \cite{Li:2022pcy} to a three-parameter expansion. Following \cite{Chandrasekhar_1983, Sopuerta:2009iy}, we then re-derived the null geodesics on the equatorial plane, from which we found the NP tetrad for slowly rotating BHs in dCS gravity up to $\mathcal{O}(\chi)$. The resulting tetrad is the Kinnersly tetrad expanded to $\mathcal{O}(\chi)$, with an additional adjustment accounting for the dCS correction. This tetrad is the same as the one in \cite{Sopuerta:2009iy}. Since BHs in dCS gravity are non-Ricci-flat, this direct extension of the Kinnersly tetrad leads to some nonzero background Weyl scalars $\Psi_1$ and $\Psi_3$, so we performed additional tetrad rotations to remove them and computed all the background NP quantities in this rotated tetrad. 

The source terms of the modified Teukolsky equation for $\Psi_{0,4}$ arise from two distinct contributions. Some of them originate from the homogeneous component of certain Bianchi and Ricci identities, so they only rely on the corrections to the background geometry. For Petrov type D spacetimes, these stationary corrections only couple to the perturbations of $\Psi_{0,4}$, so we evaluated them using the NP quantities in the dCS background and the solutions to the Teukolsky equation in GR. The other source terms stem from the stress tensor associated with corrections to the Einstein-Hilbert action. In dCS gravity, these source terms couple the scalar field with the metric in GR. Thus, to completely evaluate them, we need to solve for the dynamical scalar field. In this case, we first evaluated the scalar field equation and used the same methodology to guide our calculations for $\Psi_{0,4}$.

Since the scalar field is driven by dynamical metric perturbations in GR, one needs to first reconstruct the metric associated with curvature perturbations in GR. In this work, we chose to follow the CCK procedures developed in~\cite{Cohen_Kegeles_1975, Chrzanowski:1975wv, Kegeles_Cohen_1979, Lousto_Whiting_2002, Ori_2003, Whiting_Price_2005, Yunes_Gonzalez_2006, Keidl_Friedman_Wiseman_2007, Keidl_Shah_Friedman_Kim_Price_2010}, where the perturbed metric is obtained from the Hertz potential, though other procedures in \cite{Chandrasekhar_1983, Loutrel_Ripley_Giorgi_Pretorius_2020} may also apply. Using the reconstructed metric in \cite{Cohen_Kegeles_1975, Chrzanowski:1975wv, Kegeles_Cohen_1979, Lousto_Whiting_2002, Ori_2003, Whiting_Price_2005, Yunes_Gonzalez_2006, Keidl_Friedman_Wiseman_2007, Keidl_Shah_Friedman_Kim_Price_2010}, we then computed all the perturbed NP quantities in GR following the approach in \cite{Campanelli_Lousto_1999, Loutrel_Ripley_Giorgi_Pretorius_2020}. Since we also chose the gauge that the perturbations of $\Psi_{1,3}$ vanish in both GR and dCS gravity \cite{Li:2022pcy}, we performed additional tetrad rotations to transform all the perturbed NP quantities into this gauge. In the end, projecting the scalar field equation onto the NP basis, we used the reconstructed NP quantities to express all the source terms as differential operators acting on the Hertz potential.

The Hertz potential can be obtained from the perturbations of $\Psi_{0,4}$ in GR, which are solutions to the Teukolsky equations in GR. Decomposing the Hertz potential into spin-weighted spheroidal harmonics, we presented the source terms of the scalar field equation in  Boyer-Lindquist coordinates explicitly. The radial function of the Hertz potential was then determined from the radial function of the perturbed $\Psi_{0,4}$ in GR following \cite{Ori_2003}. 

In the IRG, the above steps led to three coupled, partial differential equations for the $(\ell,m)$ and $(\ell,-m)$ modes of the Weyl scalar perturbation in dCS $\Psi_0^{(1,1)}$ and the $(\ell,m)$ mode of the scalar field perturbation $\vartheta^{(1,1)}$ [the $(\ell,m)$ and $(\ell,-m)$ modes of $\vartheta^{(1,1)}$ are redundant since $\vartheta^{(1,1)}$ is real], that we refer to as master equations. Similarly, in the ORG, we obtained three coupled, partial differential equations for the $(\ell,m)$ and $(\ell,-m)$ modes of the Weyl scalar perturbation in dCS $\Psi_4^{(1,1)}$ and the $(\ell,m)$ mode of $\vartheta^{(1,1)}$. More explicitly, the master equation of $\Psi_0^{(1,1)}$ (or $\Psi_4^{(1,1)}$) consists of the GR Teukolsky operator for a spin $2$ (or spin $-2$) field acting on $\Psi_0^{(1,1)}$ (or $\Psi_4^{(1,1)}$), as well as a source term that depends on $\vartheta^{(1,1)}$ and the Weyl scalar perturbation in GR $\Psi_0^{(0,1)}$ (or $\Psi_4^{(0,1)}$). Similarly, the master equation of $\vartheta^{(1,1)}$ consists of the GR Teukolsky operator for a scalar field acting on $\vartheta^{(1,1)}$ and a source term that depends on $\Psi_0^{(0,1)}$ (or $\Psi_4^{(0,1)}$).

To separate these master equations into radial and angular \textit{ordinary} differential equations, we exploited the orthogonality properties of spin-weighted spheroidal harmonics and performed a harmonic decomposition to eliminate all angular dependence of the source terms. The homogeneous part of the scalar field equation naturally separates, so we obtain a purely radial differential equation [i.e., Eq.~\eqref{eq:scalar_radial_IRG} in the IRG and Eq.~\eqref{eq:scalar_radial_ORG} in the ORG]. Similar procedures were then implemented for the modified Teukolsky equations of $\Psi_{0,4}$. The source terms of the modified Teukolsky equations were expressed in terms of the Hertz potential and the dynamical scalar field. We then projected the source terms into the radial direction by integrating them over spin-weighted spheroidal harmonics. The homogeneous part of these equations separates in the same way as the Teukolsky equations in GR, so we also obtained two radial differential equations for $\Psi_{0}$ [i.e., Eq.~\eqref{eq:psi0_mastereqn_radial}] and $\Psi_4$ [i.e., Eq.~\eqref{eq:psi4_mastereqn_radial}], respectively.

Through these procedures, we obtained three coupled, ordinary (radial) differential equations for $\left\{{}_{2}R_{\ell m}^{(1,1)}(r),\,{}_{2}R_{\ell -m}^{(1,1)}(r),\,\Theta_{\ell m}^{(1,1)}(r)\right\}$ (or $\left\{{}_{-2}R_{\ell m}^{(1,1)}(r),\,{}_{-2}R_{\ell -m}^{(1,1)}(r),\,\Theta_{\ell m}^{(1,1)}(r)\right\}$ in the ORG), where the first two are radials functions of $\Psi_0^{(1,1)}$ (or $\rho^{-4}\Psi_4^{(1,1)}$), and the last one is the radial function of $\vartheta^{(1,1)}$. All of these equations have the same structure. The left-hand side is the radial Teukolsky operator for particles of spin $2$ ($\Psi_0^{(1,1)}$), spin $-2$ ($\Psi_4^{(1,1)}$), or spin $0$ ($\vartheta^{(1,1)}$). For the radial master equation of $\Psi_0^{(1,1)}$ (or $\Psi_4^{(1,1)}$), the right-hand side contains source terms that depend on $\left\{\Theta_{\ell m}^{(1,1)}(r),\,{}_{2}R_{\ell m}^{(0,1)}(r),\,{}_{2}R_{\ell -m}^{(0,1)}(r)\right\}$ (or $\left\{\Theta_{\ell m}^{(1,1)}(r),\,{}_{-2}R_{\ell m}^{(0,1)}(r),\,{}_{-2}R_{\ell -m}^{(0,1)}(r)\right\}$), where the last two are radial functions of $\Psi_0^{(0,1)}$ (or $\rho^{-4}\Psi_4^{(0,1)}$). For the radial master equation of $\vartheta^{(1,1)}$, the right-hand side contains source terms that depend on $\left\{{}_{2}R_{\ell m}^{(0,1)}(r),\,{}_{2}R_{\ell -m}^{(0,1)}(r)\right\}$ (or $\left\{{}_{-2}R_{\ell m}^{(0,1)}(r),\,{}_{-2}R_{\ell -m}^{(0,1)}(r)\right\}$). The coupled system therefore forms a (Sturm-Liouville) eigenvalue problem that should be amenable to standard procedures to find the eigenvectors and eigenvalues, i.e., the QNM and scalar frequencies.

The primary objective of this study was to apply the modified Teukolsky formalism in \cite{Li:2022pcy} to investigate perturbations of BHs in some specific modified theories of gravity. To illustrate this, we considered the case of slowly rotating BHs to leading order in spin within the framework of dCS gravity. Although the slow rotation approximation may not provide highly accurate results for more realistic BHs (with spins $\chi \sim 0.6$), it is a simplified problem for testing the newly developed formalism. Incorporating additional degrees of freedom associated with dCS gravity, coupled with the intricacies introduced by the metric reconstruction procedures, renders this calculation complex. Yet, in this work, we successfully demonstrated that the modified Teukolsky equation in \cite{Li:2022pcy} does not only decouple Weyl scalars $\Psi_{0,4}$ from other NP quantities but also admits a separation into radial and angular parts, a key advantage of the Teukolsky equation in GR, especially for rapidly rotating BHs. Although this paper focused on the first order in the slow rotation expansion, the separation of the modified Teukolsky equation should hold for any spin since the orthogonality properties of spin-weighted spheroidal harmonics we have used to separate the equation apply for a general spin. Thus, this calculation is an ideal initial step toward determining the QNM spectra for BHs with general spin in modified gravity.

This work creates a new path to directly calculate the corrections to the QNM frequencies for slowly rotating perturbed BHs in dCS gravity. Having obtained the master equations for the perturbed Weyl scalars $\Psi_{0,4}$ and the perturbed scalar field, we can now integrate these equations using numerical integration schemes, such as the eigenvalue perturbation method in \cite{Zimmerman:2014aha, Mark_Yang_Zimmerman_Chen_2015, Hussain:2022ins} to find the QNM spectra. Moreover, the QNM spectra obtained using the modified Teukolsky formalism can then be compared to the results from the metric perturbation approach~\cite{Cardoso:2009pk, Molina:2010fb, Pani_Cardoso_Gualtieri_2011, Wagle:2021tam, Srivastava:2021imr} and numerical relativity~\cite{Okounkova_Stein_Scheel_Hemberger_2017, Okounkova_Scheel_Teukolsky_2019, Okounkova:2019dfo, Okounkova_Stein_Moxon_Scheel_Teukolsky_2020}. Notice that, in higher-derivative gravity, Refs.~\cite{Cano:2023jbk, Cano:2023tmv} have followed our formalism to compute the QNMs in the slow-rotation expansion of BHs in that theory, and they obtained results valid for $\chi\lesssim 0.7$. Nonetheless, the dCS case we have focused on is more complicated due to the coupling to the scalar field equation. As discussed above, we have also presented in detail the angular dependence of the master equations of the perturbed $\Psi_{0,4}$ and $\vartheta$ and showed explicitly that they are separable, while Refs.~\cite{Cano:2023jbk, Cano:2023tmv} only briefly discussed using the orthogonality of spin-weighted spheroidal harmonics to extract the radial equations. 

An additional aspect worth exploring is the phenomenon of isospectrality breaking in the QNM spectra. In GR, odd and even parity modes oscillate and decay at the same rate. However, certain modified theories of gravity have been shown to exhibit a breaking of isospectrality, e.g., dCS gravity \cite{Cardoso:2009pk, Molina:2010fb, Pani_Cardoso_Gualtieri_2011, Wagle:2021tam, Srivastava:2021imr}, EdGB gravity \cite{Blazquez-Salcedo:2016enn, Blazquez-Salcedo_Khoo_Kunz_2017, Pierini:2021jxd}, and higher-derivative gravity \cite{Cano_Fransen_Hertog_Maenaut_2021}. The investigation of isospectrality breaking has, so far, primarily focused on metric perturbations, as the Zerilli-Moncrief and Regge-Wheeler equations naturally separate metric perturbations into even- and odd-parity sectors~\cite{Zerilli:1971wd, Regge:PhysRev.108.1063}. However, for BHs with arbitrary spin, there are no known extensions of the Zerilli-Moncrief and the Regge-Wheeler equations, so we need to use the modified Teukolsky equation to study isospectrality breaking. In another study \cite{paritybreaking} involving all the authors, the definite-parity modes of curvature perturbations in modified gravity were found, and the features in these bGR theories that result in isospectrality breaking were revealed and demonstrated in several simple cases. Nonetheless, a direct mapping from the Zerilli-Moncrief and Regge-Wheeler functions to the modified Teukolsy equations of these definite-parity modes still remains unknown. The implementation of the modified Teukolsky equation in a concrete bGR theory has opened up possibilities for addressing these questions and more.

Building upon the insights gained from the present study, further investigations can be pursued involving more complex systems within various gravitational theories. As part of our collaborative effort, we are currently engaged in extending this calculation to derive the master equations and QNM spectra for BHs with arbitrary spin in dCS gravity, where the BH spacetime is Petrov type I. In addition, we are also actively involved in computing the master equations for rotating Petrov type I BHs within the framework of EdGB gravity. For the first time, we can explore the QNM spectra for BHs with general spin in a wide range of gravitational theories and spacetime geometries, which can then be compared with real observation data to scrutinize these possible deviations from GR.

%%%%%%%%%%%%%%%%%%%%%%%%%%%%%%%%%%%%%%%%%%%%%%%%%%%%%%%%%%%%
\section{Acknowledgements} 
\label{sec:acknowledgements}
\appendix
	
N. Y. and P. W. acknowledge support from the Simons Foundation through Award No. 896696 and National Science Foundation (NSF) Grant No. PHY-2207650. Y. C. and D. L. acknowledge support from the Brinson Foundation, the Simons Foundation (Award No. 568762), and NSF Grants No. PHY-2011961 and No. PHY-2011968. Some of our algebraic work used the package {\sc xAct}~\cite{xact} for Mathematica.
	
%%%%%%%%%%%%%%%%%%%%%%%%%%%%%%%%%%%%%%%%%%%%%%%%%%%%%%%%%%%%
%%%%%%%%%%%%%%%%%%%%%%%%%%%%%%%%%%%%%%%%%%%%%%%%%%%%%%%%%%%%
\section{Principal tetrad, spin coefficients, and some auxiliary functions} 
\label{appendix:background_NP_more}
	
In Sec.~\ref{sec:NP_quantities_background}, we performed tetrad rotations to set $\Psi_{1,3}^{(1,0)}=0$. As discussed in \cite{Chandrasekhar_1983}, these tetrad rotations preserving the orthogonality conditions of the NP tetrad can be divided into three types, 
\begin{subequations} \label{eq:tetrad_rotations}
\begin{align}
    \begin{split} \label{eq:rotate1}
        \text{\rom{1}}: 
        & \;l\rightarrow l\,,\; m\rightarrow m+al\,,\;
        \bar{m}\rightarrow \bar{m}+\bar{a}l\,,\; \\
        & \;n\rightarrow n+\bar{a}m+a\bar{m}+a\bar{a}l\,.
    \end{split} \\
    \begin{split} \label{eq:rotate2}
        \text{\rom{2}}: 
        & \;n\rightarrow n\,,\; m\rightarrow m+bn\,,\;
        \bar{m}\rightarrow \bar{m}+\bar{b}n\,,\; \\
        & \;l\rightarrow l+\bar{b}m+b\bar{m}+b\bar{b}n\,.
    \end{split} \\
    \begin{split} \label{eq:rotate3}
        \text{\rom{3}}: 
        & \;l\rightarrow A^{-1}l\,,\; n\rightarrow An\,,\;
        m\rightarrow e^{i\varphi}m\,,\; \\
        & \bar{m}\rightarrow e^{-i\varphi}\bar{m}\,,
    \end{split}
\end{align}		
\end{subequations}
where $a$ and $b$ are complex functions while $A$ and $\varphi$ are real functions. The transformations of Weyl scalars and spin coefficients under the tetrad rotations in Eq.~\eqref{eq:tetrad_rotations} can be found in \cite{Chandrasekhar_1983}. The tetrad rotations above are precise, but when these rotation parameters are small, for example at $\mathcal{O}(\zeta^1,\epsilon^0)$, the rotations of the tetrad at $\mathcal{O}(\zeta^1,\epsilon^0)$ simplify into
\begin{equation} \label{eq:tetrad_rotations_10}
\begin{aligned}
    & l^{(1,0)}\rightarrow 
    l^{(1,0)}+\bar{b}^{(1,0)}m+b^{(1,0)}\bar{m}-\delta A^{(1,0)}l\,, \\
    & n^{(1,0)}\rightarrow 
    n^{(1,0)}+\bar{a}^{(1,0)}m+a^{(1,0)}\bar{m}+\delta A^{(1,0)}n\,, \\
    & m^{(1,0)}\rightarrow m^{(1,0)}+a^{(1,0)}l+b^{(1,0)}n+i\varphi^{(1,0)}m\,,
\end{aligned}		
\end{equation}
where we defined $\delta A=A-1$ and combined the three types of tetrad rotations. Then, the Weyl scalars at $\mathcal{O}(\zeta^1,\epsilon^0)$ transform as
\begin{equation} \label{eq:Weyl_scalars_rotated_10}
\begin{aligned}
    & \Psi_{0,2,4}^{(1,0)}\rightarrow 0\,, \\
    & \Psi_1^{(1,0)}\rightarrow \Psi_1^{(1,0)}+3b^{(1,0)}\Psi_2\,, \\
    & \Psi_3^{(1,0)}\rightarrow \Psi_3^{(1,0)}+3\bar{a}^{(1,0)}\Psi_2\,,
\end{aligned}
\end{equation}
where we used that the background at $\mathcal{O}(\zeta^0,\epsilon^0)$ is Petrov type D, so $\Psi_{0,1,3,4}^{(0,0)}=0$. Since the spin coefficients at $\mathcal{O}(\zeta^1,\epsilon^0)$ after the rotations can be easily computed from the rotated tetrad, e.g., Eq.~\eqref{eq:principal_tetrad} in this work, we do not provide their general transformations under the tetrad rotations here. 
    
Using the tetrad rotations in Eq.~\eqref{eq:tetrad_rotations} and the results in Eq.~\eqref{eq:Weyl_scalars_rotated_10}, we set $\Psi_{1,3}^{(1,0)}=0$ and found the principal tetrad to be
\begin{subequations} \label{eq:principal_tetrad}
\begin{align}
    \begin{split} \label{eq:principal_l}
        l^{\mu}=
        & \;\Bigg(\frac{r}{r-r_s},\,1,\,0,\,\frac{\chi M}{r(r-r_s)}
        +\frac{\zeta\chi\tilde{G}(r)}{2r(r-r_s)} \\
        & \;-\frac{\zeta\chi A_1(r)}{16Mr^7} \Bigg)\,,
    \end{split} \\
    \begin{split} \label{eq:principal_n}
        n^{\mu}=
        & \;\tilde{N}(r)\Bigg(\frac{r}{r-r_s},\,-1,\,0,\,\frac{\chi M}{r(r-r_s)}
        +\frac{\zeta\chi\tilde{G}(r)}{2r(r-r_s)} \\
        & \;+\frac{\zeta\chi A_3(r)}{16Mr^7(r-r_s)}\Bigg)\,,
    \end{split} \\
    \begin{split} \label{eq:principal_m}
        m^{\mu}=
        & \;\frac{1}{\sqrt{2}r}\Bigg(i\chi M\left(1
        +\zeta\frac{A_3(r)-A_1(r)(r-r_s)}{32M^2r^5(r-r_s)}\right)\sin\theta\,, \\
        &\;i\zeta\chi\left(\frac{A_3(r)
            +A_1(r)(r-r_s)}{32Mr^6}\right)\sin\theta\,, \\
        & \;1-\frac{i\chi M\cos\theta}{r}\,,\; 
        i\left(1-\frac{i\chi M\cos\theta}{r}\right)\csc\theta\Bigg)\,,
    \end{split}
\end{align}
\end{subequations}
where
\begin{subequations}
\begin{align}
    & A_1(r)=306M^7+140M^6r+55M^5r^2\,, \label{eq:A1} \\
    & A_2(r)=18M^7+10M^6r+5M^5r^2\,, \label{eq:A2} \\
    & A_3(r)=612M^8-26M^7r-30M^6r^2-55M^5r^3\,. \label{eq:A3}
\end{align}	
\end{subequations}

In the principal tetrad Eq.~\eqref{eq:principal_tetrad}, the spin coefficients at background are
\begin{align} \label{eq:spin_coefs_background}
    \sigma
    =& \;\lambda=0\,, \nonumber\\
    \kappa
    =& \;-\left(\frac{1}{\tilde{N}(r)}\right)^2\nu
    =-\frac{5i\zeta\chi B_1(r)}{16\sqrt{2}r^7}\sin\theta\,, \nonumber\\
    \rho
    =& \;\frac{1}{\tilde{N}(r)}\mu 
    =-\frac{1}{r}-\frac{iM\chi}{r^2}\left(1
    -\frac{\zeta B_2(r)}{16r^5}\right)\cos\theta\,, \nonumber\\
    \tau
    =& \;-\pi
    =-\frac{iM\chi}{\sqrt{2}r^2}\left(1
    +\frac{\zeta B_3(r)}{32r^6}\right)\sin\theta\,, \nonumber\\
    \varepsilon
    =& \;\frac{iM\zeta\chi B_2(r)}{32r^7}\cos\theta\,, \nonumber\\
    \gamma
    =& \;\mu+\frac{r-M}{2r^2}\,, \nonumber\\
    \alpha
    =& \;\frac{iM\chi}{\sqrt{2}r^2}\left(1
    +\frac{\zeta B_4(r)}{16r^6}\right)\sin\theta-\bar{\beta} \nonumber\\
    =& \;-\frac{1}{2\sqrt{2}r}\bigg[\cot\theta-
    \frac{iM\chi}{r}\bigg(\left(3+
    \frac{\zeta B_4(r)}{16r^6}\right)\sin\theta \nonumber\\
    & \;-\csc\theta\bigg)\bigg]\,, 
\end{align}
with
\begin{subequations} 
\begin{align}
    B_1(r)=& \;306M^6+112M^5r+33M^4r^2\,, \label{eq:B1}\\
    B_2(r)=& \;306M^5+140M^4r+55M^3r^2\,, \label{eq:B2}\\
    \begin{split} \label{eq:B3}
        B_3(r)=& \;4680M^6-302M^5r-240M^4r^2 \\
        & \;-275M^3r^3\,, 
    \end{split} \\
    B_4(r)=& \;810M^6+54M^5r-5M^4r^2-55M^3r^3\,. 
    \label{eq:B4}
\end{align}	
\end{subequations}

Using the above NP quantities at $\mathcal{O}(\zeta^1,\epsilon^0)$, we computed the correction to the Teukolsky operators in Sec.~\ref{sec:modified_Teuk_operator}. In Eq.~\eqref{eq:H0_110}, these radial functions are defined to be
\begin{subequations} \label{eq:Ceq}
\begin{align}
    \begin{split} \label{eq:C1}
        C_1(r)=& \;57960M^4-39316M^3r-694M^2r^2 \\
        & \;-1050Mr^3+2345r^4 \,,
    \end{split} \\
    C_2(r)=&\;189M^3+120M^2r+70Mr^2\,, \label{eq:C2}\\
    \begin{split} \label{eq:C3}
        C_3(r)=& \;1602M^3-1056M^2r-515Mr^2 \\
        & \;-255r^3\,,
    \end{split} \\
    C_4(r)=& \;954M^2+440Mr+175r^2\,, \label{eq:C4}\\
    C_5(r)=& \;4680M^3-518M^2r-360Mr^2-335r^3\,. \label{eq:C5}
\end{align}
\end{subequations}
The radial functions in Eq.~\eqref{eq:H4_110} are given by
% \begin{subequations}
% \begin{align}
%     \begin{split} \label{eq:D1}
%         D_1(r)=& \;238140 M^4 - 190626 M^3 r - 1110 M^2 r^2 \\
%         & \; - 4970  M r^3 +  13125 r^4 \,,
%     \end{split} \\
%     D_2(r)=&\;756 M^3 + 480 M^2 r + 280 M r^2\,, \label{eq:D2}\\
%     \begin{split} \label{eq:D3}
%         D_3(r)=& \; -33840 M^4 + 19972  M^3 r + 634  M^2 r^2 \\
%         & \; + 1310  M r^3 -  1195  r^4
%     \end{split} \\
%     D_4(r)=& \;612 M^3 + 1270 M^2 r + 570 M r^2 + 185 r^3\,, \label{eq:D4}\\
%     D_5(r)=& \;4680 M^3 - 374 M^2 r - 280 M r^2 - 295 r^3\,. \label{eq:D5}
% \end{align}
% \end{subequations}
\begin{subequations} \label{eq:Deq}
\begin{align}
    \begin{split} \label{eq:Dall}
        D_i(r)=& ~C_i(r) \quad i \neq 3 \,,
    \end{split} \\
    D_3(r)=&\;306 M^3 + 28 M^2 r - 15 M r^2 - 95 r^3\,.\label{eq:D3}
\end{align}
\end{subequations}

\onecolumngrid
%%%%%%%%%%%%%%%%%%%%%%%%%%%%%%%%%%%%%%%%%%%%%%%%%%%%%%%%%%%%
%%%%%%%%%%%%%%%%%%%%%%%%%%%%%%%%%%%%%%%%%%%%%%%%%%%%%%%%%%%%
\section{Reconstructed NP quantities} 
\label{appendix:metric_reconstruction_more}
	
In this appendix, we provide the explicit expressions of these reconstructed NP quantities in Sec.~\ref{sec:metric_reconstruction}. As discussed in Sec.~\ref{sec:metric_reconstruction}, one can write these structure constants $C_{ab}{}^{c}$ in terms of spin coefficients using Eq.~\eqref{eq:commutation} and the definition of spin coefficients in terms of Ricci rotation coefficients
\begin{align} \label{eq:spin_coefs}
    & \kappa=\gamma_{131}\,,\quad
    \pi=-\gamma_{241}\,,\quad
    \varepsilon=\frac{1}{2}(\gamma_{121}-\gamma_{341})\,,\quad
    \rho=\gamma_{134}\,,\quad
    \lambda=-\gamma_{244}\,,\quad
    \alpha=\frac{1}{2}(\gamma_{124}-\gamma_{344})\,,
    \nonumber \\
    & \sigma=\gamma_{133}\,,\quad
    \mu=-\gamma_{243}\,,\quad
    \beta=\frac{1}{2}(\gamma_{123}-\gamma_{343})\,,\quad
    \tau=\gamma_{132}\,,\quad
    \nu=-\gamma_{242}\,,\quad
    \gamma=\frac{1}{2}(\gamma_{122}-\gamma_{342})\,.
\end{align}
It was found in \cite{Chandrasekhar_1983} that
\begin{equation} \label{eq:C_spin_coefs}
    \begin{aligned}
        & C_{12}{}^{1}=-(\gamma+\bar{\gamma})\,,\;
        C_{12}{}^{2}=-(\varepsilon+\bar{\varepsilon})\,,\;
        C_{12}{}^{3}=\bar{\tau}+\pi\,,\;
        C_{13}{}^{1}=-\bar{\alpha}-\beta+\bar{\pi}\,,\;
        C_{13}{}^{2}=-\kappa\,,\;
        C_{13}{}^{3}=\bar{\rho}+\varepsilon-\bar{\varepsilon}\,,\;
        C_{13}{}^{4}=\sigma\,, \\
        & C_{23}{}^{1}=\bar{\nu}\,,\;
        C_{23}{}^{2}=-\tau+\bar{\alpha}+\beta\,,\;
        C_{23}{}^{3}=-\mu+\gamma-\bar{\gamma}\,,\;
        C_{23}{}^{4}=-\bar{\lambda}\,,\;
        C_{34}{}^{1}=\mu-\bar{\mu}\,,\;
        C_{34}{}^{2}=\rho-\bar{\rho}\,,\;
        C_{34}{}^{3}=\bar{\beta}-\alpha\,,
    \end{aligned}
\end{equation}
and the other components can be found by using complex conjugation and that $C_{ab}{}^{c}$ is antisymmetric in its first two indices. Solving the above equation, one can also express spin coefficients in terms of $C_{ab}{}^{c}$,
\begin{equation} \label{eq:spin_coefs_C}
    \begin{aligned}
        & \kappa=C^{31}{}_{2}\,,\;
        \sigma=-C^{31}_{4}\,,\;
        \lambda=C^{42}{}_{3}\,,\;
        \nu=-C^{42}{}_{1}\,,\; \\
        & \rho
        =-\frac{1}{2}\left(C_{31}{}^{3}+C_{41}{}^{4}+C_{43}{}^{2}\right)\,,\;
        \mu
        =\frac{1}{2}\left(C_{32}{}^{3}+C_{42}{}^{4}-C_{43}{}^{1}\right)\,, \\
        & \pi
        =-\frac{1}{2}\left(C_{41}{}^{1}+C^{42}_{2}+C_{21}{}^{3}\right)\,,\;
        \tau
        =\frac{1}{2}\left(C_{31}{}^{1}+C_{32}{}^{2}-C_{21}{}^{4}\right)\,, \\
        & \varepsilon
        =\frac{1}{4}\left(C_{41}{}^{4}-C_{31}{}^{3}+2C_{21}{}^{2}
        +\rho-\bar{\rho}\right)\,,\;
        \gamma
        =\frac{1}{4}\left(C_{42}{}^{4}-C_{32}{}^{3}+2C_{21}{}^{1}
        +\mu-\bar{\mu}\right)\,, \\
        & \alpha
        =\frac{1}{4}\left(C_{41}{}^{1}-C_{42}{}^{2}+2C_{43}{}^{3}
        +\bar{\tau}-\pi\right)\,,\;
        \beta
        =\frac{1}{4}\left(C_{31}{}^{1}-C_{32}{}^{2}+2C_{43}{}^{4}
        +\tau-\bar{\pi}\right)\,.
    \end{aligned}
\end{equation}
Then following the procedures in Sec.~\ref{sec:metric_reconstruction}, one finds the spin coefficients at $\mathcal{O}(\zeta^0,\epsilon^1)$ to be
\begin{subequations}  \label{eq:perturbed_spin_coefs}
\begin{align}
    & \begin{aligned}
         \kappa^{(0,1)}
        =& \;\frac{1}{2}\delta_{[-2,-2,1,1]}h_{ll}^{(0,1)}
        -D_{[-2,0,0,-1]}h_{lm}^{(0,1)}\,,
    \end{aligned} \\
    & \begin{aligned}
        \sigma^{(0,1)}
        =& \;-\frac{1}{2}D_{[-2,2,1,-1]}h_{mm}^{(0,1)}
        +(\bar{\pi}+\tau)h_{lm}^{(0,1)}\,,
    \end{aligned} \\
    & \begin{aligned}
        \lambda^{(0,1)}
        =& \;(\pi+\bar{\tau})h_{n\bar{m}}^{(0,1)}
        +\frac{1}{2}\Delta_{[-1,1,2,-2]}h_{\bar{m}\bar{m}}^{(0,1)}\,,
    \end{aligned} \\
    & \begin{aligned}
        \nu^{(0,1)}
        =& \;-\frac{1}{2}\bar{\delta}_{[2,2,-1,-1]}h_{nn}^{(0,1)}
        +\boldsymbol{\Delta}_{[0,1,2,0]}h_{n\bar{m}}^{(0,1)}\,,
    \end{aligned} \\
    & \begin{aligned}
        \epsilon^{(0,1)}
        =& \;\frac{1}{4}\Big[\boldsymbol{\Delta}_{[-1,1,0,-2]}h_{ll}^{(0,1)}
        -2D_{[0,0,\frac{1}{2},-\frac{1}{2}]}h_{ln}^{(0,1)}
        -\bar{\delta}_{[-2,0,-3,-2]}h_{lm}^{(0,1)}
        +\delta_{[-2,0,1,2]}h_{l\bar{m}}^{(0,1)}
        -(\rho-\bar{\rho})h_{m\bar{m}}^{(0,1)}\Big]\,,
    \end{aligned} \\
    & \begin{aligned}
        \rho^{(0,1)}
        =& \;\frac{1}{2}\Big[-\mu h_{ll}^{(0,1)}
        -(\rho-\bar{\rho})h_{ln}^{(0,1)}
        -\bar{\delta}_{[-2,0,-1,0]}h_{lm}^{(0,1)}
        +\delta_{[-2,0,1,2]}h_{l\bar{m}}^{(0,1)}
        -D_{[0,0,1,-1]}h_{m\bar{m}}^{(0,1)}\Big]\,,
    \end{aligned} \\
    & \begin{aligned}
        \mu^{(0,1)}
        =& \;\frac{1}{2}\Big[-\rho h_{nn}^{(0,1)}
        -\bar{\delta}_{[0,2,-2,-1]}h_{nm}^{(0,1)}
        +\delta_{[0,2,0,1]}h_{n\bar{m}}^{(0,1)}
        +(\mu+\bar{\mu})h_{ln}^{(0,1)}
        +\boldsymbol{\Delta}_{[-1,1,0,0]}h_{m\bar{m}}^{(0,1)}\Big]\,,
    \end{aligned} \\
    & \begin{aligned}
        \gamma^{(0,1)}
        =& \;\frac{1}{4}\Big[-D_{[0,2,1,-1]}h_{nn}^{(0,1)}
        -\bar{\delta}_{[0,2,-2,-1]}h_{nm}^{(0,1)}
        +\delta_{[0,2,2,3]}h_{n\bar{m}}^{(0,1)}
        -(\mu-\bar{\mu}-4\gamma)h_{ln}^{(0,1)}
        -(\mu-\bar{\mu})h_{m\bar{m}}^{(0,1)}\Big]\,,
    \end{aligned} \\
    & \begin{aligned}
        \alpha^{(0,1)}
        =& \;\frac{1}{4}\Big[-D_{[-2,0,-1,-2]}h_{n\bar{m}}^{(0,1)}
        +\delta_{[-2,0,1,1]}h_{\bar{m}\bar{m}}^{(0,1)}
        -\bar{\delta}_{[0,0,-1,-1]}h_{ln}^{(0,1)}
        +\boldsymbol{\Delta}_{[-2,1,4,-2]}h_{l\bar{m}}^{(0,1)}
        -\bar{\delta}_{[2,0,-1,-1]}h_{m\bar{m}}^{(0,1)}\Big]\,,
    \end{aligned} \\
    & \begin{aligned}
        \beta^{(0,1)}
        =& \;\frac{1}{4}\Big[-D_{[-4,2,2,-1]}h_{nm}^{(0,1)}
        -\bar{\delta}_{[0,2,-1,-1]}h_{mm}^{(0,1)}
        -\delta_{[0,0,-1,-1]}h_{ln}^{(0,1)}
        +\boldsymbol{\Delta}_{[1,2,2,0]}h_{lm}^{(0,1)}
        +\delta_{[0,-2,1,1]}h_{m\bar{m}}^{(0,1)}\Big]\,,
    \end{aligned} \\
    & \begin{aligned}
        \pi^{(0,1)}
        =& \;\frac{1}{2}\Big[D_{[2,0,-1,0]}h_{n\bar{m}}^{(0,1)}
        +\tau h_{\bar{m}\bar{m}}^{(0,1)}
        -\delta_{[0,0,-1,-1]}h_{ln}^{(0,1)}
        +\boldsymbol{\Delta}_{[0,1,0,-2]}h_{l\bar{m}}^{(0,1)}
        +\bar{\tau}h_{m\bar{m}}^{(0,1)}\Big]\,,
    \end{aligned} \\
    & \begin{aligned}
        \tau^{(0,1)}
        =& \;\frac{1}{2}\Big[-D_{[0,2,0,-1]}h_{nm}^{(0,1)}
        +\pi h_{mm}^{(0,1)}
        +\delta_{[0,0,1,1]}h_{ln}^{(0,1)}
        -\boldsymbol{\Delta}_{[1,0,-2,0]}h_{lm}^{(0,1)}
        +\bar{\pi}h_{m\bar{m}}^{(0,1)}\Big]\,.
    \end{aligned}
\end{align}
\end{subequations}
	
For Weyl scalars at $\mathcal{O}(\zeta^0,\epsilon^1)$, one can use Ricci identities to retrieve them from spin coefficients. The equations below work for both vacuum and non-vacuum spacetimes since we have linearly combined Ricci identities to remove NP Ricci scalars $\Phi_{ab}$ following \cite{Campanelli_Lousto_1999, Loutrel_Ripley_Giorgi_Pretorius_2020}.
\begin{subequations} \label{eq:Weyl_scalar_Ricci}
\begin{align} 
    & \Psi_0
    =D_{[-3,1,-1,-1]}\sigma-\delta_{[-1,-3,1,-1]}\kappa\,, \\
    & \Psi_1
    =D_{[0,1,0,-1]}\beta-\delta_{[-1,0,1,0]}\varepsilon
    -(\alpha+\pi)\sigma+(\gamma+\mu)\kappa\,, \\
    & \begin{aligned}
        \Psi_2
        =& \;\frac{1}{3}\Big[\bar{\delta}_{[-2,1,-1,-1]}\beta
        -\delta_{[-1,0,1,1]}\alpha
        +D_{[1,1,1,-1]}\gamma
        -\boldsymbol{\Delta}_{[-1,1,-1,-1]}\varepsilon
        +\bar{\delta}_{[-1,1,-1,-1]}\tau
        -\boldsymbol{\Delta}_{[-1,1,-1,-1]}\rho
        +2(\nu\kappa-\lambda\sigma)\Big]\,,
    \end{aligned} \\
    & \Psi_3
    =\bar{\delta}_{[0,1,0,-1]}\gamma
    -\boldsymbol{\Delta}_{[0,1,0,-1]}\alpha
    +(\varepsilon+\rho)\nu-(\beta+\tau)\lambda\,, 
    \label{eq:Weyl_scalar_Ricci_Psi3}\\
    & \Psi_4
    =\bar{\delta}_{[3,1,1,-1]}\nu
    -\boldsymbol{\Delta}_{[1,1,3,-1]}\lambda\,.
\end{align}
\end{subequations}
The equations above are precise, so one needs to linearize them when extracting the Weyl scalars at $\mathcal{O}(\zeta^0,\epsilon^1)$ using the perturbed tetrad and spin coefficients at $\mathcal{O}(\zeta^0,\epsilon^1)$. 
	
In Refs.~\cite{Kegeles_Cohen_1979, Keidl_Friedman_Wiseman_2007}, they also computed the perturbed Weyl scalars in the IRG and expressed them in terms of the Hertz potential,
\begin{subequations} \label{eq:Weyl_scalar_Hertz}
\begin{align}
    & \Psi_0^{(0,1)}
    =-\frac{1}{2}D_{[-3,1,0,-1]}D_{[-2,2,0,-1]}h_{mm}^{(0,1)}\,, \\
    & \begin{aligned}
        \Psi_1^{(0,1)}
        &=-\frac{1}{8}\Big[2D_{[-1,1,1,-1]}D_{[0,2,1,-1]}h_{nm}^{(0,1)}
        +D_{[-1,1,1,-1]}\delta_{[-2,2,-2,-1]}h_{mm}^{(0,1)}
        +\bar{\delta}_{[-3,1,-3,-1]}D_{[-2,2,0,-1]}h_{mm}^{(0,1)}\Big]\,,
    \end{aligned} \\
    & \begin{aligned}
        \Psi_2^{(0,1)}
        &=-\frac{1}{12}\Big[D_{[1,1,2,-1]}D_{[2,2,2,-1]}h_{nn,1}^{(0,1)}
        +2\left(D_{[1,1,2,-1]}\bar{\delta}_{[0,2,-1,-1]}\right.
        \left.+\bar{\delta}_{[-1,1,-2,-1]}D_{[0,2,1,-1]}\right)h_{nm}^{(0,1)} \\
        & \quad\,+\bar{\delta}_{[-1,1,-2,-1]}\bar{\delta}_{[-2,2,-2,-1]}h_{mm}^{(0,1)}\Big]\,,
    \end{aligned} \label{eq:Psi2_Hertz} \\
    & \begin{aligned}
        \Psi_3^{(0,1)}
        &=-\frac{1}{8}\Big[\left(D_{[3,1,3,-1]}
        \bar{\delta}_{[2,2,0,-1]}\right.
        \left.+\bar{\delta}_{[1,1,-1,-1]}D_{[2,2,,2,-1]}\right)h_{nn,1}^{(0,1)} +\bar{\delta}_{[1,1,-1,-1]}\bar{\delta}_{[0,2,-1,-1]}h_{nm}^{(0,1)}\Big]\,,
    \end{aligned} \\
    & \begin{aligned}
        \Psi_4^{(0,1)}
        &=-\frac{1}{2}\Big[\bar{\delta}_{[3,1,0,-1]}
        \bar{\delta}_{[2,2,0,-1]}h_{nn,1}^{(0,1)}
        +3\Psi_2\left(\tau\bar{\delta}_{[4,0,0,0]}
        -\rho\boldsymbol{\Delta}_{[0,0,4,0]}\right.
        \left.-\mu D_{[4,0,0,0]}+\pi\delta_{[0,4,0,0]}
        +2\Psi_2\right)\bar{\Psi}_{\Hertz}\Big]\,,
    \end{aligned}
\end{align}
\end{subequations}
where $h_{nn,1}^{(0,1)}$ is the piece of $h_{nn}^{(0,1)}$ proportional to $\bar{\Psi}_{\Hertz}$ in Eq.~\eqref{eq:metricpert_irg}, i.e., $h_{nn,1}^{^{(0,1)}}=\bar{\delta}_{[1,3,0,-1]}\bar{\delta}_{[0,4,0,3]}\bar{\Psi}_{\Hertz}$. As we discussed in Sec.~\ref{sec:metric_reconstruction}, our results using either the Ricci identities or direct computation from the linearized Riemann tensor agree for $\Psi_{0,1,2,4}^{(0,1)}$ but not for $\Psi_{3}^{(0,1)}$ due to different choices of the perturbed tetrad. Thus, we can use Eq.~\eqref{eq:Weyl_scalar_Hertz} for $\Psi_{0,1,2,4}^{(0,1)}$ and Eq.~\eqref{eq:Weyl_scalar_Ricci_Psi3} for $\Psi_{3}^{(0,1)}$.

For Schwarzschild, the equations above simplify into
\begin{subequations} \label{eq:Weyl_scalar_Hertz_Schw}
\begin{align}
    & \Psi_0^{(0,1)}=-\frac{1}{2}D^4\bar{\Psi}_{\Hertz}\,, \\
    & \Psi_1^{(0,1)}=-\frac{1}{2}D^3(\bar{\delta}-4\alpha)\bar{\Psi}_{\Hertz}\,, \\
    & \Psi_2^{(0,1)}=-\frac{1}{2}D^2
    (\bar{\delta}-2\alpha)(\bar{\delta}-4\alpha)\bar{\Psi}_{\Hertz}\,, \\
    & \Psi_3^{(0,1)}=-\frac{1}{2}D\bar{\delta}
    (\bar{\delta}-2\alpha)(\bar{\delta}-4\alpha)\bar{\Psi}_{\Hertz}
     +\frac{3}{2}\Psi_{2}h_{n\bar{m}}\,, \\
    & \begin{aligned}
        \Psi_4^{(0,1)}
        &=-\frac{1}{2}(\bar{\delta}+2\alpha) \bar{\delta}
        (\bar{\delta}-2\alpha)(\bar{\delta}-4\alpha)\bar{\Psi}_{\Hertz}
        +\frac{3}{2}\Psi_2\left[\mu D+\rho(\boldsymbol{\Delta}+4\gamma)
        -2 \Psi_2\right]\Psi_{\Hertz}\,,
    \end{aligned}
\end{align}
\end{subequations}
where we have added the correction term $\frac{3}{2}\Psi_{2}h_{n\bar{m}}^{(0,1)}$ to $\Psi_{3}^{(0,1)}$.

To use a consistent gauge with the one in Sec.~\ref{sec:modified_Teuk_eqn}, we need to rotate the tetrad to remove $\Psi_{1,3}^{(0,1)}$. Under type I and II tetrad rotations at $\mathcal{O}(\zeta^0,\epsilon^1)$, the tetrad becomes
\begin{equation} \label{eq:tetrad_rotated_01}
\begin{aligned}
    & l^{(0,1)}\rightarrow 
    l^{(0,1)}+\bar{b}^{(0,1)}m+b^{(0,1)}\bar{m}-\delta A^{(0,1)}l\,,\quad
    n^{(0,1)}\rightarrow 
    n^{(0,1)}+\bar{a}^{(0,1)}m+a^{(0,1)}\bar{m}+\delta A^{(0,1)}n\,, \\
    & m^{(0,1)}\rightarrow m^{(0,1)}+a^{(0,1)}l+b^{(0,1)}n+i\varphi^{(0,1)}m\,,
\end{aligned}
\end{equation}
and the Weyl scalars transform as 
\begin{equation} \label{eq:Weyl_scalars_rotated_01}
\begin{aligned}
    \Psi_{0,2,4}^{(0,1)}\rightarrow 0\,,\quad
    \Psi_1^{(0,1)}\rightarrow \Psi_1^{(0,1)}+3b^{(0,1)}\Psi_2\,,\quad
    \Psi_3^{(0,1)}\rightarrow \Psi_3^{(0,1)}+3\bar{a}^{(0,1)}\Psi_2\,.
\end{aligned}
\end{equation}
The rotation coefficients $a^{(0,1)}$ and $b^{(0,1)}$ are given by Eq.~\eqref{eq:rotate_coefs_01}. For spin coefficients, due to the complication of the reconstructed tetrad, instead of computing the spin coefficients from the rotated tetrad directly, we chose to use the transformation of spin coefficients under tetrad rotations in \cite{Chandrasekhar_1983}. In this case, the spin coefficients transform as
\begin{equation} \label{eq:spin_coefs_rotated_01}
    \begin{aligned}
        & \kappa^{(0,1)}\rightarrow\kappa^{(0,1)}
        +b^{(0,1)}\rho-Db^{(0,1)}\,,\quad
        && \sigma^{(0,1)}\rightarrow\sigma^{(0,1)}
        +b^{(0,1)}(2\beta+\tau)-\delta b^{(0,1)}\,, \\
        & \lambda^{(0,1)}\rightarrow\lambda^{(0,1)}
        +\bar{a}^{(0,1)}(2\alpha+\pi)+\bar{\delta}\bar{a}\,,\quad
        && \nu^{(0,1)}\rightarrow\nu^{(0,1)}
        +\bar{a}^{(0,1)}(\mu+2\gamma)
        +\boldsymbol{\Delta}\bar{a}^{(0,1)}\,, \\
        & \varepsilon^{(0,1)}\rightarrow\varepsilon^{(0,1)}
        +\bar{b}^{(0,1)}\beta+b^{(0,1)}(\alpha+\pi)\,,\quad
        && \rho^{(0,1)}\rightarrow\rho^{(0,1)}
        +\bar{b}^{(0,1)}\tau+2b^{(0,1)}\alpha
        -\bar{\delta}b^{(0,1)}\,, \\
        & \mu^{(0,1)}\rightarrow\mu^{(0,1)}
        +a^{(0,1)}\pi+2\bar{a}^{(0,1)}\beta+\delta \bar{a}^{(0,1)}\,,\quad
        && \gamma^{(0,1)}\rightarrow\gamma^{(0,1)}
        +a^{(0,1)}\alpha+\bar{a}^{(0,1)}(\beta+\tau)\,, \\
        & \alpha^{(0,1)}\rightarrow\alpha^{(0,1)}
        +\bar{a}^{(0,1)}\rho+\bar{b}^{(0,1)}\gamma\,,\quad
        && \beta^{(0,1)}\rightarrow\beta^{(0,1)}
        +b^{(0,1)}(\mu+\gamma)\,, \\
        & \pi^{(0,1)}\rightarrow\pi^{(0,1)}
        +\bar{b}^{(0,1)}\mu+D\bar{a}^{(0,1)}\,,\quad
        && \tau^{(0,1)}\rightarrow\tau^{(0,1)}
        +a^{(0,1)}\rho+2b^{(0,1)}\gamma-\boldsymbol{\Delta}b^{(0,1)}\,.
    \end{aligned}
\end{equation}
Furthermore, besides the tetrad rotations, when one performs coordinate transformations $x^{\mu}\rightarrow x^{\mu}+\xi^{\mu}$ at $\mathcal{O}(\zeta^0,\epsilon^1)$, one finds
\begin{equation} \label{eq:Weyl_scalars_coord_01}
    \Psi^{(0,1)}\rightarrow
    \Psi^{(0,1)}+\xi^{\mu(0,1)}\partial_{\mu}\Psi^{(0,0)}
    +\xi^{\mu(0,1)}\partial_{\mu}\Psi^{(0,0)}\,. 
\end{equation}
for the scalar-type NP quantities such as Weyl scalars and spin coefficients \cite{Campanelli_Lousto_1999}.

%%%%%%%%%%%%%%%%%%%%%%%%%%%%%%%%%%%%%%%%%%%%%%%%%%%%%%%%%%%%
%%%%%%%%%%%%%%%%%%%%%%%%%%%%%%%%%%%%%%%%%%%%%%%%%%%%%%%%%%%%
\section{Expression of $\Phi_{ij}$} 
\label{appendix:source_terms_detail}

In Sec.~\ref{sec:corrections_from_stress}, we want to rewrite the Ricci tensor in Eqs.~\eqref{eq:R_ab} and \eqref{eq:R_ab_components} in terms of Weyl scalars, spin coefficients, and directional derivatives. Since $\gamma_{abc}$ is antisymmetric in the first two indices, it has $24$ independent components. These $24$ components can be further reduced to $14$ components using complex conjugation, which can then be expressed in terms of spin coefficients using the definition in Eq.~\eqref{eq:spin_coefs},
\begin{equation} \label{eq:Ricci_spin}
    \begin{aligned}
        & \gamma_{121}=\varepsilon+\bar{\varepsilon}\,,\;
        \gamma_{122}=\gamma+\bar{\gamma}\,,\;
        \gamma_{123}=\bar{\alpha}+\beta\,,\;
        \gamma_{131}=\kappa\,,\;\gamma_{132}=\tau\,,\;
        \gamma_{133}=\sigma\,,\;\gamma_{134}=\rho\,, \\
        & \gamma_{231}=-\bar{\pi}\,,\;\gamma_{232}=-\bar{\nu}\,,\;
        \gamma_{233}=-\bar{\lambda}\,,\;\gamma_{234}=-\bar{\mu}\,,\;
        \gamma_{341}=\bar{\varepsilon}-\varepsilon\,,\;
        \gamma_{342}=\bar{\gamma}-\gamma\,,\;
        \gamma_{343}=\bar{\alpha}-\beta\,,
    \end{aligned}
\end{equation}
For Riemann tensor or Weyl tensor, it is antisymmetric within its first pair and second pair of indices and symmetric under the exchange of the first and the second pair of indices, so the total number of independent components reduce to $21$ using these symmetries. Besides these symmetries, $C_{a[bcd]}=0$ and $C_{abcd}$ is traceless in the vacuum, which further give us the following relations in \cite{Chandrasekhar_1983},
\begin{equation}	
    \begin{aligned}
        & C_{1314}=C_{1323}=C_{1424}=C_{2324}=0\,, \\
        & C_{1334}=-C_{1213}\,,\;C_{2334}=C_{1223}\,,\;
        C_{3434}=C_{1212}\,,\;C_{1342}=\frac{1}{2}\left(C_{1212}-C_{1234}\right)\,,
    \end{aligned}
\end{equation}
which reduce the number of independent components of $C_{abcd}$ to $10$. These $10$ independent components can be represented by $5$ Weyl scalars and their conjugates. With complex conjugation, we can find all the components of the Weyl tensor using the symmetries above and the components below,
\begin{equation} \label{eq:Riemann_Weyl}
    \begin{aligned}
        & C_{1212}=\Psi_2+\bar{\Psi}_2\,,\;C_{1213}=\Psi_1\,,\;
        C_{1223}=-\bar{\Psi}_3\,,\;C_{1234}=\bar{\Psi}_2-\Psi_2\,,\;
        C_{1313}=\Psi_0\,,\;C_{2323}=\bar{\Psi}_4\,.
    \end{aligned}
\end{equation}

With Eqs.~\eqref{eq:R_ab} and \eqref{eq:R_ab_components} and \eqref{eq:Ricci_spin}--\eqref{eq:Riemann_Weyl}, we find Eq.~\eqref{eq:R_ab} in the NP basis to be
\begin{subequations} \label{eq:R_ab_detail}
\begin{align}
    & \begin{aligned} \label{eq:R_11}
        R_{11} 
        =& \;i\mathcal{R}_1\bigg\{(D\vartheta)\Big[\lambda\Psi_0-\bar{\lambda}
        \bar{\Psi}_0-(\alpha+\bar{\beta}+\pi)\Psi_1
        +(\bar{\alpha}+\beta+\bar{\pi})\bar{\Psi}_1
        +(\varepsilon+\bar{\varepsilon})(\Psi_2-\bar{\Psi}_2)\Big] \\
        & \;-(\boldsymbol{\Delta}\vartheta)\Big[\bar{\sigma}\Psi_0-\sigma\bar{\Psi}_0
        -\bar{\kappa}\Psi_1+\kappa\bar{\Psi}_1\Big] \\
        & \;+(\delta\vartheta)\Big[(\bar{\alpha}-\beta)\bar{\Psi}_0+\bar{\sigma}
        \Psi_1+(\varepsilon-\bar{\varepsilon}-\bar{\rho})\bar{\Psi}_1
        -\bar{\kappa}(\Psi_2-\bar{\Psi}_2)\Big] \\
        & \;-(\bar{\delta}\vartheta)\Big[(\alpha-\bar{\beta})\Psi_0-(\varepsilon
        -\bar{\varepsilon}+\rho)\Psi_1+\sigma\bar{\Psi}_1
        +\kappa(\Psi_2-\bar{\Psi}_2)\Big] \\
        & \;-\frac{1}{2}\Psi_0\{\bar{\delta},\bar{\delta}\}\vartheta
        +\frac{1}{2}\bar{\Psi}_0\{\delta,\delta\}\vartheta
        +\Psi_1\{D,\bar{\delta}\}\vartheta
        -\bar{\Psi}_1\{D,\delta\}\vartheta
        -(\Psi_2-\bar{\Psi}_2)\frac{1}{2}\{D,D\}\vartheta\bigg\}
        +\mathcal{R}_2(D\vartheta)(D\vartheta)\,,
    \end{aligned} \\
    & \begin{aligned} \label{eq:R_12}
        R_{12}
        =& \;\frac{i}{2}\mathcal{R}_1\bigg\{(D\vartheta)\Big[\nu\Psi_1-\bar{\nu}\bar{\Psi}_1
        -(\gamma+\bar{\gamma}+\mu+\bar{\mu})(\Psi_2-\bar{\Psi}_2)
        +\left(\bar{\alpha}+\beta+\bar{\pi}\right)\Psi_3
        -\left(\alpha+\bar{\beta}+\pi\right)\bar{\Psi}_3\Big] \\
        & \;-(\boldsymbol{\Delta}\vartheta)\Big[(\alpha+\bar{\beta}+\bar{\tau})\Psi_1
        -(\bar{\alpha}+\beta+\tau)\bar{\Psi}_1
        -(\varepsilon+\bar{\varepsilon}+\rho+\bar{\rho})(\Psi_2-\bar{\Psi}_2)
        +\kappa\Psi_3-\bar{\kappa}\bar{\Psi}_3\Big] \\
        & \;+(\delta\vartheta)\Big[\lambda\Psi_1
        -(\gamma-\bar{\gamma}+\mu)\bar{\Psi}_1-(\alpha-\bar{\beta}+\pi-\bar{\tau})
        (\Psi_2-\bar{\Psi}_2)+(\varepsilon-\bar{\varepsilon}-\bar{\rho})
        \Psi_3+\bar{\sigma}\bar{\Psi}_3\Big] \\
        & \;-(\bar{\delta}\vartheta)\Big[(\gamma-\bar{\gamma}-\bar{\mu})\Psi_1
        +\bar{\lambda}\bar{\Psi}_1+(\bar{\alpha}-\beta+\bar{\pi}-\tau)
        (\Psi_2-\bar{\Psi}_2)+\sigma\Psi_3-(\varepsilon-\bar{\varepsilon}
        +\rho)\bar{\Psi}_3\Big] \\
        & \;-\Psi_1\{\boldsymbol{\Delta},\bar{\delta}\}\vartheta
        +\bar{\Psi}_1\{\boldsymbol{\Delta},\delta\}\vartheta
        +(\Psi_2-\bar{\Psi}_2)\Big[\{D,\boldsymbol{\Delta}\}
        +\{\delta,\bar{\delta}\}\Big]\vartheta
        -\Psi_3\{D,\delta\}\vartheta+\bar{\Psi}_3\{D,\bar{\delta}\}\vartheta
        \bigg\}+\mathcal{R}_2(D\vartheta)(\boldsymbol{\Delta}\vartheta)\,,
    \end{aligned} \\
    & \begin{aligned} \label{eq:R_13}
        R_{13}
        =& \;\frac{i}{2}\mathcal{R}_1\bigg\{(D\vartheta)\Big[\nu\Psi_0
        -(\gamma+\bar{\gamma}+\mu+\bar{\mu})\Psi_1
        -2\bar{\lambda}\bar{\Psi}_1+(\bar{\alpha}+\beta+\bar{\pi})
        (\Psi_2+2\bar{\Psi}_2)-2(\varepsilon+\bar{\varepsilon})\bar{\Psi}_3\Big] \\
        & \;-(\boldsymbol{\Delta}\vartheta)\Big[(\alpha+\bar{\beta}+\bar{\tau})\Psi_0
        -(\varepsilon+\bar{\varepsilon}+\rho+\bar{\rho})\Psi_1
        -2\sigma\bar{\Psi}_1+\kappa(\Psi_2+2\bar{\Psi}_2)\Big] \\
        & \;+(\delta\vartheta)\Big[\lambda\Psi_0-(\alpha-\bar{\beta}
        +\pi-\bar{\tau})\Psi_1+2(\bar{\alpha}-\beta)\bar{\Psi}_1
        +(\varepsilon-\bar{\varepsilon}-\bar{\rho})
        (\Psi_2+2\bar{\Psi}_2)+2\bar{\kappa}\bar{\Psi}_3\Big] \\
        & \;-(\bar{\delta}\vartheta)\Big[(\gamma-\bar{\gamma}-\bar{\mu})\Psi_0
        +(\bar{\alpha}-\beta+\bar{\pi}-\tau)\Psi_1
        +\sigma(\Psi_2+2\bar{\Psi}_2)-2\kappa\bar{\Psi}_3\Big] \\
        & \;-\Psi_0\{\boldsymbol{\Delta},\bar{\delta}\}\vartheta
        +\Psi_1\Big[\{D,\boldsymbol{\Delta}\}+\{\delta,\bar{\delta}\}\Big]\vartheta
        +\bar{\Psi}_1\{\delta,\delta\}\vartheta
        -(\Psi_2+2\bar{\Psi}_2)\{D,\delta\}\vartheta
        +\bar{\Psi}_3\{D,D\}\vartheta\bigg\}
        +\mathcal{R}_2(D\vartheta)(\delta\vartheta)\,,
    \end{aligned} \\
    & \begin{aligned} \label{eq:R_22}
        R_{22}
        =& \;i\mathcal{R}_1\bigg\{-(D\vartheta)\Big[\bar{\nu}\Psi_3
        -\nu\bar{\Psi}_3-\bar{\lambda}\Psi_4+\lambda\bar{\Psi}_4\Big]
        -(\boldsymbol{\Delta}\vartheta)\Big[(\gamma+\bar{\gamma})(\Psi_2-\bar{\Psi}_2)
        -(\bar{\alpha}+\beta+\tau)\Psi_3 \\
        & \;+(\alpha+\bar{\beta}+\bar{\tau})\bar{\Psi}_3
        +\sigma\Psi_4-\bar{\sigma}\bar{\Psi}_4\Big]
        +(\delta\vartheta)\Big[\nu(\Psi_2-\bar{\Psi}_2)-(\gamma-\bar{\gamma}
        +\mu)\Psi_3+\lambda\bar{\Psi}_3+(\bar{\alpha}-\beta)\Psi_4\Big] \\
        & \;+(\bar{\delta}\vartheta)\Big[\bar{\nu}(\Psi_2-\bar{\Psi}_2)
        -\bar{\lambda}\Psi_3-(\gamma-\bar{\gamma}-\bar{\mu})\bar{\Psi}_3
        +(\alpha-\bar{\beta})\bar{\Psi}_4\Big] \\
        & \;-\frac{1}{2}(\Psi_2-\bar{\Psi}_2)\{\boldsymbol{\Delta},\boldsymbol{\Delta}\}\vartheta
        +\Psi_3\{\boldsymbol{\Delta},\delta\}\vartheta
        -\bar{\Psi}_3\{\boldsymbol{\Delta},\bar{\delta}\}\vartheta
        -\frac{1}{2}\Psi_4\{\delta,\delta\}\vartheta
        +\frac{1}{2}\bar{\Psi}_4\{\bar{\delta},\bar{\delta}\}\vartheta
        \bigg\}+\mathcal{R}_2(\boldsymbol{\Delta}\vartheta)(\boldsymbol{\Delta}\vartheta)\,,
    \end{aligned} \\
    & \begin{aligned} \label{eq:R_23}
        R_{23}
        =& \;\frac{i}{2}\mathcal{R}_1\bigg\{-(D\vartheta)\Big[\bar{\nu}
        (2\Psi_2+\bar{\Psi}_2)-2\bar{\lambda}\Psi_3
        -(\gamma+\bar{\gamma}+\mu+\bar{\mu})\bar{\Psi}_3
        +(\alpha+\bar{\beta}+\pi)\bar{\Psi}_4\Big] \\
        & \;-(\boldsymbol{\Delta}\vartheta)\Big[2(\gamma+\bar{\gamma})\Psi_1
        -(\bar{\alpha}+\beta+\tau)(2\Psi_2+\bar{\Psi}_2)+2\sigma\Psi_3
        +(\varepsilon+\bar{\varepsilon}+\rho+\bar{\rho})
        \bar{\Psi}_3-\bar{\kappa}\bar{\Psi}_4\Big] \\
        & \;+(\delta\vartheta)\Big[2\nu\Psi_1-(\gamma-\bar{\gamma}+\mu)
        (2\Psi_2+\bar{\Psi}_2)-2(\bar{\alpha}-\beta)\Psi_3
        +(\alpha-\bar{\beta}+\pi-\bar{\tau})\bar{\Psi}_3
        +\bar{\sigma}\bar{\Psi}_4\Big] \\
        & \;+(\bar{\delta}\vartheta)\Big[2\bar{\nu}\Psi_1-\bar{\lambda}
        (2\Psi_2+\bar{\Psi}_2)+(\bar{\alpha}-\beta+\bar{\pi}-\tau)\bar{\Psi}_3
        +(\varepsilon-\bar{\varepsilon}+\rho)\bar{\Psi}_4\Big] \\
        & \;-\Psi_1\{\boldsymbol{\Delta},\boldsymbol{\Delta}\}\vartheta+(2\Psi_2+\bar{\Psi}_2)
        \{\boldsymbol{\Delta},\delta\}\vartheta-\Psi_3\{\delta,\delta\}\vartheta
        -\bar{\Psi}_3\Big[\{D,\boldsymbol{\Delta}\}+\{\delta,\bar{\delta}\}\Big]
        \vartheta+\bar{\Psi}_4\{D,\bar{\delta}\}\vartheta\bigg\}
        +\mathcal{R}_2(\boldsymbol{\Delta}\vartheta)(\delta\vartheta)\,,
    \end{aligned} \\
    & \begin{aligned} \label{eq:R_33}
        R_{33}
        =& \;i\mathcal{R}_1\bigg\{-(D\vartheta)\Big[\bar{\nu}\Psi_1-\bar{\lambda}
        (\Psi_2-\bar{\Psi}_2)
        -(\bar{\alpha}+\beta+\bar{\pi})\bar{\Psi}_3 +(\varepsilon+\bar{\varepsilon})\bar{\Psi}_4\Big] \\
        & \;-(\boldsymbol{\Delta}\vartheta)\Big[(\gamma+\bar{\gamma})\Psi_0-(\bar{\alpha}+\beta
        +\tau)\Psi_1
        +\sigma(\Psi_2-\bar{\Psi}_2)+\kappa\bar{\Psi}_3\Big] \\
        & \;+(\delta\vartheta)\Big[\nu\Psi_0-(\gamma-\bar{\gamma}+\mu)\Psi_1
        -(\bar{\alpha}-\beta)(\Psi_2-\bar{\Psi}_2)
        +(\varepsilon-\bar{\varepsilon}-\bar{\rho})+\bar{\kappa}
        \bar{\Psi}_4\Big] \\
        & \;+(\bar{\delta}\vartheta)\Big[\bar{\nu}\Psi_0-\bar{\lambda}\Psi_1
        -\sigma\bar{\Psi}_3+\kappa\bar{\Psi}_4\Big] \\
        & \;-\frac{1}{2}\Psi_0\{\boldsymbol{\Delta},\boldsymbol{\Delta}\}\vartheta+\Psi_1
        \{\boldsymbol{\Delta},\delta\}\vartheta-\frac{1}{2}(\Psi_2-\bar{\Psi}_2)
        \{\delta,\delta\}\vartheta
        -\bar{\Psi}_3\{D,\delta\}\vartheta+
        \frac{1}{2}\bar{\Psi}_4\{D,D\}\vartheta\bigg\}
        +\mathcal{R}_2(\delta\vartheta)(\delta\vartheta)\,,
    \end{aligned} \\
    & \begin{aligned} \label{eq:R_34}
        R_{34}
        =& \;\frac{i}{2}\mathcal{R}_1\bigg\{(D\vartheta)\Big[\nu\Psi_1-\bar{\nu}\bar{\Psi}_1
        -(\gamma+\bar{\gamma}+\mu+\bar{\mu})(\Psi_2-\bar{\Psi}_2)
        +(\bar{\alpha}+\beta+\bar{\pi})\Psi_3
        -(\alpha+\bar{\beta}+\pi)\bar{\Psi}_3\Big] \\
        & \;-(\boldsymbol{\Delta}\vartheta)\Big[(\alpha+\bar{\beta}+\bar{\tau})\Psi_1
        -(\bar{\alpha}+\beta+\tau)\bar{\Psi}_1
        -(\varepsilon+\bar{\varepsilon}+\rho+\bar{\rho})
        (\Psi_2-\bar{\Psi}_2)+\kappa\Psi_3-\bar{\kappa}\bar{\Psi}_3\Big] \\
        & \;+(\delta\vartheta)\Big[\lambda\Psi_1-(\gamma-\bar{\gamma}+\mu)
        \bar{\Psi}_1-(\alpha-\bar{\beta}+\pi-\bar{\tau})
        (\Psi_2-\bar{\Psi}_2)+(\varepsilon-\bar{\varepsilon}
        -\bar{\rho})\Psi_3+\bar{\sigma}\bar{\Psi}_3\Big] \\
        & \;-(\bar{\delta}\vartheta)\Big[(\gamma-\bar{\gamma}-\bar{\mu})\Psi_1
        +\bar{\lambda}\bar{\Psi}_1+(\bar{\alpha}-\beta+\bar{\pi}-\tau)
        (\Psi_2-\bar{\Psi}_2)+\sigma\Psi_3-(\varepsilon-\bar{\varepsilon}
        +\rho)\bar{\Psi}_3\Big] \\
        & \;-\Psi_1\{\boldsymbol{\Delta},\bar{\delta}\}\vartheta+\bar{\Psi}_1\{\boldsymbol{\Delta},\delta\}
        \vartheta+(\Psi_2-\bar{\Psi}_2)\Big[\{D,\boldsymbol{\Delta}\}+\{\delta,\bar{\delta}\}
        \Big]\vartheta
        -\Psi_3\{D,\delta\}\vartheta+\bar{\Psi}_3\{D,\bar{\delta}\}
        \vartheta\bigg\}+\mathcal{R}_2(\delta\vartheta)(\bar{\delta}\vartheta)
        \,,
    \end{aligned}
\end{align}
\end{subequations}
where 
\begin{equation} \label{eq:R_coefs}
    \mathcal{R}_1\equiv-\left(\frac{1}{\kappa_g}
    \right)^{\frac{1}{2}}M^2\,,
    \quad\mathcal{R}_2\equiv\frac{1}{2\kappa_g\zeta}\,.
\end{equation}
and the remaining components of $R_{ab}$ can be found by exchanging the indices or complex conjugation. 

The Ricci NP scalars $\Phi_{ij}$ are related to the Ricci tensor via
\begin{equation} \label{eq:Ricci_NP_scalars}
    \begin{aligned}
        & \Phi_{00}=\frac{1}{2}R_{11}\,,\;
        \Phi_{01}=\frac{1}{2}R_{13}\,,\;
        \Phi_{02}=\frac{1}{2}R_{33}\,,\;
        \Phi_{10}=\frac{1}{2}R_{14}\,,\;
        \Phi_{11}=\frac{1}{4}(R_{12}+R_{34})\,,\; \\
        & \Phi_{12}=\frac{1}{2}R_{23}\,,\; 
        \Phi_{20}=\frac{1}{2}R_{44}\,,\;
        \Phi_{21}=\frac{1}{2}R_{24}\,,\;
        \Phi_{22}=\frac{1}{2}R_{22}\,,\;
        \Lambda=R/24 \,. 
    \end{aligned}
\end{equation}
Using the projection of the Ricci tensor onto the NP basis in Eq.~\eqref{eq:R_ab_detail}, the stationary scalar field in Eq.~\eqref{eq:scalar_stationary}, and the NP quantities in Schwarzschild [setting $\zeta=\chi=0$ in Eqs.~\eqref{eq:principal_tetrad} and \eqref{eq:spin_coefs_background}], we find the $\mathcal{O}(\zeta^1,\chi^{1},\epsilon^0)$ contributions to $\Phi_{ij}$
\begin{subequations} \label{eq:phi_ij_10}
    \begin{align}
        & \Phi_{00}^{(1,1,0)}=\Phi_{02}^{(1,1,0)}= \Phi_{11}^{(1,1,0)}= \Phi_{20}^{(1,1,0)}= \Phi_{22}^{(1,1,0)}=0 \,,\\
        & \Phi_{01}^{(1,1,0)}
        =\bar{\Phi}_{10}^{(1,1,0)}= -\frac{15iM^5\left(18 M^2+8 M r+3 r^2\right)\sin\theta}{16 \sqrt{2}r^9} \,, \\
        & \Phi_{12}^{(1,1,0)}
        =\bar{\Phi}_{21}^{(1,1,0)}
        = -\frac{15iM^5\left(r-2M\right)\left(18 M^2+8 M r+3 r^2\right)\sin\theta }{32\sqrt{2}r^{10}}\,,
    \end{align}
\end{subequations}
which are used in Sec.~\ref{sec:source_Teuk} to evaluate $S_{IA}^{(1,1,1)}$ and $S_{IIA}^{(1,1,1)}$. From Eq.~\eqref{eq:phi_ij_10}, one can also find $S_{1,2}^{(1,0)}$ to be
\begin{equation}
    S_{1}^{(1,1,0)}=-\frac{45i M^5(42M^2+16Mr+5r^2)\sin{\theta}}{16\sqrt{2}r^{10}}\,,\quad
    S_{2}^{(1,1,0)}=0\,.
\end{equation}

As discussed in Sec.~\ref{sec:source_Teuk}, to compute $\mathcal{S}_{IIB}$, one needs to evaluate $\Phi_{00}^{(1,1)}$, $\Phi_{01}^{(1,1)}$, and $\Phi_{00}^{(1,1)}$. When $\vartheta$ is stationary, since $\vartheta^{(1,0,0)}=0$, $\Phi_{ij}^{(1,0,1)}=0$, and we have only contributions from $\Phi_{ij}^{(1,1,1)}$ at $\mathcal{O}(\zeta^1,\chi^1,\epsilon^1)$. Based on our classifications in in Sec.~\ref{sec:source_Teuk}, we find the first type of contributions to be
\begin{subequations} \label{eq:Phi_A}
    \begin{align}
        & \Phi_{00,A}^{(1,1,1)}
        =-\frac{i\mathcal{R}_1}{2}
        \left[\Psi_0^{(0,0,1)}\left(\bar{\delta}^2
        +2\alpha\bar{\delta}\right)\vartheta^{(1,1,0)}
        -\bar{\Psi}_0^{(0,0,1)}\left(\delta^2
        +2\alpha\delta\right)\vartheta^{(1,1,0)}
        +(\Psi_2^{(0,0,1)}-\bar{\Psi}_2^{(0,0,1)})
        D^2\vartheta^{(1,1,0)}\right]\,, \\
        & \Phi_{01,A}^{(1,1,1)}
        =-\frac{i\mathcal{R}_1}{4}\left[
        \Psi_0^{(0,0,1)}\left(\{\boldsymbol{\Delta},\bar{\delta}\}
        -\mu\bar{\delta}\right)\vartheta^{(1,1,0)}
        +(\Psi_2^{(0,0,1)}+2\bar{\Psi}_2^{(0,0,1)})
        \left(\{D,\delta\}+\rho\delta\right)\vartheta^{(1,1,0)}\right]\,, \\
        & \Phi_{02,A}^{(1,1,1)}
        =-\frac{i\mathcal{R}_1}{2}\left[
        \Psi_0^{(0,0,1)}(\boldsymbol{\Delta}^2
        +2\gamma\boldsymbol{\Delta})\vartheta^{(1,1,0)}
        +(\Psi_2^{(0,0,1)}-\bar{\Psi}_2^{(0,0,1)})
        (\delta^2+2\alpha\delta)\vartheta^{(1,1,0)}
        -\bar{\Psi}_4^{(0,0,1)}D^2\vartheta^{(1,1,0)}\right]\,,
    \end{align}
\end{subequations}
where we used that in Schwarzschild,
\begin{equation} \label{eq:spin_coefs_Schw}
    \begin{aligned}
        \bar{\alpha}^{(0,0,0)}=\alpha^{(0,0,0)}=-\beta^{(0,0,0)}\,,\quad
        \bar{\rho}^{(0,0,0)}=\rho^{(0,0,0)}\,,\quad
        \bar{\mu}^{(0,0,0)}=\mu^{(0,0,0)}\,,\quad
        \bar{\gamma}^{(0,0,0)}=\gamma^{(0,0,0)}\,,
    \end{aligned}
\end{equation}
and other spin coefficients at $\mathcal{O}(\zeta^0,\chi^0,\epsilon^0)$ vanish with the gauge choice $\Psi_1^{(0,1)}=\Psi_3^{(0,1)}=0$. For simplicity, we have also dropped the superscripts of all the terms at $\mathcal{O}(\zeta^0,\epsilon^0)$. For the second type of contributions, we have
\begin{subequations} \label{eq:Phi_B}
    \begin{align}
        & \Phi_{00,B}^{(1,1,1)}=0\,, \\
        & \begin{aligned}
             \Phi_{01,B}^{(1,1,1)}
            =& \;\frac{3i\mathcal{R}_1}{8}\left\{
            2\left[(D+\rho)a^{(0,0,1)}+2(\gamma+\mu)b^{(0,0,1)}\right]D\vartheta^{(1,1,0)}
            +2(D-\rho)b^{(0,0,1)}\boldsymbol{\Delta}\vartheta^{(1,1,0)}\right. \\
            & \;\left.+2\left[2\alpha b^{(0,0,1)}
            +(\delta-4\alpha)\bar{b}^{(0,0,1)}\right]\delta\vartheta^{(1,1,0)}
            +\left[2(\delta+2\alpha)b^{(0,0,1)}+Dh_{mm}^{(0,0,1)}\right]\bar{\delta}
            \vartheta^{(1,1,0)}\right\}\,,
        \end{aligned} \\
        & \Phi_{02,B}^{(1,1,1)}=0\,,
    \end{align}
\end{subequations}
where we used that in Schwarzschild,
\begin{equation} \label{eq:Weyl_scalars_Schw}
    \Psi_{0,1,3,4}^{(0,0,0)}=0\,,\quad\bar{\Psi}_2^{(0,0,0)}=\Psi_2^{(0,0,0)}\,.
\end{equation}
The parameters $a^{(0,0,1)}$ and $b^{(0,0,1)}$ are rotation parameters given by Eq.~\eqref{eq:rotate_coefs_01}. For the third type of contributions, we find
\begin{subequations} \label{eq:Phi_C}
    \begin{align}
        & \Phi_{00,C}^{(1,1,1)}=0\,, \\
        & \begin{aligned}
            \Phi_{01,C}^{(1,1,1)}
            =& \;-\frac{3i\mathcal{R}_1\Psi_2}{8}\left\{
            2\left[(D+\rho)h_{nm}^{(0,0,1)}+(D+\rho)a^{(0,0,1)}\right]D\vartheta^{(1,1,0)}
            +2(D+\rho)b^{(0,0,1)}\boldsymbol{\Delta}\vartheta^{(1,1,0)}\right. \\
            & \;\left.+2\delta\bar{b}^{(0,0,1)}\delta\vartheta^{(1,1,0)}
            -\left[(D+\rho)h_{mm}^{(0,0,1)}-2\delta b^{(0,0,1)}\right]
            \bar{\delta}\vartheta^{(1,1,0)}\right. \\
            & \;\left.+4\left(h_{nm}^{(0,0,1)}+a^{(0,0,1)}\right)
            D^2\vartheta^{(1,1,0)} 
            +4\bar{b}^{(0,0,1)}\delta^2\vartheta^{(1,1,0)} \right.\\
            & \;\left.+2b^{(0,0,1)}\left(\{D,\boldsymbol{\Delta}\}
            +\{\delta,\bar{\delta}\}\right)\vartheta^{(1,1,0)}
            -h_{mm}^{(0,0,1)}\{D,\bar{\delta}\}\vartheta^{(1,1,0)}
            \right\}\,,
        \end{aligned} \\
        & \Phi_{02,C}^{(1,1,1)}=0\,,
    \end{align}
\end{subequations}
where we used Eqs.~\eqref{eq:spin_coefs_Schw} and \eqref{eq:Weyl_scalars_Schw}. 

For the last type of contributions, since $\vartheta^{(1,1)}$ can have both contributions from $\vartheta^{(1,0,1)}$ and $\vartheta^{(1,1,1)}$, the background metric we need is generally up to $\mathcal{O}(\zeta^0,\chi^1,\epsilon^0)$. This is also true for the operators converting $\Phi_{ij}$ to $\mathcal{S}$. For this reason, we will not expand the expression below explicitly in $\chi$ but do the expansion at the end when plugging in the coordinate-based values of the NP quantities. Then, at $\mathcal{O}(\zeta^1,\epsilon^1)$, we find
\begin{subequations} \label{eq:Phi_D}
    \begin{align}
        & \Phi_{00,D}^{(1,1)}
        =\mathcal{R}_{2}D\vartheta^{(1,0)}D\vartheta^{(1,1)}
        -\frac{i\mathcal{R}_1}{2}(\Psi_2-\bar{\Psi}_2)D^2\vartheta^{(1,1)}\,, \\ 
        & \begin{aligned}
            \Phi_{01,D}^{(1,1)}
            =& \;\frac{1}{4}\left[
            2\mathcal{R}_2(\delta\vartheta^{(1,0)}D
            +D\vartheta^{(1,0)}\delta)\vartheta^{(1,1)}
            +i\mathcal{R}_1(\bar{\alpha}+\beta+\bar{\pi})
            (\Psi_2+2\bar{\Psi}_2)D\vartheta^{(1,1)}
            -i\mathcal{R}_1(\Psi_2+2\bar{\Psi}_2)
            (\{D,\delta\}+\bar{\rho}\delta)\vartheta^{(1,1)}\right]\,,
        \end{aligned} \\
        & \Phi_{02,D}^{(1,1)}
        =\frac{1}{2}\left[2\mathcal{R}_2\delta\vartheta^{(1,0)}
        +i\mathcal{R}_1(\beta-\bar{\alpha})(\Psi_2-\bar{\Psi}_2)\right]
        \delta\vartheta^{(1,1)}
        -\mathcal{R}_1(\Psi_2-\bar{\Psi}_2)\delta^2\vartheta^{(1,1)}\,.
    \end{align}
\end{subequations}
Notice, since we do not do an expansion in $\chi$ above, Eq.~\eqref{eq:Phi_D} works for fully rotating BHs in dCS gravity. In addition, due to the expansion convention we used, where we have absorbed an additional $\zeta^{\frac{1}{2}}$ into the expansion of $\vartheta$, there are terms taking the form $\sim\mathcal{R}_2\vartheta^{(1,0)}\vartheta^{(1,1)}$ at $\mathcal{O}(\zeta^1,\epsilon^1)$. These terms come from the usual pseudoscalar action with minimal coupling. In our expansion convention, we have inserted $\zeta^{-1}$ before these terms [i.e., Eqs.~\eqref{eq:R_ab} and \eqref{eq:R_coefs}], so their contribution is still at $\mathcal{O}(\zeta^1,\epsilon^1)$. In the equations above, we have also dropped the superscript for terms at $\mathcal{O}(\zeta^0,\epsilon^0)$ for simplicity.

%%%%%%%%%%%%%%%%%%%%%%%%%%%%%%%%%%%%%%%%%%%%%%%%%%%%%%%%%%%%
%%%%%%%%%%%%%%%%%%%%%%%%%%%%%%%%%%%%%%%%%%%%%%%%%%%%%%%%%%%%
\section{An approach to compute projection coefficients in Eqs.~\eqref{eq:angular_coeffs} and \eqref{eq:angular_coeffs_conj}} 
\label{appendix:angular_projection}

In this section, we present an approach to compute the projection coefficients in Eqs.~\eqref{eq:angular_coeffs} and \eqref{eq:angular_coeffs_conj} using the series representation of spin-weighted spherical harmonics ${}_{s}Y_{\ell m}(\theta,\phi)$ in Eq.~\eqref{eq:spherical_formlua}. From Eq.~\eqref{eq:spherical_formlua}, we can see that the integrals in Eqs.~\eqref{eq:angular_coeffs} and \eqref{eq:angular_coeffs_conj} become a series sum over $q_1$ and $q_2$ of
\begin{align} \label{eq:projection_coeff}
    & \mathcal{Q}(a,b,c,d)\equiv\int d\theta\;\sin^{a}\left(\frac{\theta}{2}\right)
    \cot^{b}\left(\frac{\theta}{2}\right)\sin^{1+c}{\theta}\cos^d{\theta}\,, \\
    & a=2(\ell_1+\ell_2)\,,\nonumber\\
    & b=2(q_1+q_2)+s_1+s_2-m_1-m_2\,, \nonumber\\
    & c,d\in\{0,1\}\,,\nonumber
\end{align}
multiplied by the remaining constants dependent on $(s_1,\ell_1,m_1)$ and $(s_2,\ell_2,m_2)$ in Eq.~\eqref{eq:spherical_formlua}. The integral in Eq.~\eqref{eq:projection_coeff} can be evaluated analytically in terms of Gamma functions, i.e.,
\begin{subequations}
\begin{align}
    & \mathcal{Q}(a,b,0,0)
    =\frac{2\Gamma\left(1+\frac{a-b}{2}\right)
    \Gamma\left(1+\frac{b}{2}\right)}
    {\Gamma\left(2+\frac{a}{2}\right)}\,, \\
    & \mathcal{Q}(a,b,0,1)
    =\frac{(2b-a)\Gamma\left(1+\frac{a-b}{2}\right)
    \Gamma\left(1+\frac{b}{2}\right)}
    {\Gamma\left(3+\frac{a}{2}\right)}\,, \\
    & \mathcal{Q}(a,b,1,0)
    =\frac{4\Gamma\left(\frac{3+a-b}{2}\right)
    \Gamma\left(\frac{3+b}{2}\right)}
    {\Gamma\left(3+\frac{a}{2}\right)}\,,
\end{align}
\end{subequations}
so we can express the coefficients in Eqs.~\eqref{eq:angular_coeffs} and \eqref{eq:angular_coeffs_conj} as a series sum of Gamma functions, which are much faster to evaluate than direct integration for large $\ell_{1,2}$. More specifically, we get 
\begin{subequations}
\begin{align}
& \begin{aligned}
    \mathbf{\Lambda}^{\ell_1\ell_2 m}_{s_1 s_2}(\alpha,\beta)
    =& \;\frac{1}{2}\sqrt{\frac{(\ell_1+m)!(\ell_1-m)!(2\ell_1+1)}
    {(\ell_1+s_1)!(\ell_1-s_1)!}}
    \sqrt{\frac{(\ell_2+m)!(\ell_2-m)!(2\ell_2+1)}
    {(\ell_2+s_2)!(\ell_2-s_2)!}} \\
    & \;\times\sum_{q=0}^{\ell-s}\bigg[
    \left(\begin{array}{c}\ell_1-s_1 \\ q_1\end{array}\right)
    \left(\begin{array}{c}\ell_1+s_1 \\ q_1+s_1-m\end{array}\right)
    \left(\begin{array}{c}\ell_2-s_2 \\ q_2\end{array}\right)
    \left(\begin{array}{c}\ell_2+s_2 \\ q_2+s_2-m\end{array}\right) \\
    & \;(-1)^{(\ell_1+\ell_2-s_1-s_2+q_1+q_2)}
    \mathcal{Q}(2\ell_1+2\ell_2,2q_1+2q_2+s_1+s_2-2m,\alpha,\beta)\bigg]\,,
\end{aligned} \\
& \begin{aligned}
    \mathbf{\Lambda}^{\dagger\ell_1\ell_2 m}_{s_1 s_2}(\alpha,\beta)
    =& \;\frac{1}{2}\sqrt{\frac{(\ell_1+m)!(\ell_1-m)!(2\ell_1+1)}
    {(\ell_1+s_1)!(\ell_1-s_1)!}}
    \sqrt{\frac{(\ell_2-m)!(\ell_2+m)!(2\ell_2+1)}
    {(\ell_2+s_2)!(\ell_2-s_2)!}} \\
    & \;\times\sum_{q=0}^{\ell-s}\bigg[
    \left(\begin{array}{c}\ell_1-s_1 \\ q_1\end{array}\right)
    \left(\begin{array}{c}\ell_1+s_1 \\ q_1+s_1-m\end{array}\right)
    \left(\begin{array}{c}\ell_2-s_2 \\ q_2\end{array}\right)
    \left(\begin{array}{c}\ell_2+s_2 \\ q_2+s_2+m\end{array}\right) \\
    & \;(-1)^{(\ell_1+\ell_2-s_1-s_2+q_1+q_2)}
    \mathcal{Q}(2\ell_1+2\ell_2,2q_1+2q_2+s_1+s_2,\alpha,\beta)\bigg]\,,
\end{aligned}
\end{align}
\end{subequations}
where
\begin{subequations}
\begin{align}
    & \mathbf{\Lambda}^{\ell_1\ell_2 m}_{s_1 s_2}(0,0)
    =\Lambda^{\ell_1\ell_2 m}_{s_1 s_2}\,,\quad
    \mathbf{\Lambda}^{\ell_1\ell_2 m}_{s_1 s_2}(0,1)
    =\Lambda^{\ell_1\ell_2 m}_{s_1 s_2 c}\,,\quad
    \mathbf{\Lambda}^{\ell_1\ell_2 m}_{s_1 s_2}(1,0)
    =\Lambda^{\ell_1\ell_2 m}_{s_1 s_2 s}\,, \\
    & \mathbf{\Lambda}^{\dagger\ell_1\ell_2 m}_{s_1 s_2}(0,0)
    =\Lambda^{\dagger\ell_1\ell_2 m}_{s_1 s_2}\,,\quad
    \mathbf{\Lambda}^{\dagger\ell_1\ell_2 m}_{s_1 s_2}(0,1)
    =\Lambda^{\dagger\ell_1\ell_2 m}_{s_1 s_2 c}\,,\quad
    \mathbf{\Lambda}^{\dagger\ell_1\ell_2 m}_{s_1 s_2}(1,0)
    =\Lambda^{\dagger\ell_1\ell_2 m}_{s_1 s_2 s}\,.
\end{align}
\end{subequations}
We have provided this series-sum representation of the coefficients in Eqs.~\eqref{eq:angular_coeffs} and \eqref{eq:angular_coeffs_conj} in a Mathematica notebook as Supplementary Material~\cite{Pratikmodteuk}.
\twocolumngrid

\bibliographystyle{apsrev4-1}
\bibliography{reference}
	
\end{document}